\documentclass[nofootinbib,twocolumn,superscriptaddress,preprintnumbers,letterpaper,natbib,nofootinbib]{revtex4-1}
\pdfoutput = 1
\usepackage{amssymb}
\usepackage{amsmath}
\usepackage{epsfig}
\usepackage[usenames,dvipsnames]{xcolor}
\usepackage{hyperref}
\usepackage{comment}
\usepackage{cleveref}
\usepackage{multirow}
\usepackage{capt-of}
\usepackage{siunitx}
\usepackage{graphicx}
\usepackage{slashed}
\usepackage{tabularx}
\usepackage{enumitem}
\usepackage{multirow}
\usepackage{braket}
\usepackage{amsfonts}
\usepackage[compat=1.1.0]{tikz-feynman}
\tikzfeynmanset{warn luatex=false}
\usepackage{scalerel}
\usepackage[toc,page]{appendix}
\usepackage[skins,theorems,many]{tcolorbox}
\tcbset{highlight math style={enhanced,
  colframe=gray,colback=white,arc=2pt,boxrule=1pt}}
  
\usepackage[normalem]{ulem}

\hypersetup{
    colorlinks=true,       
    linkcolor=Cyan,          
    citecolor=Magenta}

\newcommand{\br}{\text{Br}(B\to \psi \, \mathcal{B} \,\mathcal{M})}
\newcommand{\Mpl}{M_{\mathrm{pl}}}
\newcommand{\diff}{\mathrm{d}}
\newcommand{\MeV}{\,\mathrm{MeV}}
\newcommand{\GeV}{\,\mathrm{GeV}}
\newcommand{\TeV}{\,\mathrm{TeV}}
\newcommand{\half}{\frac{1}{2}}
\newcommand{\Gammasc}{\Gamma_{{\mathrm{e}^{\pm}\! B}}}

\newcommand{\nocontentsline}[3]{}
\newcommand{\tocless}[2]{\bgroup\let\addcontentsline=\nocontentsline#1{#2}\egroup}

\makeatletter

\begin{document}

\title{
Collider Signals of Baryogenesis and Dark Matter from $B$ Mesons: \\ A Roadmap to Discovery
}

\author{Gonzalo Alonso-{\'A}lvarez}
\email{galonso@physics.mcgill.ca}
\thanks{ORCID: \href{https://orcid.org/0000-0002-5206-1177}{0000-0002-5206-1177}}

\affiliation{McGill University Department of Physics \& McGill Space Institute, 3600 Rue University, Montr\'eal, QC, H3A 2T8, Canada}
\affiliation{Institut f{\"u}r Theoretische Physik, Universit{\"a}t Heidelberg, Philosophenweg 16, 69120 Heidelberg, Germany}

\author{Gilly Elor}
\email{gelor@uw.edu}
\thanks{ORCID: \href{https://orcid.org/0000-0003-2716-0269}{0000-0003-2716-0269}}
\affiliation{Department of Physics, University of Washington, Seattle, WA 98195, USA}

\author{Miguel Escudero}
\email{miguel.escudero@tum.de}
\thanks{ORCID: \href{https://orcid.org/0000-0002-4487-8742}{0000-0002-4487-8742}}
\affiliation{Physik-Department, Technische Universit{\"{a}}t, M{\"{u}}nchen, James-Franck-Stra{\ss}e, 85748 Garching, Germany}

\begin{abstract} 

\hypersetup{linkcolor=black,urlcolor=black,citecolor=black}

\noindent Low-scale baryogenesis could be discovered at $B$-factories and the LHC. In the $B$-Mesogenesis paradigm [G. Elor, M. Escudero, and A. E. Nelson, \href{https://journals.aps.org/prd/abstract/10.1103/PhysRevD.99.035031}{PRD 99, 035031 (2019)}, \href{https://arxiv.org/abs/1810.00880}{arXiv:1810.00880}], the CP violating oscillations and subsequent decays of $B$ mesons in the early Universe simultaneously explain the origin of the baryonic and the dark matter of the Universe. This mechanism for baryo- and dark matter genesis from $B$ mesons gives rise to distinctive signals at collider experiments, which we scrutinize in this paper. We study CP violating observables in the $B^0_q-\bar{B}_q^0$ system, discuss current and expected sensitivities for the exotic decays of $B$ mesons into a visible baryon and missing energy, and explore the implications of direct searches for a TeV-scale colored scalar at the LHC and in meson-mixing observables. Remarkably, we conclude that a combination of measurements at BaBar, Belle, Belle II, LHCb, ATLAS and CMS can fully test $B$-Mesogenesis.

\end{abstract}

\preprint{TUM-HEP 1299/20}
  
\maketitle

{
  \hypersetup{linkcolor=black}
 \setlength\parskip{0.0pt}
  \setlength\parindent{0.0pt}
\tableofcontents
}

\hypersetup{linkcolor=Cyan,urlcolor=Magenta,citecolor=Magenta}

\section{Introduction}
The Standard Model of particle physics (SM) is a highly successful framework describing the known particles and their interactions, and it has been tested to great precision at collider experiments. 
Meanwhile, astrophysical and cosmological data are consistent with the standard cosmological model in which a very hot early Universe is preceded by a stage of inflationary expansion, describing the birth and evolution of our Universe.
Strikingly, these two highly successful models are in conflict with one another, leaving many unanswered questions and manifesting the need for physics beyond the Standard Model (BSM).

Precision measurements of the Cosmic Microwave Background (CMB) by the Planck satellite~\cite{planck} show that only about 16\% of the matter content of the Universe corresponds to known baryonic matter, while the remaining 84\% is due to yet undiscovered \emph{dark matter}. 
The SM does not contain a candidate dark matter particle~\cite{Bertone:2016nfn,Peebles:2017bzw}. Furthermore, a hot Big Bang governed only by SM physics leads to a Universe with equal parts of matter and antimatter~\cite{Kolb:1990vq,Gorbunov:2011zz}, thereby necessitating a dynamical mechanism, dubbed \emph{baryogenesis}, to generate a primordial asymmetry of baryonic matter over antimatter. 

The mysteries of dark matter and baryogenesis have been at the forefront of both the  experimental and theoretical particle physics programs for many years. 
The existence and properties of dark matter have been gravitationally inferred (for instance, the CMB anisotropies point towards a particle description of dark matter), but its actual nature is yet to be unveiled. Since the firm observational establishment of dark matter in the 1970's~\cite{Rubin:1970zza,Freeman:1970mx}, several well-motivated candidates have been proposed: weakly interacting massive particles (WIMPs)~\cite{Jungman:1995df,Bertone:2004pz}, axions~\cite{Preskill:1982cy,Dine:1982ah,Abbott:1982af}, sterile neutrinos~\cite{Dodelson:1993je,Shi:1998km,Adhikari:2016bei}, and primordial black holes~\cite{Carr:2016drx}, among others. On the experimental front, a multidisciplinary enterprise has been set forth to test these scenarios. Unfortunately, thus far no direct signal of dark matter has been found at laboratory experiments, particle colliders, or satellite missions. 

The baryon ($\mathcal{B}$) asymmetry of the Universe has been measured with sub-percent accuracy to be~\cite{planck} (see also~\cite{Mossa:2020gjc})
\begin{align}\label{eq:YBmeas}
    Y_{\mathcal{B}} = (n_{\mathcal{B}} - n_{\bar{\mathcal{B}}}) /s = (8.718\pm 0.004)\times 10^{-11}\,,
\end{align}
where $s$ is the entropy density of the Universe. The physical requirements to generate an asymmetry of matter over antimatter were outlined by Sakharov in the late 1960's~\cite{sakharov}, and since then many mechanisms of baryogenesis have been proposed~\cite{Dolgov:1991fr,Cohen:1993nk,Morrissey:2012db,Cline:2006ts,Riotto:1999yt,Dine:2003ax,Davidson:2008bu,Drewes:2017zyw}. However, they typically operate at very high energies in the early Universe, see e.g.~\cite{Dimopoulos:1978kv,Weinberg:1979bt,Affleck:1984fy,Fukugita:1986hr}, and are therefore notoriously difficult to test experimentally.

The standard lore, that baryogenesis is troublesome to confirm experimentally, has been challenged in recent years as potentially testable baryogenesis mechanisms have been put forward, see e.g.~\cite{Akhmedov:1998qx,Asaka:2005pn,Asaka:2005an,Babu:2006xc,Babu:2013yca,Aitken:2017wie,Nelson:2019fln,Elor:2020tkc}.
In this work, we concentrate on the mechanism proposed in~\cite{Elor:2018twp}, which achieves not only MeV-scale baryogenesis but also dark matter production.
This scenario relies on the CP-violating oscillations and subsequent decays of $B$ mesons into a dark sector in the early Universe\footnote{We refer to~\cite{Aitken:2017wie}, \cite{Nelson:2019fln}, and~\cite{Elor:2020tkc} for studies dealing with baryogenesis using heavy baryons, $B$ mesons oscillations, and charged $D$ meson decays, respectively. We also refer the reader to~\cite{Alonso-Alvarez:2019fym} for a Supersymmetric UV completion of the mechanism of~\cite{Elor:2018twp}.}, and as such we hereafter refer to this mechanism as \emph{$B$-Mesogenesis}\footnote{\footnotesize{The $B$ perfecter emphasizes that \emph{B}aryogenesis is achieved by leveraging the properties of $B$ mesons.}}.
The primary novelties of $B$-Mesogenesis, as envisioned in~\cite{Elor:2018twp}, are:
\begin{itemize}[leftmargin=0.7cm,itemsep=0.4pt]
    \item Contrary to typical baryogenesis scenarios, \newline $B$-Mesogenesis operates at very low temperatures, \newline $5\,\text{MeV} \lesssim T \lesssim 30\,\text{MeV}$.
    \item Baryon number of the Universe is actually \emph{conserved}. This is achieved by linking baryogenesis to dark matter, which is charged under baryon number.
    \item The baryon asymmetry is directly related to $B$-meson observables, and as such there are unique signatures at current hadron colliders and $B$ factories.
\end{itemize}

\begin{figure*}[t]
\centering
\includegraphics[width=0.9\textwidth]{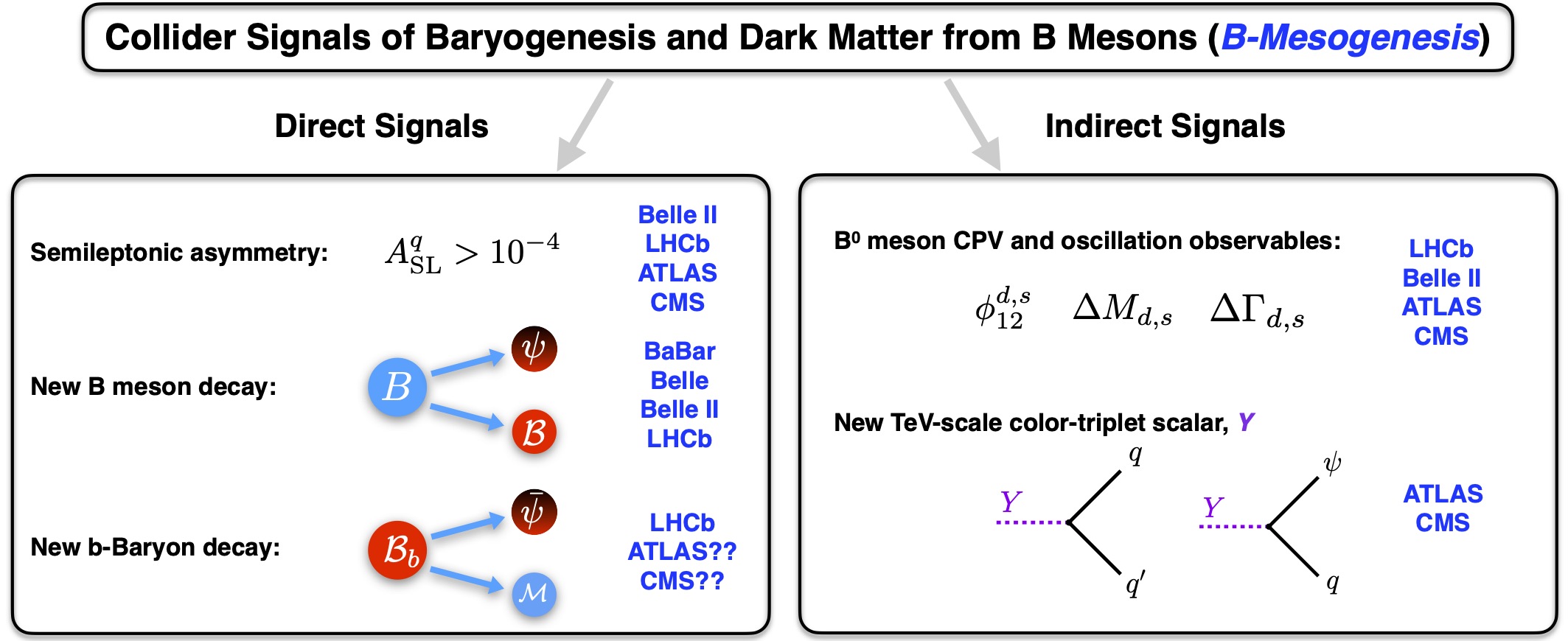}
\vspace{-0.2cm}
\caption{Summary of the collider implications of baryogenesis and dark matter from $B$ mesons~\cite{Elor:2018twp}, i.e. \textit{B-Mesogenesis}. The distinctive signals of the mechanism are: \textit{i)} the requirement that at least one of the semileptonic (CP) asymmetries in $B_q^0$ decays is $A_{\rm SL}^q > 10^{-4}$, \textit{ii)} that both neutral and charged $B$ mesons decay into a dark sector antibaryon (appearing as missing energy in the detector), a visible baryon, and any number of light mesons with $\br > 10^{-4}$, \textit{iii)} that $b$-flavored baryons should decay into light mesons and missing energy at a rate $\text{Br}(\mathcal{B}_b \to \bar{\psi} \, \mathcal{M}) > 10^{-4}$. In addition, we include as indirect signals the various oscillation observables in the $B_q^0-\bar{B}_q^0$ system as they are linked to $A_{\rm SL}^q$, and the presence of a new TeV-scale color-triplet scalar $Y$ that is needed to trigger the $B\to \psi \, \mathcal{B} \,\mathcal{M}$ decay. We also highlight the existing experiments that can probe each corresponding signal.\\
\textit{Notation:} $B:$ $B$ meson, $\mathcal{B}:$ SM baryon, $\mathcal{M}:$ any number of light mesons, $\psi:$ dark sector antibaryon (ME in the detector).}
\label{fig:Summary}
\end{figure*}

As briefly discussed in~\cite{Elor:2018twp} and as extensively explored in this work, there are three robust predictions that make the key elements of $B$-Mesogenesis testable at current experimental facilities: 
\begin{enumerate}[leftmargin=0.5cm,itemsep=0.4pt]
\item Both neutral and charged $B$ mesons should have a large branching ratio
\begin{align}
\br  \,\, >\,\,  10^{-4}
\label{eq:brfirst}
\end{align}
into a final state containing a SM baryon $\mathcal{B}$, missing energy in the form of a dark sector antibaryon $\psi$, and any number of light mesons denoted here by $\mathcal{M}$.
\item $b$-flavored baryons ($\mathcal{B}_b$) should decay into a dark baryon and mesons at a rate
\begin{align}
\text{Br}(\mathcal{B}_b \to \bar{\psi} \, \mathcal{M}) \,\, >\,\, 10^{-4}\,,
\end{align}
where again, $\mathcal{M}$ accounts for any number of light mesons and $\bar{\psi}$ would appear as missing energy in the detector. 
\item The CP violation in the neutral $B$ meson system should be such that:
\begin{align}
A_{\rm SL}^q > 10^{-4} \,,
\end{align}
where $A_{\rm SL}^q$ is the semileptonic charge asymmetry in $B_q^0-\bar{B}_q^0$ decays.
\end{enumerate}
Accompanying these unique signatures, there are two other key set of observables that are useful to indirectly constrain $B$-Mesogenesis:
\begin{enumerate}[leftmargin=0.5cm,itemsep=0.4pt]
\item The parameters describing $B_q^0-\bar{B}_q^0$ oscillations
\begin{align}
\phi_{12}^{d,s}\, , \quad  \Delta \Gamma_{d,s}\,,\quad \mathrm{and}\quad \Delta M_{d,s}\,.
\end{align}
Here and as usual, $\phi_{12}^{d,s}$ denotes the CP violating phase in $B_{d,s}^0$ oscillations, and $\Delta M_{d,s}$ and $\Delta \Gamma_{d,s}$ are the mass and width differences of the physical $B^0_q$ states. A combination of these quantities determines $A_{\rm SL}^{d,s}$ and therefore their measurements indirectly constrain the mechanism.

\item Resonant jets and jets plus missing energy transfer (MET) at TeV energies.
A color-triplet scalar, $Y$, is needed in order to trigger the exotic $B$ meson decay into a baryon and missing energy in Eq.~\eqref{eq:brfirst}. Such a scalar would also have implications for  heavy-resonance searches at the LHC.
\end{enumerate}
 In Fig.~\ref{fig:Summary}, we summarize each of these direct and indirect signals, as well as highlight the key collider experiments that can target each of them. We emphasize that these signals are general features of $B$-Mesogenesis, independent of the details of a UV model. Given a model that realizes this mechanism, there will likely be many other complementary probes (see e.g. \cite{Alonso-Alvarez:2019fym}).

Surprisingly, there have been no experimental searches for decay modes of the type $B\to \psi \, \mathcal{B}$ for which baryon number is apparently violated\footnote{There are however already several searches in development using BaBar~\cite{privateBaBarBR} and Belle/Belle II data~\cite{privateBelleBR}, and also feasibility studies at LHCb~\cite{Rodriguez:2021urv}.}. At present, a bound on such process at the $\br \lesssim 10\,\%$ level arises from inclusive decay measurements of $B$ mesons (see Sec.~\ref{sec:BR}). In addition, searches for large missing energy events from b-quark decays at LEP can be used to set constraints $\br \lesssim 10^{-4}-10^{-3}$ across some regions of parameter space. Otherwise, the current lack of dedicated searches for this $B$ meson decay mode renders $B$-Mesogenesis relatively unconstrained at present.
Given that $B$ factories have reached sensitivities of order $10^{-5}$ for exclusive decay modes involving missing-energy final states, such as $B \to K \bar{\nu}\nu$, we expect a substantial improvement on the measurement of $\br$ once this decay mode is targeted. Our estimates indicate that BaBar and Belle should be able to test large regions of the relevant parameter space, while we expect that Belle II and LHCb could be able to fully test the mechanism by searching for these processes.

$B$-Mesogenesis directly relates the matter-antimatter asymmetry of the Universe to the CP violation in the neutral $B_d^0$ and $B_s^0$ meson mixing systems. Although many BSM scenarios can lead to non-standard CP violation in the $B$ meson system, see e.g.~\cite{Artuso:2015swg}, prior to the work of~\cite{Elor:2018twp}, there existed no mechanisms that could directly connect such CP violation to the baryon asymmetry of the Universe. Therefore, $B$-Mesogenesis makes current and upcoming measurements of CP violation in the neutral $B$ meson system not only a powerful probe of BSM physics but also a potential test of the physics of baryogenesis.

Additionally and as discussed above, $B$-Mesogenesis requires the existence of a new bosonic colored mediator in order for $B$ mesons to decay into a baryon and missing energy. Thus, searches for heavy colored scalars at ATLAS and CMS lead to relevant implications for the mechanism. In particular, multi-jet and jet plus missing energy searches at the LHC have a direct connection to $\br$.

Given the exciting possibility of generating baryogenesis and dark matter from $B$ mesons and the potential for $B$-Mesogenesis to be tested at hadron colliders and $B$ factories, in this work we set up an enterprise to shape the experimental signatures of  the mechanism  proposed in~\cite{Elor:2018twp}. 
In particular, we study the reach of current and upcoming collider experiments to the new decay mode $B\to \psi \, \mathcal{B} \, \mathcal{M}$, the implications from CP violation measurements in the $B$ meson system, and the phenomenology of TeV-scale color-triplet scalars. The conclusion of this paper is that $B$-Mesogenesis could be fully confirmed at current hadron colliders and $B$ factories.
It is our intention for this work to provide a roadmap for experimental efforts directed to uncovering the mechanism responsible for baryogenesis and dark matter production.

\begin{figure*}[t]
\centering
\includegraphics[width=0.75\textwidth]{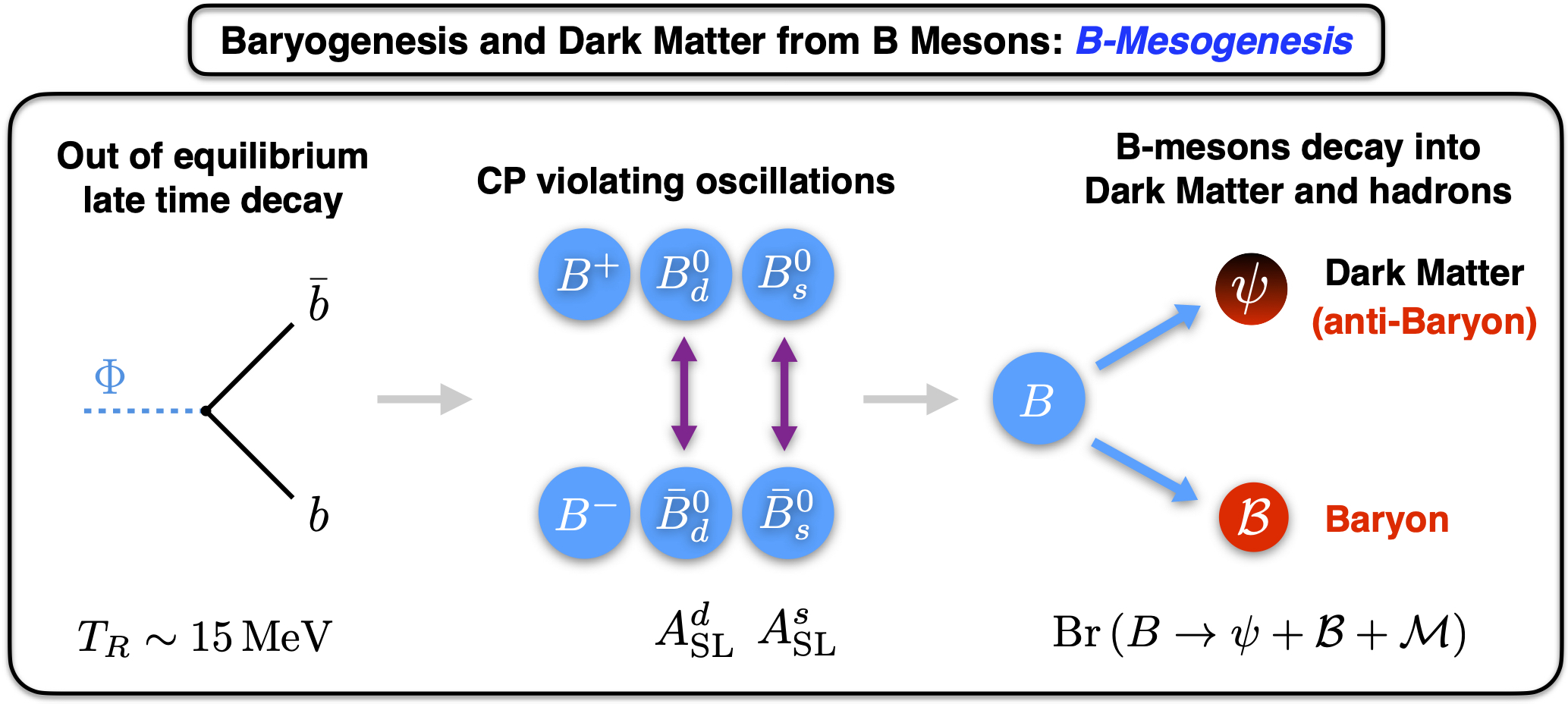}
\vspace{-0.2cm}
\caption{Summary of the mechanism of baryogenesis and dark matter from $B$ mesons~\cite{Elor:2018twp}, i.e. \textit{B-Mesogenesis}. Adapted from Figure 1 in~\cite{Elor:2018twp}.}
\label{fig:BMesogenesis}
\end{figure*}

This paper is organized as follows: in Sec.~\ref{sec:BaryogenesisnandDM} we begin by reviewing the key ingredients and features of the $B$-Mesogenesis mechanism, including an updated calculation of the early-Universe dynamics that allow us to refine the predictions for $B$-meson observables. Sec.~\ref{sec:CP_violation} is devoted to the study of the implications of current and upcoming measurements of CP violation in mixing in $B_d^0$ and $B_s^0$ mesons. In particular, we use these measurements to set a theoretical lower bound on $\br$. In Sec.~\ref{sec:BR}, we review the current experimental limits on $B\to \psi \, \mathcal{B}\,\mathcal{M}$ decays and comment on the prospects for $B$ factories and LHC experiments. Next, in Sec.~\ref{sec:TripletScalar} we consider the various collider implications of the new bosonic mediator needed to trigger the new decay mode $B\to \psi\, \mathcal{B}\,\mathcal{M}$, including dijet and jet plus MET signatures of color-triplet scalars, as well as flavor mixing constraints. After discussing the interplay of the different searches in Sec.~\ref{sec:global}, we conclude in Sec.~\ref{sec:conclusions} by summarizing our main results and outlining various avenues for future work.

\vspace{-0.2cm}

\section{\texorpdfstring{$B$}{B}-Mesogenesis}\label{sec:BaryogenesisnandDM}

\vspace{-0.15cm}

In this section, we explore how allowing $B$ mesons to decay into a baryon and dark matter can lead to the generation of the baryon asymmetry and the dark matter of the Universe. We begin by  briefly reviewing the  mechanism presented in~\cite{Elor:2018twp}, while referring to Appendix~\ref{sec:decoherence} for a detailed discussion of the cosmological dynamics.
We then continue by describing the new particles and interactions necessary for the mechanism, with the goal of finding theory-motivated benchmark points for experimental searches.
This includes the minimal particle content required to trigger the new decay mode of $B$ mesons in all its possible flavorful variations.
We refer to~\cite{Elor:2018twp} for more detailed and alternative discussions. 

\subsection{The Mechanism in a Nutshell}\label{sec:BaryogenesisnandDMmech}
In the late 1960's, Sakharov outlined the three conditions that must be satisfied in the early Universe in order for it to evolve into what we observe today: a Universe in which matter dominates over antimatter.
These \emph{Sakharov conditions}~\cite{sakharov} are: 
\textit{(i) C and CP violation}.
Matter and antimatter need to interact differently if an excess of one over the other is to be generated.
\textit{(ii) Departure from thermal equilibrium}.
In thermal equilibrium, even if C and CP are violated, the rates of particle and antiparticle production and destruction are equal. Thus, the system must be out of thermal equilibrium so that the rate of producing particles is larger than that of antiparticles.
\textit{(iii) Baryon number violation}.
One requires interactions which violate baryon number if an excess of baryons over antibaryons is to be generated in the early Universe.

The mechanism of $B$-Mesogenesis addresses each of the Sakharov conditions as follows: 
 \begin{enumerate}[label={\itshape(\roman*)},leftmargin=0.7cm,itemsep=0.4pt]
    \item 
    \textit{C and CP violation}.
     The most novel feature of~\cite{Elor:2018twp} is to leverage the C and CP violation within the oscillations in the SM neutral $B$ meson systems.
     
    \item \textit{Departure from thermal equilibrium}. It is assumed that the early Universe is dominated by a combination of radiation and a very weakly coupled scalar particle, $\Phi$, with mass $M_\Phi \gtrsim 11\,\text{GeV}$ and a lifetime of $\tau_{\Phi} \sim 10^{-3}\,\text{s}$\footnote{These values are roughly those expected for a pseudo-Goldstone boson with Planck suppressed interactions with matter, see e.g.~\cite{Nilles:1983ge}. Note that these values imply that it is hopeless to produce the $\Phi$ particle at colliders since the coupling to matter would be $\lesssim 10^{-10}$.}. The $\Phi$ particle predominantly decays into $b$ quarks, and in doing so reheats the Universe at temperatures $T_{\rm R} \sim 20 \, \text{MeV}$. Given that $T < T_{\rm QCD} \sim 200 \, \text{MeV}$ during this era, the produced $b$ quarks quickly hadronize to yield a substantial population of $B$ mesons in the early Universe: $n_B/n_\gamma \sim 10  \! \, \times \,  \! T_{\rm R}/M_{\Phi} \sim 0.02$. This population of $B$ mesons is in this way out of thermal equilibrium.
     
    \item \textit{Baryon number violation?}
    In this set up, $B$ mesons posses a nonstandard decay channel into a dark sector antibaryon ($\psi$) and a SM baryon: $B\to \psi \, \mathcal{B} \, \mathcal{M}$. This results in the generation of a baryon asymmetry in the visible sector that is exactly compensated by a dark antibaryon asymmetry. As a consequence, total baryon number is actually conserved\footnote{In a similar fashion to~\cite{Davoudiasl:2010am}.}.
\end{enumerate}
In summary, the late time out of equilibrium $B^0_q$ and $\bar{B}^0_q$ production, oscillation and subsequent decay into $\psi$ and $\mathcal{B}$, results in the generation of an excess of baryons in the visible sector and an excess of antibaryons in the dark sector. In this way, the origin of baryogenesis and that of dark matter are linked.
 
Importantly, following the chain of events described above and as depicted in Fig.~\ref{fig:BMesogenesis}, it is possible to show that the observed baryon abundance today (see Appendix~\ref{sec:decoherence}) can be directly related to two observables at collider experiments: 
\begin{equation}
\tcbhighmath[boxrule=1.3pt,drop fuzzy shadow=black] {\!\!\!\!\!\!\!\! \hspace{0.08 in} Y_\mathcal{B} \simeq 8.7\times 10^{-11} \, \frac{\br}{10^{-3}} \sum_q \alpha_q \, \frac{A^q_{\rm SL}}{10^{-3}} \, , \,\,\,\, \!\!\!\!\!\!\!\!\!\!}
\label{eq:basypara}
\end{equation}
\noindent
where we have normalized to the observed baryon asymmetry given in Eq.~\eqref{eq:YBmeas}.
 
In Eq.~\eqref{eq:basypara}, $\br$ is the inclusive branching ratio of $B$ mesons into a dark antibaryon, a visible baryon, and any number of light mesons; and $A_{\rm SL}^q$ are the semileptonic asymmetries in neutral $B_q^0$ meson decays, which measure the amount of CP violation in mixing in the $B_q^0$ systems. Finally, $\alpha_{q}$ are functions that encode the dependence on the mass and lifetime of the $\Phi$ field. They are bounded to be $0 \le \alpha_q \le 1.4$, see Fig.~\ref{fig:alpha_ds} in Appendix~\ref{sec:decoherence}. 

Although all relevant cosmological details are present in~\cite{Elor:2018twp} and detailed in Appendix~\ref{sec:decoherence}, some comments on Eq.~\eqref{eq:basypara} are in order. First, the $\alpha_q$ parameters are different for each $B_q$ meson. The physical reason is that the early Universe is a hot dense plasma in which electrons can interact with $B_q^0$ mesons and potentially decohere the $B_q^0-\bar{B}_q^0$ oscillations. Given that $B_s^0$ mesons oscillate $\sim 35$ times faster than $B_d^0$ mesons, they are more resilient against decoherence effects and $\alpha_s \gg \alpha_d$ for temperatures in which decoherence is important. In addition and although not explicitly stated in~\cite{Elor:2018twp}, it is clear that the baryon asymmetry generated by $B$-Mesogenesis is only sourced by the CP violation in the $B_q^0$ oscillation system and is totally independent of any potential CP violation in the $\Phi\, \bar{b}\, b$ coupling illustrated in Fig.~\ref{fig:BMesogenesis}. The reason is as follows: at temperatures $T<T_{\rm QCD}$, a CP asymmetry in the $B_q^0$ system appears at 1 loop since $B$ mesons are hadronized, while the CP violation arising from a CP violating $\Phi\, \bar{b}\, b$ coupling appears only at the 3-loop level, see~\cite{Ellis:1978xg,Barr:1979ye}. Given that the $\Phi \, \bar{b}\,b$ coupling is $< 10^{-10}$, any contribution to the baryon asymmetry arising from such a 3-loop process is extremely suppressed and can safely be neglected.

Eq.~\eqref{eq:basypara} clearly shows the direct connection that exists between baryogenesis, the CP violation in the $B_q^0$ system parametrized by $A_{\rm SL}^q$, and the branching fraction for the new decay $\br$. In light of this relation, current measurements of CP violation in the $B_q^0$ meson systems can constrain $B$-Mesogenesis, as we illustrate in Sec.~\ref{sec:CP_violation}. Furthermore, the aforementioned exotic decay mode of $B$ mesons has important implications for $B$ factories and the LHC, which we discuss in Sec.~\ref{sec:BR}.

Before proceeding to analyze this in detail, it is necessary to address two important points regarding the new particle content of the model, namely: \textit{i)} what are the minimal requirements of the dark sector, and \textit{ii)} what additional particles are needed to trigger the new decay mode $B\to \psi \, \mathcal{B}\,\mathcal{M}$. 

\subsection{The Dark Sector}
A key element in $B$-Mesogenesis is that $B$ mesons can decay into a SM (visible) baryon and a dark sector antibaryon ($\psi$). However, this antibaryon $\psi$ cannot represent the dark matter of the Universe. The reason for this is that the decay $B \to \psi\,\mathcal{B}$ can only proceed if $\psi$ interacts with 3 quarks. The same operator that mediates such an interaction can also allow the GeV-scale $\psi$ to decay into light antibaryons, thereby washing out the generated asymmetry. Thus, successful baryogenesis demands that the $\psi$ state must rapidly decay into other stable dark sector particles. This requirement is trivially met provided the existence of two additional states coupled to the dark fermionic antibaryon: a SM singlet scalar antibaryon $\phi$, and a SM singlet Majorana fermion $\xi$, coupling to $\psi$ via the Yukawa interaction
\begin{equation}
{\cal L}\,\, \supset \,\, -y_d\bar\psi \phi \xi +{\rm h.c.}\,.
\label{eq:Ldarksector}
\end{equation}
To write this Lagrangian, we have assumed the existence of a $\mathbb{Z}_2$ symmetry that stabilizes both $\phi$ and $\xi$. 

Given the interaction in Eq.~\eqref{eq:Ldarksector}, and provided that $m_\psi > m_\phi + m_\xi$, $\psi$ rapidly decays into $\phi$ and $\xi$ states in the early Universe. In this way, a combination of number densities of $\phi$ and $\xi$ particles can be such that their present-day abundance matches the dark matter density measured by the Planck satellite, $\Omega_{\rm DM} h^2 = 0.1200\pm0.0012$~\cite{planck}.

With the knowledge that the dark sector is made up of at least 3 states (the dark Dirac antibaryon $\psi$, the dark scalar baryon $\phi$, and the dark Majorana fermion $\xi$), relevant limits on the mass of these particles can be set given the requirements of baryogenesis in conjunction with several other kinematic constraints. First, the $B\to \psi\, \mathcal{B} $ decay can only occur if
\begin{align}
m_\psi < m_B - m_p \simeq 4.34 \,\text{GeV}\,.
\end{align}
Second, proton stability requires
\begin{align}
m_\psi > m_p - m_e \simeq 937.8 \,\text{MeV} \,.
\end{align}
Third, the $\phi$ and $\xi$ states need to be both stable. Otherwise, they could decay into each other by emitting an antiproton and an electron which would erase the generated baryon asymmetry from the $B$ meson oscillations and decays. In addition to the $\mathbb{Z}_2$ symmetry, this requirement implies
\begin{align}\label{eq:stability}
|m_\xi - m_\phi | < m_p + m_e \simeq 938.8 \,\text{MeV} \,.
\end{align}
Finally, allowing for the $\psi \to \xi\phi$ decay requires
\begin{align}\label{eq:DSkin}
m_\psi > m_\phi + m_\xi \,.
\end{align}
In light of the above constraints, we can conclude that the mass of the dark antibaryon $\psi$ needs to lie within the range
\begin{equation}
\tcbhighmath[boxrule=1.3pt,drop fuzzy shadow=black]{0.94\,\text{GeV} < m_\psi  < 4.34\,\text{GeV}.}
\label{eq:psirange}
\end{equation}

We note that since both $\psi$ and $\phi$ carry baryon number, they can potentially be produced in high-density environments, even if they interact very weakly with regular matter.
One of the most favourable environments for this is the interior of a neutron star, where light baryons can be produced provided that they are lighter than the chemical potential of a neutron $\mu_n \sim 1.2\,\text{GeV}$ within the star~\cite{McKeen:2018xwc} (see also~\cite{McKeen:2020oyr,Berezhiani:2020zck,Ellis:2018bkr,Motta:2018rxp,Baym:2018ljz,Goldman:2013qla}). In~\cite{Elor:2018twp}, this motivated the additional restriction to only consider masses above this threshold, i.e. $m_\psi > m_\phi > 1.2\,\text{GeV}$ at face value. However, there is in fact no dedicated study of these neutron stars constraints on light baryons like the ones considered here, which may not have interactions involving only light quarks. Furthermore, these bounds are subject to several uncertainties, such as the unknown equation of state of neutron stars. In this situation, we opt to take the conservative approach of not applying this further restriction to the masses of $\psi$ and $\phi$ in this work.
That said, we note that the regime of light mass $m_\psi \lesssim 1.5\,\text{GeV}$ may be subject to these astrophysical bounds once a dedicated study addressing the aforementioned uncertainties is performed. 

An important observation at this stage is that the energy density in asymmetric dark matter that is produced through $B$-Mesogenesis cannot be larger than $m_\psi/m_p\times \Omega_\mathcal{B}$.
In light of Eq.~\eqref{eq:stability}, this means that the full observed dark matter abundance cannot be generated solely via this process.
That said, other processes such as annihilation freeze-out can yield the correct amount of dark matter. In fact, and as discussed in~\cite{Elor:2018twp}, dark sector interactions are crucial in order to allow the annihilation of any symmetric abundance of dark sector particles that would result in an overproduction of dark matter. The details of these processes depend on the exact matter content and dynamics of the dark sector. If we restrict the dark matter to be comprised of $\phi$ and $\xi$ particles, then there could also be additional detection signals of the dark matter. For instance, the dark matter could annihilate into SM neutrinos, as can occur in supersymmetric versions of the mechanism~\cite{Alonso-Alvarez:2019fym}. Although this annihilation channel is at present unconstrained~\cite{Frankiewicz:2017trk,Campo:2017nwh}, it could be tested in upcoming neutrino experiments such as Hyper-Kamiokande~\cite{Bell:2020rkw,Arguelles:2019ouk}. Alternatively, the dark matter could annihilate into sterile neutrinos. If the dark sector states are much heavier than the sterile neutrinos, annihilation would be p-wave~\cite{Escudero:2016ksa,Escudero:2016tzx} and would not yield any relevant signals for CMB or neutrino experiments. Finally, as was discussed in~\cite{Elor:2018twp} (see also~\cite{Davoudiasl:2011fj}), the dark matter particles considered here do not yield typical WIMP-like scattering signals at underground laboratories, as the dark matter states $\xi$ and $\phi$ do not interact directly with SM fermions.

Of course, $\psi$, $\phi$, and $\xi$ need not be the only dark sector states. Indeed, one may expect many other particles to be present in a full realization of the dark sector. A theoretically motivated example of an additional dark species is a SM gauge singlet $\mathcal{A}$ carrying the opposite baryon number of a quark, i.e. $-1/3$~\cite{Elor:2018twp}. The $\mathcal{A}$ state can be stabilized via a $\mathbb{Z}_2$ symmetry and represent the entirety of the dark matter and could provide an explanation for why the dark matter and baryon energy densities are observed to be so similar, $\Omega_{\rm DM}/\Omega_{\mathcal B} \simeq 5$. Since baryon number is conserved in our setup, $n_\mathcal{A} = 3n_\mathcal{B}$ if $\mathcal{A}$ is the only stable dark sector antibaryon. With this, assuming that the asymmetric $\mathcal{A}$ population makes up the entirety of the dark matter, we can solve for $m_\mathcal{A}$ to find
\begin{align}
m_\mathcal{A} =\frac{m_{\mathcal{B}}}{3}  \frac{\Omega_{\rm DM} h^2}{\Omega_\mathcal{B} h^2} = 1680\pm 20\,\text{MeV},
\end{align}
where $m_{\mathcal{B}}$ is the average mass per baryon~\cite{Steigman:2006nf}, and where we have used the current errors from Planck~\cite{planck} on $\Omega_{\rm DM} h^2$ and $\Omega_\mathcal{B} h^2$.

$\mathcal{A}$ scalars can be produced in the early Universe via interactions of the type $\phi +\phi^\star \to \mathcal{A}+\mathcal{A}^\star$, while the asymmetry in the dark sector can be transferred via processes of the type $\phi +\mathcal{A}^\star \to \mathcal{A} +\mathcal{A}$. For these processes to  be active in the early Universe requires $m_\phi > m_\mathcal{A}$. Since $m_\psi > m_\phi > m_\mathcal{A}$, then for masses of $m_\psi > 1.7\,\text{GeV}$ $B$-Mesogenesis with this extended dark sector could provide an explanation for the observed dark matter-to-baryon density ratio. 

Motivated by the above discussion, we suggest adopting the following benchmark value for the mass of $\psi$:
\begin{equation}
m_\psi  = 2 \,\text{GeV} \,\, (\text{Benchmark}) \,.
\label{eq:Benchmark}
\end{equation}
\noindent For this benchmark, the dark matter could be fully composed of antibaryons $\mathcal{A}$ with baryon number $-1/3$,  thereby providing an explanation for the observed ratio $\Omega_{\rm DM}/\Omega_{\mathcal{B}} \simeq 5$. Needless to say, while this benchmark is particularly theoretically appealing, the entire range~\eqref{eq:psirange} is very well motivated as it can lead to an understanding of baryogenesis and dark matter generation.

\subsection{Exotic \texorpdfstring{$B$}{B} Meson Decays}\label{sec:BaryonDM_decays}
As discussed in the Introduction, one of the key predictions of $B$-Mesogenesis is the presence of a new decay mode of $B$ mesons into a dark antibaryon $\psi$, a visible baryon $\mathcal{B}$ and any number of light mesons with a branching fraction $\br \gtrsim 10^{-4}$. 

In order for the $B\to \psi \, \mathcal{B} \,\mathcal{M}$ decay to exist, a new BSM TeV-scale bosonic mediator is needed. In particular, this state should be a color-triplet scalar $Y$ which couples to $\psi$ and SM quarks. The LHC and flavor observables set relevant constraints on the mass and couplings of this color-triplet scalar which we discuss in detail in Sec.~\ref{sec:TripletScalar}. This heavy mediator can be integrated out to yield a low energy Lagrangian of the form $\mathcal{L}_{\rm eff} = \sum_{i,j} \mathcal{O}_{u_id_j} \frac{y_{ij}^2}{M_Y^2}$, with $y_{i j}^2$ being the product of the two relevant dimensionless couplings. The four possible flavor combination  operators $\mathcal{O}_i$ of interest for $B$ meson decays are
\begin{subequations}\label{eq:LflavorIR}
\begin{align}
\mathcal{O}_{ud} &= \psi \, b \, u \, d  \,, \\
\mathcal{O}_{us} &= \psi \, b \, u \, s\,,\label{eq:bus}\\
\mathcal{O}_{cd} &= \psi \, b \, c \, d\,, \\
\mathcal{O}_{cs} &= \psi \, b \, c \, s \,,
\end{align}
\end{subequations}
where all fermions are assumed to be right-handed\footnote{In principle, operators of the form $\psi\, d\,Q_L\,Q_L'$, mediated by a color-triplet vector in the fundamental of $SU(2)$, are also possible. Although for simplicity we do not expand on this possibility here, they constitute another viable option.} and color indices are contracted in a totally antisymmetric way.
These operators can induce the decay of the $\bar{b}$ quark within the $B$ meson into two light quarks and a dark antibaryon $\psi$. The resulting possible hadronic processes are summarized in Table~\ref{tab:hadronmasses} for the different operators in Eq.~\eqref{eq:LflavorIR}. 
Matrix elements involving the operators in Eq.~\eqref{eq:LflavorIR} depend on the precise pairing of the spinors. Each of the operators can come in three different versions: ``type-1'' $\mathcal{O}^1_{ij} = (\psi b)(u_id_j)$,  ``type-2'' $\mathcal{O}^2_{ij} = (\psi d_j)(u_ib)$ and ``type-3'' $\mathcal{O}^3_{ij} = (\psi u_i)(d_jb)$. This distinction becomes relevant for some of the constraints discussed in the next sections.

As we will see in Sec.~\ref{sec:TripletScalar}, flavor constraints on the $Y$ triplet scalar imply that \emph{only one of these operators can be active in the early Universe}. In practice, this means that we only expect one dominant flavor combination of these possible operators at collider experiments and not a combination of the above.
Therefore, only one of the sets of decay channels listed in Table~\ref{tab:hadronmasses} is expected to have a sizeable branching ratio, while all others should be suppressed.

In view of the form of the effective operators in Eq.~\eqref{eq:LflavorIR}, it is important to note that all $B$ mesons should decay at a very similar rate given that $m_{B^\pm} \simeq m_{B_d^0} \simeq m_{B_s^0}$. Additionally, $b$-flavored baryons (generically denoted by $\mathcal{B}_b$) should also posses a branching fraction with a size $ \text{Br}\left(\mathcal{B}_b\to \bar{\psi} \,\mathcal{M}\right) \sim \br $, again given that the masses of all the $b$-flavored hadrons are fairly similar to the $B$ mesons ones.

\begin{table}[t]
\renewcommand{\arraystretch}{1.2}
  \setlength{\arrayrulewidth}{.25mm}
\centering
\small
\setlength{\tabcolsep}{0.18 em}
\begin{tabular}{ |c || c | c | c  |}
    \hline\hline
    Operator 			&  \,\, Initial \,\,  &  Final 				&   $\,\, \Delta M$   \,\,\,    \\ 
       \,\,\, and Decay \,\, 			&  State &   State				& (MeV)       \\ \hline
        \hline
					  	            &   	$B_d$   &  $\psi + n \,(udd)$		        &4340.1	  \\ 
$\mathcal{O}_{ud} = \psi \, b\, u\, d$	 &   	$B_s$    &   $\psi + \Lambda \,(uds)$		&4251.2	  \\ 
{$\bar{b}\to \psi \, u\, d$} 		  &   	$B^+$    &  $\psi + p \,(duu)$	  		& 4341.0       \\ 
    		   				&   	$\Lambda_b$   &  $\bar{\psi} + \pi^0 $	& 5484.5	 \\  \hline\hline 
						&   	$B_d$   &  $\psi + \Lambda \,(usd)$		&4164.0	  \\ 
$\mathcal{O}_{us} = \psi \, b\, u\, s$  &   	$B_s$    &   $\psi + \Xi^0 \,(uss)$		&4025.0		   \\ 
 {$\bar{b}\to \psi \, u\, s$}   	 & 	$B^+$    &  $\psi + \Sigma^+ \,(uus)$	  	& 4090.0		    \\ 
    		  				&   	$\Lambda_b$   &  $\bar{\psi} + K^0 $	        &5121.9		   \\  \hline\hline 
						 &   	$B_d$   &  $\psi + \Lambda_c+ \pi^- \,(cdd)$&2853.6		 \\ 
$\mathcal{O}_{cd} = \psi \, b\, c\, d$   &   	$B_s$    &   $\psi + \Xi_c^0 \,(cds)$		&2895.0		   \\ 
 {$\bar{b}\to \psi \, c\, d$}		 &   	$B^+$    &  $\psi + \Lambda_c^+ \,(dcu)$	  & 2992.9		   \\ 
		  				 &   	$\Lambda_b$   &  $\bar{\psi} + \overline{D}^0 $ &3754.7	 \\ \hline\hline
						 &   	$B_d$   &  $\psi + \Xi_c^0 \,(csd)$		&2807.8			   \\ 
$\mathcal{O}_{cs} = \psi \, b\, c\, s$    &   	$B_s$    &   $\psi + \Omega_c \,(css)$	&2671.7		   	 \\ 
 {$\bar{b}\to \psi \, c\, s$}	          &   	$B^+$    &  $\psi + \Xi^+_c \,(csu)$	  	& 2810.4		 \\ 
    		 				  &   	$\Lambda_b$   &  $\bar{\psi} + D^-+ K^+$  &3256.2	 \\   \hline\hline

\end{tabular}
\caption{The lightest final state resulting from the new decay of $b$ quarks as necessary to give rise to baryogenesis and dark matter production. We list each of the possible flavorful operators that can equally lead to $B$-Mesogenesis, see Eq.~\eqref{eq:LflavorIR}. For a given operator, the rate of each decay is fairly similar given that $m_{B^\pm} \simeq m_{B_d^0} \simeq m_{B_s^0} \sim m_{\Lambda_b}$. $\Delta M$ refers to the difference in mass between the initial and final SM hadron. Note that additional light mesons can be present in the final state, which act to decrease $\Delta M$ by their corresponding masses.}
\label{tab:hadronmasses}
\end{table}

\section{CP Violation in the \texorpdfstring{$B$}{B} Meson System}
\label{sec:CP_violation}
As described in the previous sections, $B$-Mesogenesis~\cite{Elor:2018twp} directly relates the CP violation in the neutral $B$ meson systems to the observed baryon asymmetry of the Universe. In this section, we discuss how current measurements of CP violating observables in $B^0_q-\bar{B}_q^0$ mixing constrain the mechanism. In particular, as clearly seen from Eq.~\eqref{eq:basypara}, there is a correlation between $\br$ and the CP asymmetries in the $B^0_q$ systems. Thus, current measurements on the CP violation of $B^0_q$ mesons  set a lower bound to $\br$ which we find to be $\sim 10^{-4}$. 

To set the stage, we first review the origin of CP violation in $B^0_q$ mesons and the associated observables. The oscillations of neutral $B_{s}^0$ and $B_{d}^0$ mesons are described by the mass ($M_{12}^q$) and decay ($\Gamma_{12}^q$) mixing amplitudes between the flavor eigenstates $B_q^0$ and ${\bar B_q^0}$ --- see~\cite{pdg} for reviews on CP violation in the quark sector and $B^0_q-\bar{B}^0_q$ oscillations. CP violation can be present in these systems and manifests itself as a relative phase between $M_{12}^q$ and $\Gamma_{12}^q$. The observables that connect the dynamics of $B^0_q$ mesons in the early Universe and observations in laboratory experiments are the semileptonic asymmetries, defined as
\begin{align}
A_{\rm SL}^q & = \frac{\Gamma(\bar{B}_q^0\to {B}_q^0 \to f) - \Gamma(B_q^0\to \bar{B}_q^0\to \bar{f})}{\Gamma(\bar{B}_q^0\to {B}_q^0 \to f )+\Gamma(B_q^0\to \bar{B}_q^0\to \bar{f})}\,, \label{eq:semi_clear}
\end{align}
where $A_{\rm SL}^{s,d}$ is the semileptonic asymmetry in $B^0_{s,d}$ decays, $f$ is a CP eigenstate that is only accessible to the $B_q^0$ meson and $\bar{f}$ is its CP conjugate. Note that $A_{\rm SL}^q$ is called semileptonic asymmetry simply because the decays used to quantify it are semileptonic (e.g. $\bar{b} \to c \bar{\nu}_\ell \ell^-$). In terms of the mass and decay mixing amplitudes, Eq.~\eqref{eq:semi_clear} can be written as
\begin{align}
	A_{\rm SL}^q     &=  -\frac{\Delta \Gamma_q}{\Delta M_q} \tan \left(\phi_{12}^q \right) \,.
		    \label{eq:semi_12}
\end{align}
Here, $\Delta M_q$ and $\Delta \Gamma_q$ represent the physical mass and width differences of the  $B_q^0$ eigenstates, which are related to the amplitudes by $\Delta M_q = 2|M_{12}|$ and $\Delta \Gamma_{q} = 2|\Gamma_{12}|\cos\phi_{12}^q$, where $\phi_{12}^q$ is the relative phase between $M_{12}^q$ and $\Gamma_{12}^q$.

From Eq.~\eqref{eq:semi_clear}, we can see why the semileptonic asymmetries play a fundamental role in the mechanism and appear in Eq.~\eqref{eq:basypara}: they precisely control how often more matter than antimatter is generated from the decays of $B_q^0$ and $\bar{B}_q^0$ mesons. As pointed out in~\cite{Elor:2018twp}, since the Universe is made out of only matter, one of the key predictions of the mechanism is that at least one of the semileptonic asymmetries should be positive and larger than $\sim 10^{-4}$.

\subsection{Current Measurements and Implications}\label{sec:CP_today}

\textit{Semileptonic Asymmetries}. The current world averages for $A_{\rm SL}^q$, as reported in the PDG 2020~\cite{pdg} and as prepared by the \href{https://hflav.web.cern.ch/}{HFLAV}~\cite{Amhis:2019ckw} group, at 68\% CL, read
\begin{subequations}\label{eq:semiL_ave}
\begin{align}
    A_{\rm SL}^d &= \left(-2.1 \pm 1.7\right)\times 10^{-3} \,, \label{eq:semiL_ave_d} \\
    A_{\rm SL}^s &= \left(-0.6 \pm 2.8\right)\times 10^{-3} \, \label{eq:semiL_ave_s} \,. 
\end{align}
\end{subequations}
Meanwhile, the SM prediction for these quantities, which we take from~\cite{Lenz:2019lvd}\footnote{Note that Ref.~\cite{Lenz:2020efu} has recently pointed out that renormalization-scale dependent effects could change these predictions by up to a factor of two.}, are
\begin{subequations}\label{eq:semiL_SM}
\begin{align}
    A_{\rm SL}^d|_{\rm SM} &= \left(-4.7 \pm 0.4\right)\times 10^{-4} \,, \label{eq:semiL_SM_d} \\
    A_{\rm SL}^s|_{\rm SM} &= \left(2.1 \pm 0.2\right)\times 10^{-5} \, \label{eq:semiL_SM_s} \,.
\end{align}
\end{subequations}
Comparing Eqs.~\eqref{eq:semiL_ave} and~\eqref{eq:semiL_SM}, it is clear that the current experimental error bars are about four times larger than the SM prediction for the $B^0_d$ meson, and about $30$ times larger for the $B_s^0$ system. 

Existing measurements of $A_{SL}^q$ are already useful for constraining the parameter space of the mechanism. As highlighted in Eq.~\eqref{eq:basypara}, the baryon asymmetry of the Universe is proportional to both $\br$ and $A_{\rm SL}^q$. Thus, given the observed baryon asymmetry, a constraint on $A_{\rm SL}^q$ can indeed be used to indirectly set a lower limit on $\br$. In particular, the fact that $A_{\rm SL}^q \lesssim 10^{-3}$ implies that
\begin{align}\label{eq:BR_constraint_ALL_meas}
\br \,\, \gtrsim \,\,  10^{-4} \,.
\end{align}

\begin{figure}[t!]
\centering
\hspace{-0.4cm}\includegraphics[width=0.50\textwidth]{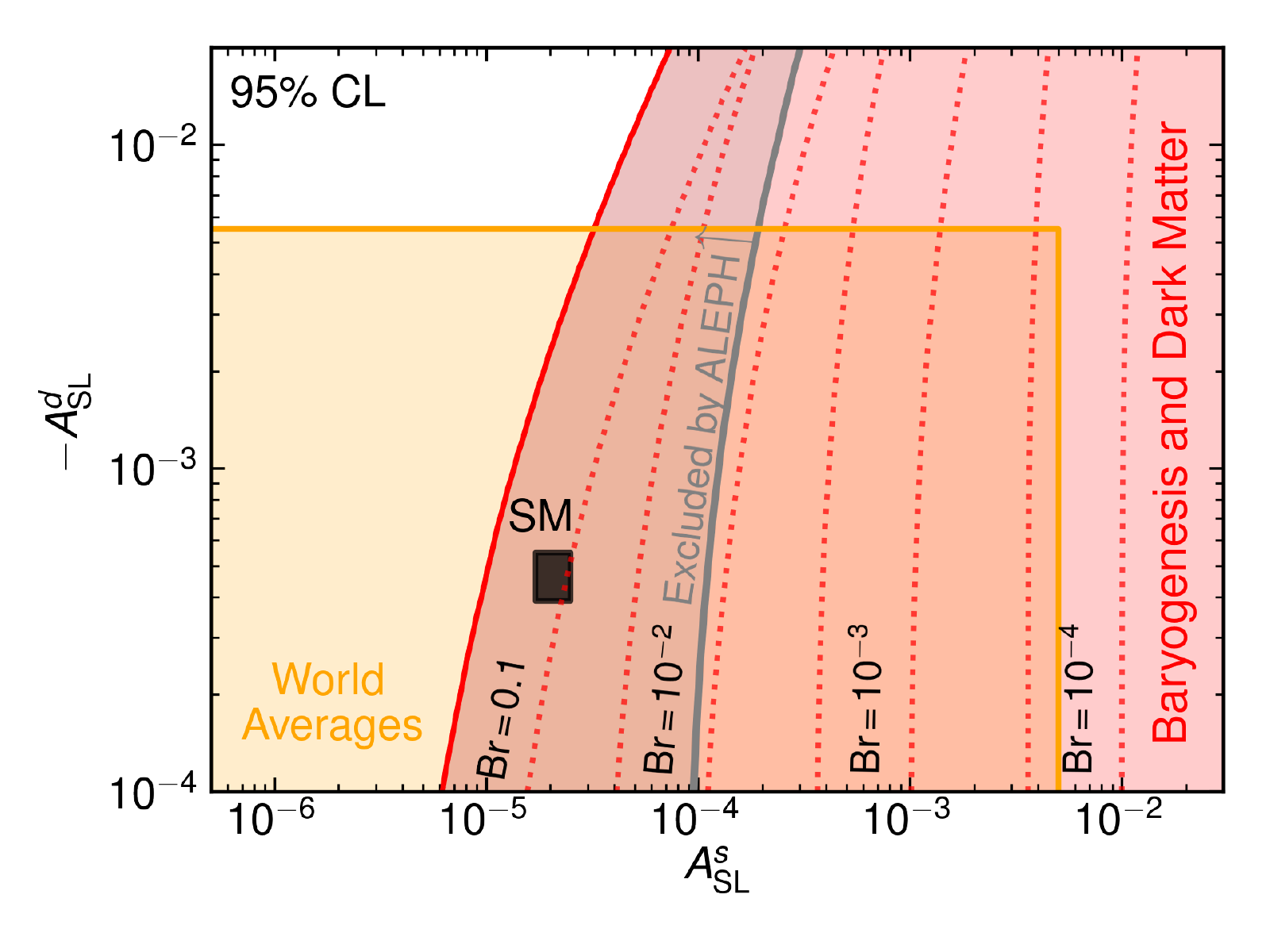} 
\vspace{-0.65cm}
\caption{Contour lines for the minimum $\br$ required for baryogenesis and dark matter generation as a function of the semileptonic asymmetries in $B_ q^0$ meson decays, $A_{\rm SL}^q$. In red, we show the relevant parameter space in which baryogenesis can successfully occur. The dashed lines delineate the cosmological uncertainties of our predictions (see text for more details). The black rectangle corresponds to the SM prediction for the semileptonic asymmetries~\cite{Lenz:2019lvd}, while the orange contour corresponds to the current world averages for experimental measurements of these quantities~\cite{pdg}. The grey line highlights the region of parameter space corresponding to $\br > 0.5\%$, which is disfavoured by an ALEPH search as discussed in Sec.~\ref{sec:BR_current_LEP}. All contours are shown at $95\%$ CL.
This figure showcases that, given current measurements of the semileptonic asymmetries, a branching ratio $\br \gtrsim 10^{-4}$ is required for successful baryogenesis. Similarly, in light of the ALEPH constraint, $A_{\rm SL}^q > 10^{-4}$ is necessary in order to explain the observed baryon asymmetry of the Universe.}
\label{fig:semileptonic_asymmetries}
\end{figure}

Given the direct correlation between $A_{\rm SL}^q$ and $\br$ in generating the baryon asymmetry of the Universe, a lower limit on $A_{\rm SL}^q$ can be established given an upper limit on $\br$. This correlation is clearly seen in Fig.~\ref{fig:semileptonic_asymmetries}. As is discussed in Sec.~\ref{sec:BR_current}, at present we know that $\br < 0.5\%$, which implies that in order for $B$-Mesogenesis to explain the observed baryon asymmetry of the Universe, at least one semileptonic asymmetry should satisfy
\begin{align}\label{eq:ALL_constraint_br_meas}
A_{\rm SL}^q \,\, \gtrsim \,\,  10^{-4} \,.
\end{align}

\textit{CP violation in interference and global fits:} In addition to the direct observation of $A_{\rm SL}^q$, one can also attempt to place indirect limits on $A_{\rm SL}^q$ by taking into account constraints on other relevant $B$ meson observables. As seen from Eq.~\eqref{eq:semi_12}, the sign of the semileptonic asymmetries is characterized by the phase $\phi_{12}^q$. As such, the value of $\phi_{12}^q$ is  critical to baryogenesis.
Recall that this phase depends only on the difference between $\phi^q_M$ (the phase of $M^q_{12}$) and $\phi^q_\Gamma$ (the phase of $\Gamma^q_{12}$).
Of these two individual phases, $\phi^q_M$ can be reconstructed by measuring the interference between $B_q^0-\bar{B}_q^0$ oscillations and their decay into CP eigenstates. In particular, the most favourable decays are the so-called golden modes in $b \to c \bar{c} s$ transitions, which are $B_d^0 \to J/\psi\, K_S^0 $ and $B_s^0 \to J/\psi\, \phi$\footnote{The $J/\psi$ and $\phi$ mesons are not to be confused with the dark Dirac antibaryon $\psi$ and the dark scalar antibaryon $\phi$ of $B$-Mesogenesis.}. The extraction of $\phi_q^M$ from these modes depends on the experimental uncertainties in the measurements along with potential contamination from penguin diagrams. At present, the theoretical uncertainty on the SM prediction from penguin diagrams is dominant for $\phi^d_M$ while the one for $\phi^s_M$ is dominated by current experimental uncertainties (for a recent discussion see~\cite{Barel:2020jvf}). Including all relevant measurements, the~\href{https://hflav-eos.web.cern.ch/hflav-eos/osc/PDG_2020/}{HFLAV} collaboration~\cite{Amhis:2019ckw} reports an average of
\begin{align}
    \phi_s^{c\bar{c}s} \, =\, -0.050 \pm 0.019 = (-2.9\pm 1.1)^\circ \, .
\end{align}
This number is to be compared with the SM prediction which, neglecting contamination from penguin diagrams, is $\phi_s^{c\bar{c}s}|_{\rm SM}= -2\beta_s = -0.037\pm 0.001 $~\cite{Artuso:2015swg}.

\vspace{0.4cm}

In order to gauge the relevance of these measurements for $B$-Mesogenesis, let us write the two phases in the most general case where new physics can modify $M_{12}$, $\Gamma_{12}$, as well as the penguin contributions, as
\vspace{0.1cm}
\begin{align}
    \phi_{12}^s &= \phi_{12}^{s,\mathrm{SM}} + \phi_M^{s,\mathrm{NP}} + \phi_\Gamma^{s,\mathrm{NP}},\\[7pt]
    \phi_s^{c\bar{c}s} &= -2\beta_s + \phi_M^{s,\mathrm{NP}} + \delta_{\rm pen}^{SM} + \delta_{\rm pen}^{NP}.
\end{align}

\noindent The lack of control over SM penguin diagrams as well as the possible NP effects on them translate into an uncertainty of $\sim 1^\circ$ in $\delta_{\rm pen}^{SM} + \delta_{\rm pen}^{NP}$~\cite{Artuso:2015swg}.
However, the most important uncertainty comes from the potential new physics in tree-level decays, which are largely unconstrained and can modify $\phi_\Gamma^{s,\mathrm{NP}}$ and thus $\phi^s_{12}$ without affecting $\phi_s^{c\bar{c}s}$. In fact, it has been recently shown in~\cite{Lenz:2019lvd,Jager:2019bgk} via a global fit that allowing for modifications in tree level decays such as $b\to c \bar{c} s$ can lead to large values for the $\phi^d_\Gamma$ and $\phi^s_\Gamma$ phases. We note that in our scenario such tree-level decays are naturally expected to be modified, see Sec.~\ref{sec:Decays}. Depending upon the exact flavor structure of the $b\to u_i {\bar u}_i q_j$ decay the large modifications to $\phi_\Gamma^q$ in turn allow values for the semileptonic asymmetry that can be as large as $|A_{\rm SL}^q| = 1.5\times 10^{-3}$, which are comparable to current direct constraints on these quantities, see Eq.~\eqref{eq:semiL_ave}. Thus, while CP-violation in interference represents an important test of $\phi^q_M$, current measurements allow for large values of the semileptonic asymmetries via modifications of tree-level $b$ decays. 

\vspace{0.4cm}

\textit{Cosmological uncertainties}. The predictions shown in Fig.~\ref{fig:semileptonic_asymmetries} incorporate a number of uncertainties arising in our calculation of the generation of the baryon asymmetry in the early Universe. These uncertainties are discussed in detail in Appendix~\ref{sec:decoherence}, and are primarily dominated by the uncertainty in the fragmentation ratios of $\Phi$ decays to $B^\pm$, $B_s^0$, and $B_d^0$ mesons. Clearly, these fragmentation ratios are unknown and the band simply covers the range of fragmentation ratios as measured in other environments, namely in $Z$-boson decays: $f_s/f_d = 0.25\pm 0.02$, at $p\bar{p}$ collisions at Tevatron: $f_s/f_d = 0.33\pm 0.04$, at at $pp$ collisions at the LHC: $f_s/f_d = 0.247\pm 0.009$, and at $e^+\,e^-$ collisions at the $\Upsilon(5S)$ resonance:  $f_s/f_d = 0.26^{+0.05}_{-0.04}$ (see~\href{https://hflav-eos.web.cern.ch/hflav-eos/osc/PDG_2020/}{HFLAV}~\cite{Amhis:2019ckw}). In order to generate the band in Fig.~\ref{fig:semileptonic_asymmetries}, we take $f_s/f_d \in [0.22-0.37]$. 

\vspace{0.4cm}

Another source of uncertainty arises from the fact that the charge distributions within the $B_q^0$ mesons are not precisely known. Neutral $B$ mesons interact with the plasma in the early Universe through this charge distribution; these interactions act to decohere the CP violating $B_q^0 - \bar{B}_q^0$ oscillations, therefore hindering the production of a baryon asymmetry.
Since the $B_q^0$ system is spinless and chargeless, its electromagnetic interactions can be described by an effective charge radius for which we only have theoretical estimates that range within a factor of two, see~\cite{Hwang:2001th,Becirevic:2009ya,Das:2016rio}.
This uncertainty, corresponding to the width of the bands in Fig.~\ref{fig:alpha_ds}, effectively translates into a $\lesssim 20\%$ uncertainty on our prediction of $Y_B$ and therefore of $A_{\rm SL}^q$ and $\br$.

\vspace{0.4cm}

Finally, the baryon asymmetry depends on the early Universe cosmology via the mass of the $\Phi$ field that reheats the Universe to a temperature $T_R$. Since $M_\Phi$ and $T_R$ are free parameters that can vary over a certain range, in Fig.~\ref{fig:semileptonic_asymmetries} we have marginalized over them so as to maximize the baryon asymmetry of the Universe for each given $A_{\rm SL}^q$. Thus, the red contours in Fig.~\ref{fig:semileptonic_asymmetries} should be read as a theoretical lower limit on $\br$. 

\subsection{Future Prospects at the LHC and Belle II}

The sensitivity of upcoming studies at the LHC and Belle II to CP violation in the $B_q^0$ systems are promising~\cite{Cerri:2018ypt,Bediaga:2018lhg,privateBelleIIAlld}. In particular, the projected $1\sigma$ sensitivities for the semileptonic asymmetries are\footnote{The sensitivity for $A_{\rm SL}^d$ at Belle II is not in the Belle II physics book~\cite{Kou:2018nap}. We have obtained it from~\cite{privateBelleIIAlld}. We also note that ATLAS and CMS also have the potential to measure $A_{\rm SL}^{d,\,s}$, but the projected sensitivities are not available in the literature.}
\begin{subequations}\label{eq:ASL_sensitivities_future}
\begin{align}
    \delta A_{\rm SL}^s &= 10\times 10^{-4} \,\,\, [\text{LHCb}\, (33\,\text{fb}^{-1}) -2025]\,,\\
        \delta A_{\rm SL}^s &= 3\times 10^{-4} \,\,\,\,\,\, [\text{LHCb}\, (300\,\text{fb}^{-1})-2040]\,,\\
            \delta A_{\rm SL}^d &= 8\times 10^{-4} \,\,\,\,\,\, [\text{LHCb}\, (33\,\text{fb}^{-1})-2025]\,,\\
        \delta A_{\rm SL}^d &= 2\times 10^{-4} \,\,\,\,\,\, [\text{LHCb}\, (300\,\text{fb}^{-1})-2040]\,,\\
        \delta A_{\rm SL}^d &= 5\times 10^{-4} \,\,\,\,\,\, [\text{Belle II}\, (50\,\text{ab}^{-1})-2025]\,.
\end{align}
\end{subequations}
Comparing these numbers with the parameters necessary for baryogenesis shown in Fig.~\ref{fig:semileptonic_asymmetries}, one can appreciate the great potential that upcoming measurements from LHCb or Belle II have for helping to confirm (or refute) $B$-Mesogenesis. For instance, if all measurements are compatible with the SM predictions by the year 2025, we will know that $\br \gtrsim 6\times 10^{-4}$. On the other hand,  positive semileptonic asymmetries potentially reported by future measurements would be a clear signal in favour of the mechanism and could point towards somewhat smaller branching fractions $10^{-4} \gtrsim \br \gtrsim 10^{-3}$. As such, $B$-Mesogenesis provides a complementary source of motivation for the LHCb Upgrade II~\cite{Bediaga:2018lhg}, as measurements with 300 $\text{fb}^{-1}$ would be extremely useful in constraining the relevant parameter space.

\subsection[Baryogenesis with the SM CP Violation?]{Baryogenesis with the SM CP Violation?}\label{sec:OnlySMCPviolation}

According to common lore, the CP violation within the SM is too small to generate a matter-dominated Universe as we observe it. For instance, in electroweak baryogenesis~\cite{Cohen:1993nk,Morrissey:2012db}, which can occur during a $1^\text{st}$-order electroweak phase transition, the relevant quark CP violating invariant is such that $(n_\mathcal{B}-n_{\bar{\mathcal{B}}})/n_\gamma < 10^{-20}$~\cite{Gavela:1993ts,Gavela:1994dt}. This indeed requires new BSM sources of CP violation to generate the observed baryon asymmetry of the Universe at the $(n_\mathcal{B}-n_{\bar{\mathcal{B}}})/n_\gamma \sim 10^{-10}$ level. However, it is important to note that in the SM, the CP asymmetries in the $B_q^0$ mesons are not small compared with $10^{-10}$. In particular, $A_{\rm SL}^d|_{\rm SM} \simeq -4.7\times 10^{-4}$ and $A_{\rm SL}^s|_{\rm SM} \simeq 2.1\times 10^{-5}$~\cite{Lenz:2019lvd}. As can be seen from Fig.~\ref{fig:semileptonic_asymmetries}, $B$-Mesogenesis could in principle account for the observed baryon asymmetry of the Universe with only the SM CP violation provided that $\br> 2.5\%$.
Although this possibility is disfavored given the constraints on the new $b$ decay modes from ALEPH, which tell us that $\br < 0.5\%$, it is important to note that, contrary to scenarios such as electroweak baryogenesis, the generated baryon asymmetry with only the SM CP violation is off by just an order of magnitude rather than by 10 or more. The reason for this ultimately stems from the fact that $B$-Mesogenesis proceeds at much lower temperatures than other baryogenesis scenarios -- temperatures at which the SM CP violation is not suppressed.

\begin{figure*}[t]
\centering
\includegraphics[width=0.97 \textwidth]{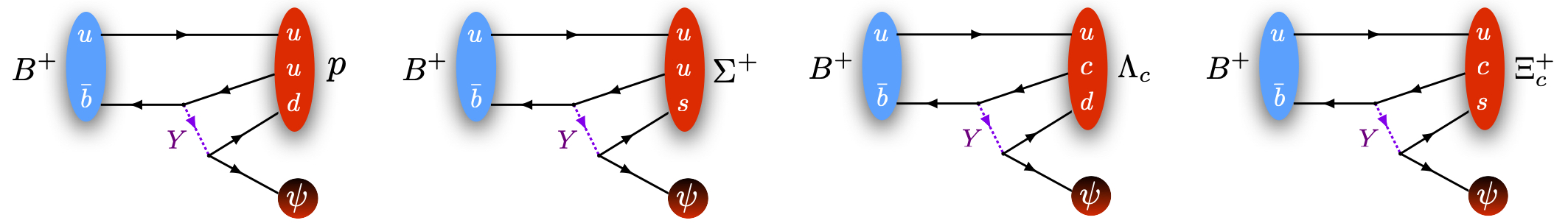}
\vspace{-0.3cm}
\caption{The decay of the $B^+$ meson to the lightest possible baryon as triggered by the four different flavor operators given in Eq.~\eqref{eq:LflavorIR}. Note that any of the four can lead to successful baryogenesis and dark matter production. As usual, the light dark sector antibaryon $\psi$ would appear as missing energy in the detector, and $Y$ is a heavy color-triplet scalar mediator with $M_Y > 1.2\,\text{TeV}$ (see Sec.~\ref{sec:TripletScalar}). }
\label{fig:decays}
\end{figure*}

\section{Searches for \texorpdfstring{$B\to \psi +\text{Baryon}$}{BtopsiBaryon} }\label{sec:BR}
In $B$-Mesogenesis, the baryon asymmetry of the Universe depends upon the inclusive rate of $B$ mesons decaying into a dark sector antibaryon ($\psi$) and a SM  baryon ($\mathcal{B}$) plus any number of light mesons, ($\mathcal{M}$) i.e. $\br$, see Eq.~\eqref{eq:basypara}.
Recall that any of the four flavors of operators in Eq.~\eqref{eq:LflavorIR} can yield such an inclusive rate; the contributing lightest final flavor states in each case are summarized in Table~\ref{tab:hadronmasses}.
To illustrate these flavorful variations, in Fig.~\ref{fig:decays} we display possible decays of $B^+$ mesons for each of the distinct operators. 

Experimental searches for $B$ meson decays into missing energy and a SM baryon have arguably been overlooked in experimental programs to date. In this section, we first summarize in Sec.~\ref{sec:BR_current} the current state of constraints on the branching fractions for these processes as relevant for baryogenesis. In Sec.~\ref{sec:BR_Bfactories} and Sec.~\ref{sec:BR_LHC}, we then discuss the potential reach of $B$ factories and the LHC respectively, to \emph{exclusive decays} of $B$ mesons involving missing energy and a baryon in the final state (these are easier to target than inclusive modes including mesons). Finally, in Sec.~\ref{sec:excl_vs_inclu} we perform a primitive phase space analysis to relate the inclusive decay rate $\br$ to the exclusive one $\text{Br}\left(B\to \psi\, \mathcal{B} \right)$, the result of which suggests that BaBar, Belle and especially Belle II and LHCb have the potential to test wide regions of parameter space of the baryogenesis and dark matter mechanism of~\cite{Elor:2018twp}.

\subsection{Current Limits}\label{sec:BR_current}
There exists no current dedicated search for $B$ meson decays into a visible baryon and missing energy plus any number of light mesons at any experimental facility. In~\cite{Elor:2018twp}, a loose bound $\br < 10\%$ was set on such a decay by taking into account the inclusive measurements of $B\to (c+\text{anything})$. This is only applicable provided that the baryon in the $B\rightarrow\psi\,\mathcal{B}\,\mathcal{M}$ does not contain a charm quark--- which can certainly be the case as can be seen in the two right panels of Fig.~\ref{fig:decays}. Similarly, one can set a very loose bound on the branching fraction by comparing the predicted SM decay rate of $b$-hadrons to the measured value. Doing so leads to $\br \lesssim 40\,\%$ as a result of the large ($\mathcal{O}(20\%)$) uncertainties in the theoretical prediction of the decay rate of $b$-hadrons in the SM~\cite{Krinner:2013cja}.

In this work, we find substantially stronger bounds by \textit{i)} examining inclusive decays of $B$ mesons into baryons (which yield constraints at the $\br \lesssim  1-10\%$ level), and \textit{ii) } by recasting an inclusive ALEPH search~\cite{Barate:2000rc} for events with large missing energy arising from b-flavored hadron decays at the $Z$ peak (which yield constraints at the $\br \lesssim  10^{-4}-10^{-2}$ level for the relevant range of the dark sector antibaryon mass).

\vspace{-0.2cm}
\subsubsection{Inclusive Considerations}\label{sec:BR_current_inclusive}
The reasoning presented in~\cite{Elor:2018twp} regarding inclusive measurements of $B$ meson decays can be refined to obtain firmer bounds. Given some reasonable assumptions, these apply to the different flavor final states possible in the $\bar{b}\to \psi u_i d_j$ decay. Firstly, the PDG~\cite{pdg} reports a measured inclusive rate of 
\begin{align}
\text{Br}(B\to p/\bar{p} + \text{anything}) &=(8.0\pm 0.4) \,\%\, ,\label{eq:BR_inclu_p}
\end{align} 
where here $B$ refers to an admixture of $B^+,\,B^-,\,B^0_d$ and $\bar{B}_d^0$ mesons. If we assume that the process $B\to \psi \, \mathcal{B}\,\mathcal{M}$ produces the same number of protons as neutrons (which is reasonable based on isospin symmetry), we can use Eq.~\eqref{eq:BR_inclu_p} to find a 95\% CL upper bound
\begin{align}\label{eq:BR_bound_solid}
\text{Br}(B\to \psi + \mathrm{Baryon} + \mathrm{Mesons}) &< 8.7 \,\%\,  .
\end{align} 

Other measurements of the inclusive decay rate of $B$ mesons into baryons can be used to set appropriate upper limits on each flavor variation of these new decays modes. The relevant averages as reported by the PDG~\cite{pdg} are\footnote{These are based on measurements by ARGUS~\cite{Albrecht:1988sj} and CLEO~\cite{Crawford:1991at} for Eqs.~\eqref{eq:BR_inclu_p},~\eqref{eq:BR_inclu_pdir}, and~\eqref{eq:BR_inclu_lam}, by BaBar~\cite{Aubert:2006cp} for Eq.~\eqref{eq:BR_inclu_lamc}, and by BaBar~\cite{Aubert:2005cu} for Eq.~\eqref{eq:BR_inclu_xi0}.}
\vspace{-0.4cm}
\begin{subequations}\label{eq:PDG_inclusive}
\begin{align}
\text{Br}(B\to p/\bar{p} (\text{dir}) + \text{anything}) &=(5.5\pm 0.5) \,\%\, ,\label{eq:BR_inclu_pdir}\\
\text{Br}(B\to \Lambda/\bar{\Lambda} + \text{anything}) &=(4.0\pm0.5) \,\%\, , \label{eq:BR_inclu_lam} \\
\text{Br}(B\to \Lambda_c^+/{\overline \Lambda}_c^- + \text{anything}) &= (3.6\pm0.4) \,\% \, , \label{eq:BR_inclu_lamc} \\
\text{Br}(B\to \Xi_c^0 + \text{anything}) &= (1.4\pm0.2)\,\% \label{eq:BR_inclu_xi0} \,,
\end{align} 
\end{subequations}
where (dir) means that the secondary protons from $\Lambda$ baryon decays have been subtracted. In order to obtain Eq.~\eqref{eq:BR_inclu_xi0}, we have additionally made use of the measurement $\text{Br}(B\to \Xi_c^0 + \text{anything}) \times \text{Br}(\Xi_c^0  \to \Xi^- +\pi^+) =(1.9\pm 0.3)\times 10^{-4}$~\cite{pdg} (see also~\cite{Aubert:2005cu}) as well as the very recently measured absolute branching fraction $\text{Br}(\Xi_c^0 \to \Xi^- +\pi^+) = 1.43\pm 0.32 \,\%$~\cite{pdg} (see also the original study by Belle~\cite{Li:2018qak}).

Eq.~\eqref{eq:PDG_inclusive} can be used to constrain the operators that lead to new $B$ meson decays provided that an assumption regarding the final-state hadronization is made. A problem arises since these bounds apply to an admixture of $B$ mesons, thus the final state baryons could be produced by any of the $B$ mesons depending upon the emission of other charged light mesons. To illustrate the issue, let us take for concreteness the case of Eq.~\eqref{eq:BR_inclu_lamc}. From the third diagram in Fig.~\ref{fig:decays}, we clearly see that the $B^+$ meson can contribute to this inclusive rate. However, this is only the case if no charged pions are emitted: if a $\pi^+$ is produced, the $\Lambda$ baryon must be neutral and the decay is not included in the inclusive measurement.
However, we can assume that the probability for this charged pion to be emitted is the same as for an oppositely-charged pion to be emitted in the decay of a $B_d^0$. If this is the case, then the $B_d^0\to \psi\, \Lambda_c^+ \, \pi^- $ process contributes to the inclusive channel with a similar rate as the one that was excluded. It is reasonable to assume that the emission of these light mesons is independent of the $B$ meson that is decaying, given that $B^+$, $B^0_d$, and $B^0_s$ all have approximately the same mass, and that the masses of the relevant light charged and neutral hadrons are similar.
Under this assumption, we can indeed assume that the new decays of the $b$ and $\bar{b}$ quarks contribute to Eq.~\eqref{eq:PDG_inclusive}. Therefore, assuming the errors in Eq.~\eqref{eq:PDG_inclusive} to be Gaussian, we can obtain the following 95\% CL upper limits for the flavorful decays of $B$ mesons into baryons and missing energy as generated by the interactions in~\eqref{eq:LflavorIR}: 
\begin{subequations}\label{eq:BR_bounds}
\begin{align}
 \text{Br}\left(\bar{b}\to \psi u d\right)  &\lesssim  6\,\%\,,\,\, \text{from Eq.}~\eqref{eq:BR_inclu_pdir}\,,\\
 \text{Br}\left(\bar{b}\to \psi u s\right)   &\lesssim  5\,\% \,,\,\, \text{from Eq.}~\eqref{eq:BR_inclu_lam}\,,\\
 \text{Br}\left(\bar{b}\to \psi c d\right)   & \lesssim 4\,\% \,,\,\, \text{from Eq.}~\eqref{eq:BR_inclu_lamc}\,,\\
 \text{Br}\left(\bar{b}\to \psi c s\right)  &  \lesssim 2 \,\% \,,\,\, \text{from Eq.}~\eqref{eq:BR_inclu_xi0}\,.
\end{align} 
\end{subequations}
These bounds should, however, be taken with a word of caution given the assumptions that have been made to obtain them.
We note that the above bounds could be slightly improved by subtracting off from Eq.~\eqref{eq:PDG_inclusive} the contribution from measured exclusive $B$ meson decays with a baryon in the final state. However, the total sum of the branching fraction for such processes is still far from the upper limit and doing so would not lead to a substantial improvement with respect to the bounds quoted in~\eqref{eq:BR_bounds}.

\vspace{-0.1cm}

\subsubsection{Missing Energy from $b$ decays at LEP}\label{sec:BR_current_LEP}

\begin{figure*}[t]
\centering
\begin{tabular}{cc}
		\label{fig:Br_exclusive}
		\includegraphics[width=0.48\textwidth]{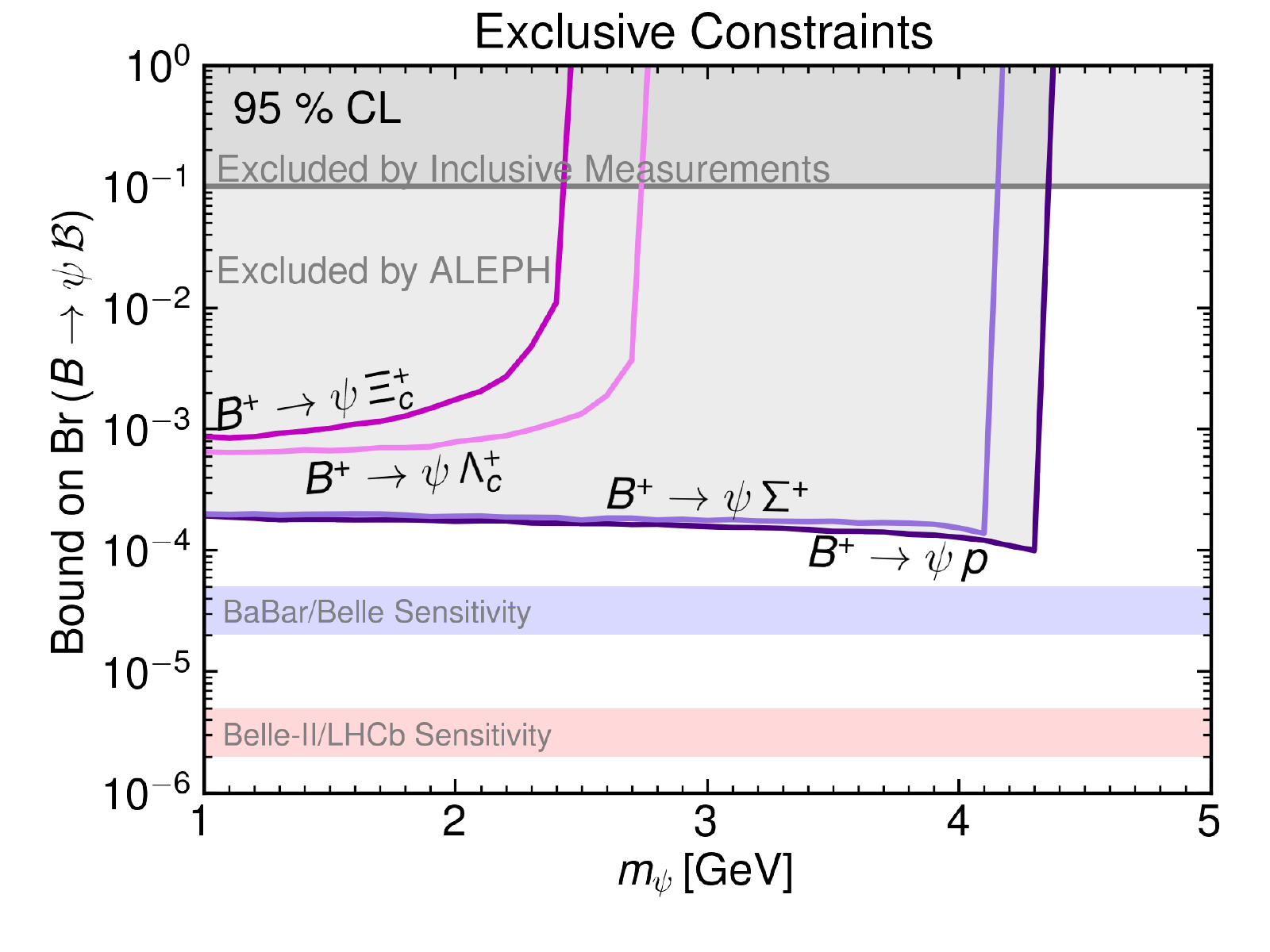}
&
	    \label{fig:Br_inclusive}
		\includegraphics[width=0.48\textwidth]{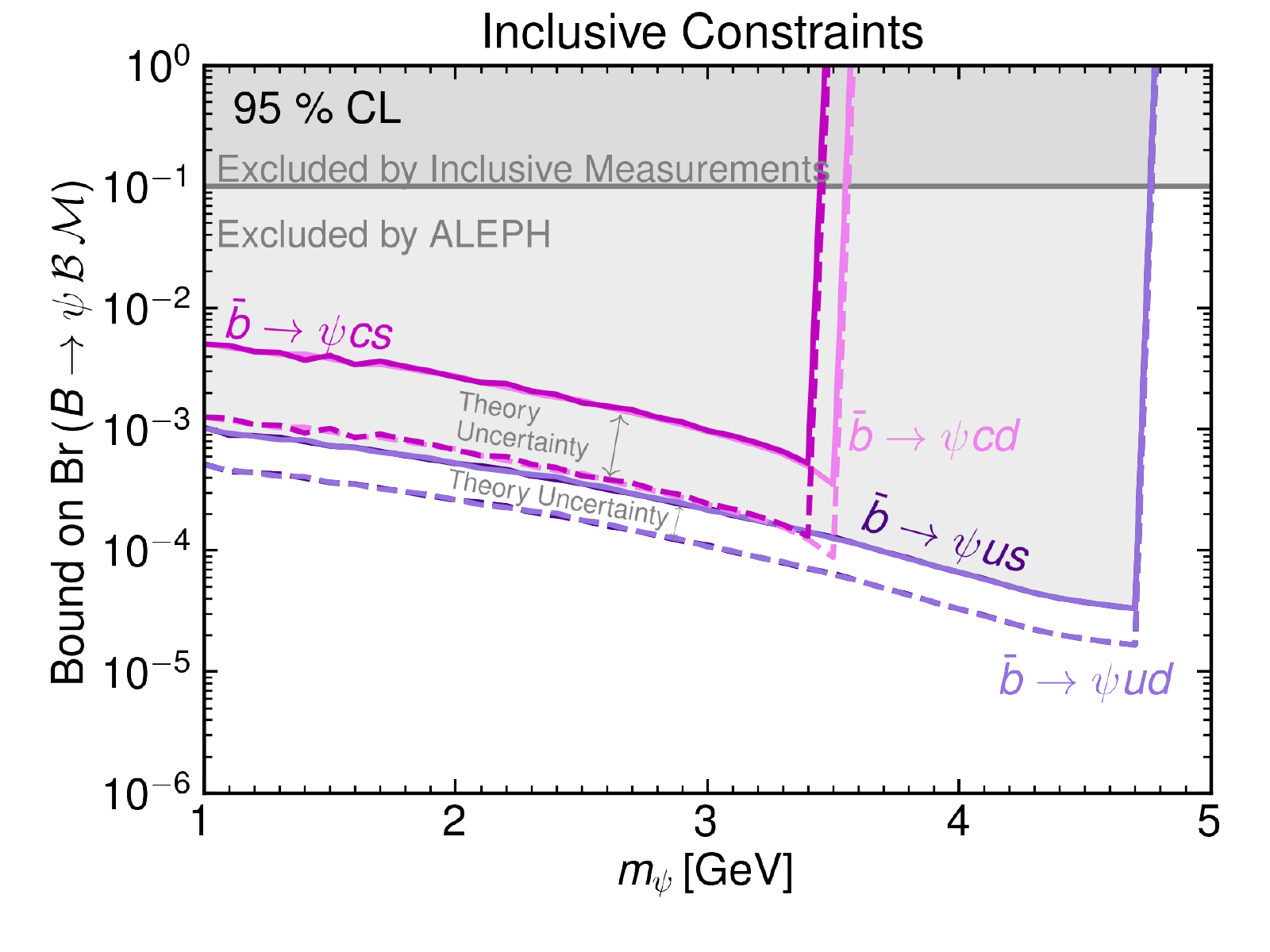}
\end{tabular}
\vspace{-0.4cm}
\caption{Constraints on $B$ mesons decays into a dark sector antibaryon $\psi$ and a baryon $\mathcal{B}$. The colored lines show the resulting constraints from our recast of a search at ALEPH for decays of b-flavored hadrons with large missing energy at LEP~\cite{Barate:2000rc}. The global bound $\text{Br} < 10\%$ arises from inclusive considerations, see Eq.~\eqref{eq:BR_bound_solid}. \textit{Left panel:} constraints on exclusive channels containing no additional mesons in the final state. We also highlight the expected reach of dedicated analyses using old BaBar/Belle data and of Belle-II with $50\,\text{ab}^{-1}$ of data, as well as that of LHCb with $15\text{fb}^{-1}$~\cite{Rodriguez:2021urv}. \textit{Right panel:} 
Constraints on the inclusive decay including any number of mesons in the final state. Note that these constraints are derived assuming a parton-level missing-energy spectrum, which may suffer from relevant modifications when effects from the $b$ quark momentum inside the $B$ meson and QCD corrections are included. We expect that including these effects will relax the constraints by a factor of 2-4 (see main text), and we thus incorporate this theoretical uncertainty in order to obtain the conservative results depicted by the solid lines. The dashed lines correspond to the bounds taking the decay kinematics and our recast of the ALEPH analysis~\cite{Barate:2000rc} at face value. }
\label{fig:LEP_constraints}
\end{figure*}

The ALEPH experiment at LEP collected and analyzed events with large missing energy arising from $b$-hadron decays~\cite{Buskulic:1992um,Buskulic:1994gj,Barate:2000rc}. These studies were originally designed to target inclusive $b \to c \, \tau^- \bar{\nu}_\tau $ decays~\cite{Buskulic:1992um}, which lead to events with large missing energy as a result of the neutrinos in the final state. Since these searches select events only based on their missing energy and do not discriminate based on the hadronic or leptonic content of the final state, they are completely inclusive. In fact, soon after~\cite{Buskulic:1992um,Buskulic:1994gj} appeared, the authors of Ref.~\cite{Grossman:1995gt} pointed out that such searches could be used to set the first constraint on inclusive $b\to s \bar{\nu}\nu$ transitions at the $\text{Br} \lesssim 10^{-3}$ level. Because of their inclusive nature, these searches can be used to constrain our decays both at the inclusive and the exclusive levels\footnote{We are grateful to our anonymous Referee for pointing out the existence of~\cite{Barate:2000rc} and for highlighting its potential relevance for our scenario.}. In what follows, we summarize the main ingredients and results from our recast of \cite{Barate:2000rc}, and point the interested reader to Appendix~\ref{sec:ALEPH_recast} where we describe the details of the simulation of the missing-energy spectrum of $B\rightarrow \psi\,\mathcal{B\, M}$ decays at LEP.

In Ref.~\cite{Barate:2000rc}, the ALEPH collaboration studied events with large missing energy at LEP arising from the decays of $b$-flavored hadrons, $Z\to \bar{b}b$. The sample of Ref.~\cite{Barate:2000rc} consists of approximately $4 \times 10^6$ hadronically decaying $Z$ bosons collected at energies near the $Z$ peak. In order to target only b-flavored hadron decays, an algorithm was used to mainly select events with a decaying particle with a lifetime similar to that of the b-quark~\cite{Buskulic:1993ka}. This requirement has an efficiency of $f_{\rm tag} = 0.567$ and reduces backgrounds significantly~\cite{Barate:2000rc}. In addition, a cut of $\cos \theta < 0.7$, where $\theta $ is the angle between the beam axis and the thrust axis, was applied. The goal of this cut is to reject events that are not well contained within the detector, for which missing energy cannot be determined reliably. Following the discussion in~\cite{Buskulic:1992um}, we believe that this likely implies an efficiency factor of $f_{\rm thrust} = 0.45$. Given these known cuts, the number of candidate $b\bar{b}$ events in the sample of~\cite{Barate:2000rc} can be estimated to be\footnote{We note that in~\cite{Barate:2000rc} several other cuts and corrections to the missing energy spectrum were applied. However, their impact on the efficiency is not stated in~\cite{Barate:2000rc} and it falls beyond the scope of our paper to analyze them. }:
\begin{align}\label{eq:NbbarLEP}
N_{b\bar{b}}^{\rm sample} &=  f_{\rm tag}\, f_{\rm thrust}\, N_Z^{\rm had}\, \frac{{\rm BR}(Z\to b\bar{b})}{{\rm BR}(Z\to {\rm hadrons})}  \,,\\
 &\simeq 2.2\times 10^5 \nonumber\,.
\end{align}
Within this sample, events with large missing energy were searched for and compared with the expected backgrounds arising from other SM processes, such as $b\to X \ell \bar{\nu}_{\ell } $. 
In the analysis for $b \to s \nu\bar{\nu}$ events, which are closest to our decays, the relevant kinematic signal region was chosen to correspond to events with $E_{\rm miss} > 30\,\text{GeV}$. In this window, the events were subdivided into three bins with the following observed and expected background counts (see Table 5 of~\cite{Barate:2000rc}):
\begin{subequations}\label{eq:Nevents_LEP}
\begin{align}
    N_{30-35}^{\rm obs} &= 31 \,,\,\,  N_{30-35}^{\rm background} = 37.0\pm 2.7 \,,\\
    N_{35-40}^{\rm obs} &= 1 \,,\,\,\,\,  N_{35-40}^{\rm background} = 2.5\pm 1.6\,, \\
    N_{>40}^{\rm obs}    &= 1 \,,\,\,\,\,  N_{>40}^{\rm background} < 1\,.
\end{align}
\end{subequations}
These can be compared with the predictions for the missing energy distribution in our decays to set a limit on their branching fractions. To this end, we simulate the missing energy spectrum of $\bar{b}\to \psi\, u_i\, d_j $ and $B \to \psi \, \mathcal{B}_{u_i d_j}$ decays. For the inclusive decays ($\bar{b} \to  \psi \, u_i\, d_j$) we calculate the missing energy spectrum from the parton-level 3-body differential decay width, while for the exclusive decays the missing energy is fixed by the 2-body decay kinematics. In addition, we simulate the initial energy of the decaying $b$ by taking into account the Peterson fragmentation function used in~\cite{Barate:2000rc}. Finally, in order to set limits, we compare the predictions from our missing energy spectrum to the observations in Eq.~\eqref{eq:Nevents_LEP} given Eq.~\eqref{eq:NbbarLEP}. We set the limits by assuming a Poisson likelihood for the measurements and by following the Bayesian procedure outlined in the PDG review for statistics for likelihoods with backgrounds, see Eq.~(40.63) of~\cite{pdg}. 

The result of this procedure is showcased in Fig.~\ref{fig:LEP_constraints}, where we show the resulting constraints from the ALEPH search at 95\% CL for exclusive and inclusive decays. In the left panel of Fig.~\ref{fig:LEP_constraints}, we can appreciate that the exclusive constraints on $B$ decays are at the $10^{-4}$ level for the final state involving hyperons or nucleons for $m_\psi \lesssim 4.2\,\text{GeV}$. For final states containing charmed baryons, the exclusive constraints are at the $\text{Br} \sim 10^{-3}$ level for $m_\psi \lesssim 2-2.5\,\text{GeV}$. These results are easy to understand with the knowledge that in $B\to \psi \,\mathcal{B}_{ud}$ decays, for $m_\psi < 4\,\text{GeV}$, roughly $20\%$ of the signal events have $E_{\rm miss}>35\,\text{GeV}$. For the case of $B\to \psi \,\mathcal{B}_{cd}$ decays and due to the heaviness of charmed baryons, only 10\% of the events have missing energy $E_{\rm miss}>30\,\text{GeV}$, which explains the comparably weaker limits (note that the number of background events is moderately large in the 30-35 GeV bin). The left panel of Fig.~\ref{fig:LEP_constraints} only shows the constraints for $B^+$ decays, but very similar results hold for $B^0$ decays. For the case of $B_s$ and $\Lambda_b$ decays the constraints are relaxed by roughly a factor of $4$ as a consequence of the smaller fragmentation ratio of these states at LEP. 

\vspace{0.1cm}

In the right panel of Fig.~\ref{fig:LEP_constraints}, we show the resulting constraints on inclusive $\bar{b} \to \psi \, u_i\, d_j $ decays. We can appreciate that the bounds are at the level of $\text{Br}\sim 3\times 10^{-5}-5\times 10^{-2}$ depending upon the exact final state flavor and the $\psi$ mass. These results are rather important in that they constrain the inclusive branching fraction $\br$ that is directly related to the baryon asymmetry of the Universe, see Eq.~\eqref{eq:basypara}. However, a word of caution is in order. We have derived these bounds by modelling the missing energy spectrum by using a tree-level parton decay of a free $b$ quark, and we have not considered the effect of hadronization or any QCD corrections in our simulations. Modeling these effects falls beyond the scope of this study but we believe that including them will yield relevant modifications to the missing energy spectrum and therefore to the constraints. Including such corrections typically leads to a softer missing-energy spectrum and therefore to weaker constraints than those derived using the free quark decay model. In fact, in the erratum of~\cite{Grossman:1995gt} it is highlighted that the inclusion of such effects does indeed lead to a relaxation of the bound on $b\to s \bar{\nu}\nu$ by a factor of 3. For our decays, one should expect a similar (and perhaps larger) uncertainty. For this reason, we have added to our constraints a relaxation factor of $4$ for the operators containing a charm quark, and a factor of $2$ for the operators containing a $u$ quark. The solid lines correspond to the conservative limits, while the limits derived directly from the free $b$ quark decay are shown as the dashed line. In addition, the derived constraints depend slightly on the exact type of operator that triggers the $b$ decay, see Sec.~\ref{sec:BaryonDM_decays}. Here we choose to show the constraints for the $\mathcal{O}_{ij}^{1}=(\psi b)(u_i d_j)$ operators and we note that the bounds for the other operators are roughly the same for $m_\psi \gtrsim 2.5\,\text{GeV}$ and that they are stronger by a factor of up to 2 for $m_{\psi} \lesssim 2\,\text{GeV}$. 

\newpage 

In summary, we find that the most stringent constraint to date on b-hadron decays involving missing energy plus a baryon in the final state arises from a search made more than twenty years ago at ALEPH~\cite{Barate:2000rc}. By recasting it to our setup, we find constraints at the $\text{Br} \lesssim 10^{-4}-10^{-2}$ level for the decays of interest for $B$-Mesogenesis. We stress once more that these bounds, and in particular those appertaining the inclusive decays, should be taken with some caution given our lack of detailed knowledge of the ALEPH analysis and of the missing energy spectrum in $b\to \psi \, u_i\, d_j$ transitions. 
In view of these uncertainties, it is plausible that the constraints derived here may be weakened by a factor of a few should a more sophisticated study be performed. In any case, these results highlight the power of inclusive searches for $b$-hadron decays with large missing energy and tell us that at 95\% CL
\begin{align}
    \br < 0.5\%\, \,\,\,\text{[ALEPH]},
\end{align}
irrespective of the value of $m_\psi$.

\vspace{0.1cm}

It is clear that searches of this sort in a future $e^+e^-$ collider at the $Z$ peak such as FCC-ee~\cite{Abada:2019lih} would be able to potentially test branching fractions as small as $10^{-7}$, and therefore make a definitive test of $B$-Mesogenesis. Similarly, an inclusive search of this kind at B factories could led to an excellent sensitivity to these decays as long as backgrounds can be kept under control.

\newpage

\subsection{Possibilities at \texorpdfstring{$B$}{B} Factories}
\label{sec:BR_Bfactories}
We now discuss the potential reach of $B$ factories when searching for decays of $B$ mesons to baryons and missing energy. Given the difficulties of making theoretical predictions that can be matched to the kinematic distributions of inclusive measurements, here we focus on exclusive decays, and mostly on the simplest 2-body ones that do not contain any number of additional light mesons.

As illustrated by Eq.~\eqref{eq:PDG_inclusive}, $B$ factories have sensitivities of less than about $0.5\%$ to inclusive $B$ meson decays with final state baryons. The kinematic data used to measure~\eqref{eq:PDG_inclusive} could in principle be used to constrain 2-body exclusive channels down to comparable levels.
This can be done by comparing the number of observed baryons as a function of momentum with the theoretically predicted distribution for an exclusive channel like $B \to \psi \, \mathcal{B}$, since the 4-momenta in this case are related by $p_\mathcal{B}^\mu = p_B^\mu - p_\psi^\mu$. This approach has been leveraged to motivate a new inclusive measurement of $B\to\Lambda_c +\text{anything} $~\cite{Rizzuto:2020zeb} using Belle (and eventually Belle II) data.

Regarding exclusive measurements, $B$ factories are extremely sensitive to $B$ meson decay modes involving missing energy in the final state, such as $B\to K \bar{\nu}\nu$. Indeed, for $B\to K \bar{\nu}\nu$, BaBar and Belle's sensitivity is about $\text{Br}\sim 10^{-5}$~\cite{Lees:2013kla,Lutz:2013ftz}, while Belle II is expected to reach a sensitivity $\sim 10^{-6}$~\cite{Kou:2018nap} with $50\,\mathrm{ab}^{-1}$ of data\footnote{See~\cite{Abudinen:2021emt} for a recent inclusive search with $63\,\text{fb}^{-1}$ of data.}. In addition to this, the BaBar collaboration has very recently reported an upper limit $\text{Br}(B^-\to \Lambda \bar{p} \bar{\nu} \nu ) < 3\times 10^{-5}$~\cite{Lees:2019lme}. This 4-body decay mode contains baryons and missing energy in the final state. Therefore, given the reach of BaBar and Belle to similar $B$ meson decays, we expect a sensitivity of
\begin{align}
   \!\!\! \text{Br}\left(B\to \psi + \text{Baryon} \right) &\sim 3 \!\times\! 10^{-5}\,\,\text{(BaBar-Belle)}\,,\label{eq:BR_Babar_sensi}\\
      \!\!\!   \text{Br}\left(B\to \psi + \text{Baryon} \right) &\sim 3\!\times\! 10^{-6}\,\,\text{(Belle II \small{[50ab$^{-1}$])}}\,,\label{eq:BR_BelleII_sensi}
\end{align}
for exclusive $B \to \psi + \text{Baryon}$ decays. 

This $10^{-6}-10^{-5}$ expected sensitivity in exclusive $B$ meson decays should be compared with the prediction from $B$-Mesogenesis for the inclusive rate $\br>10^{-4}$. In Sec.~\ref{sec:excl_vs_inclu}, the ratio between exclusive and inclusive rates is estimated to be $\sim (1-10)\%$, which implies that $B$ factories have the potential to fully test the mechanism. We therefore expect large regions of parameter space of $B$-Mesogenesis to be explored via exclusive decay searches at $B$ factories. 

It should be noted that, as discussed in Sec.~\ref{sec:BaryonDM_decays}, baryogenesis can proceed via any of the operators in Eq.~\eqref{eq:LflavorIR}.  Importantly and as we show in Sec.~\ref{sec:TripletScalar}, flavor mixing constraints imply that only one of these operators can be active in the early Universe. However, there is no \emph{a priori} way to know which operator may be the one responsible for baryogenesis. Therefore, in order to fully test the mechanism, all possible flavorful variations should be independently searched for.
Furthermore, and as discussed in Sec.~\ref{sec:BaryonDM_decays}, all $B$ mesons (including charged ones) should decay into a visible baryon ($\mathcal{B}$) and missing energy at a similar rate. This means that all the searches discussed here can in principle be performed using any of the $B$ mesons, exploiting the channels listed in Table~\ref{tab:hadronmasses}. In practice, experimentally it may be easier to target $B^\pm$ decays as they leave a charged track at each interaction point.

We conclude this section by mentioning that searches are currently under development at $B$ factories for the decay modes described in our study. Concretely, the decay $B^0_d \to \psi \,\Lambda $ is being investigated using BaBar~\cite{privateBaBarBR}, Belle~\cite{privateBelleBR}, and Belle II~\cite{privateBelleBR} data.

\subsection{Possibilities at the LHC}
\label{sec:BR_LHC}
Targeting the missing energy in the decay $B\to \psi \, \mathcal{B}$ is highly non-trivial at hadron colliders due to the fact that there is no handle on the initial energy of the decaying $B$ meson. That said, there are two reasons to consider searches for these decays at hadron colliders: first, $B_s$ mesons and $b$-flavored baryons are not produced at $B$ factories when they run at the $\Upsilon(4S)$ resonance, and second, the number of $B$ mesons produced at LHC experiments is substantially larger than that of $B$ factories. For example, at LHCb with $\mathcal{L} = 5\,\text{fb}^{-1}$, $N_B\sim \mathcal{O}(10^{12})$ have been produced~\cite{Aaij:2013noa}, while at ATLAS and CMS with $\mathcal{L} = 100\,\text{fb}^{-1}$, the figure reaches $N_B \sim \mathcal{O}(10^{14})$~\cite{Aaboud:2017vqt,Khachatryan:2011mk}. This should be compared with the number of $B$ mesons that were produced at BaBar or Belle ($N_B \sim 3\times 10^{9}$), or to those that are expected to be produced at Belle II with $\mathcal{L} = 50\,\text{ab}^{-1}$, which will be $N_{B} \sim  10^{11}$. Owing to large number of $B$ mesons produced at hadron colliders, it is possible that LHC experiments could provide relevant measurements on a new decay of the $B$ meson with a branching fraction $\br > 10^{-4}$, despite the complications related to tagging the missing energy in the decay.

\vspace{-0.15cm}
\tocless \subsubsection{Prospects and Challenges for Direct Searches}
\vspace{-0.15cm}

As discussed above, performing a dedicated search for $B$ mesons decaying into missing energy at hadron colliders is complicated. In this section, we do not present a study of the feasibility of searches for these new decay modes LHC experiments, but simply discuss some aspects that need to be addressed by any proposed search. We also summarize some recent ideas that could lead to a measurement of $\br$ at LHC experiments. Three ingredients are necessary in order to perform a direct search for $B\to \psi \,  \mathcal{B}$ decays at the LHC \textit{i)} a proper trigger, \textit{ii)} good vertex resolution, and \textit{iii)} a handle on irreducible backgrounds. Addressing points \textit{i)} and \textit{ii)} is beyond the scope of this work and would require a careful assessment by experimentalists at LHCb, ATLAS and CMS. However, we can comment on point \textit{iii)} with the goal of gaining an understanding of the potential sensitivity of LHC experiments given that the mass of the $\psi$ particle cannot be reconstructed, before points \textit{i)} and \textit{ii)} are addressed. 

Let us consider the observed exclusive $B$ meson decays rates into baryons ($\mathcal{B}$) (see~\cite{Cheng:2006bn} for a nice summary):
\begin{subequations}\label{eq:B_BaryonicSMdeay}
\begin{align}
   \text{Br}\left( B\to \mathcal{B}_c \bar{\mathcal{B}}'_c \right)&\sim 10^{-3}\,, \\
   \text{Br}\left( B\to \mathcal{B}_c \bar{\mathcal{B}}_c' {M} \right) & \sim 10^{-3}\,, \\
      \text{Br}\left( B\to \mathcal{B}_c \bar{\mathcal{B}} \right)&\sim 10^{-5}\,, \label{eq:SMirre_BcB}\\ 
            \text{Br}\left( B\to \mathcal{B}_c \bar{\mathcal{B}}  {M}\right)&\sim 10^{-5}\,, \label{eq:SMirre_B1BcM} \\ 
                  \text{Br}\left( B\to \mathcal{B} \bar{\mathcal{B}}' \right)&\lesssim 10^{-7}\,, \label{eq:SMirre_B1B2}   \\ 
      \text{Br}\left( B\to \mathcal{B} \bar{\mathcal{B}}' {M} \right)&\lesssim 10^{-5}\,.\label{eq:SMirre_B1B2M}
\end{align} 
\end{subequations}
Here, $\mathcal{B}_c$ denotes a baryon containing a $c$ quark, $\mathcal{B}$ is a baryon containing just $u,\,d,\,s$ quarks, and ${M}$ corresponds to one light meson; $\pi$, $K$ or $D$.

Since the missing energy of the $\psi$ state cannot be reconstructed, decays of the type $B\to \bar{n}\, \mathcal{B}$ yield a similar signature as the decay $B \to \psi \, \mathcal{B}$. Of course, antineutrons deposit a signal in the calorimeter, but it may prove challenging to associate this signal with the $B\to \bar{n}\, \mathcal{B}$ decay given the absence of a signature in the tracker. Thus, decays of the type $B\to \bar{n}\, \mathcal{B}$ could contribute to a significant irreducible background at the LHC. 

Given the above consideration and using Eq.~\eqref{eq:B_BaryonicSMdeay}, one finds an irreducible background at the $\text{Br}\sim \mathcal{O}(10^{-5})$ level for $B\to \psi \, \mathcal{B}_c$. If the baryon in the final state does not contain a $c$ quark, then the irreducible background would be at the $\text{Br}\sim \mathcal{O}(10^{-7})$ level. This is interesting as successful baryogenesis requires $\br \gtrsim 10^{-4}$ and so, at least in principle, the LHC could probe the relevant parameter space. Of course, one should bear in mind that these considerations are not specific; depending on the exact final state baryon that is targeted in the $B\to \psi \,\mathcal{B}$ decay, one would find a different contribution to the irreducible background. Such contributions would therefore need to be studied in a case by case basis and may very well differ from the ones we estimated above.

\tocless \subsubsection{Possible Searches}

We now highlight some recent ideas for potential LHC searches for the exotic decays of $b$-hadrons required for baryogenesis. The focus here is on the LHCb experiment~\cite{Borsato:2021aum}, but we believe that similar ideas could be pursued at ATLAS and CMS. As discussed above, targeting missing energy at hadron colliders is non trivial, but there have been some recent attempts that we find are worth discussing:
\\
\begin{itemize}[leftmargin=0.5cm,itemsep=0.4pt]
    
    \item $B\to \psi \,\mathcal{B}^\star $ Searches:
    Motivated by the excellent vertex resolution and high particle reconstruction efficiencies of the LHCb experiment, the feasibility of the direct search of $B$ mesons into a visible baryon has been recently studied~\cite{Rodriguez:2021urv}. In particular, Ref.~\cite{Rodriguez:2021urv} has considered the case of excited baryons in the final state, $B\to \psi \,\mathcal{B}^\star $.  Excited baryons decay promptly, making the decay point $B\to \psi \,\mathcal{B}^\star $ coincide with that of $\mathcal{B}^\star $ and thus allowing one to trigger on the decay. This study has shown that for modes such as $B^0 \to \psi\,\Lambda(1520)$ the sensitivity of LHCb with $15\,\text{fb}^{-1}$ of data can be at the $\text{Br} \sim (6-60)\times 10^{-7}$ level for $m_\psi \lesssim  3.5\,\text{GeV}$. In addition, for the channel $B^+ \to \psi\,\Lambda_c(1520)$ the sensitivity could reach values $\text{Br} \sim (2-4)\times 10^{-6}$ for $m_\psi \lesssim  2.5\,\text{GeV}$. Although the exact reach of the LHCb experiment to these decays will depend upon the control of several backgrounds~\cite{Rodriguez:2021urv}, these results highlight the high sensitivity of the LHCb experiment to search the decay modes required by $B$-Mesogenesis.

    \item $\mathcal{B}_b\to \bar{\psi} \, \mathcal{M}$ Searches: Within the context of $B$-Mesogenesis, $b$-flavored baryons also posses an exotic decay into missing energy and light mesons with $\text{Br}\left(\mathcal{B}_b \to \bar{\psi} \, \mathcal{M}\right)\simeq \br$. Since $b$-flavored baryons are not produced at $B$ factories, \emph{the LHC is the only current experiment potentially capable of searching for these processes}. In this context, Ref.~\cite{Stone:2014mza} proposed a method to search for $b$-flavored baryon decays involving missing energy by studying $\mathcal{B}_b$ baryons arising from the decay of resonant baryons, $\mathcal{B}_b^\star \to \mathcal{B}_b \, \pi$. In this way, the initial energy of $\mathcal{B}_b$ could be inferred from the decay kinematics of $\mathcal{B}_b^\star \to \mathcal{B}_b \, \pi$, thus offering a handle on decay channels of the $\mathcal{B}_b$ state involving missing energy. It would be interesting to study the reach of such proposals for the decay  $\mathcal{B}_b\to \bar{\psi} \, \mathcal{M}$. For example, $\Lambda_b^0 \to K^+\, \pi^-\, \bar{\psi}$ and $\Lambda_b \to D^-\,K^+\,\bar{\psi} $ decays as generated by operators $\mathcal{O}_{us}$ and  $\mathcal{O}_{cs}$ in Eq.~\eqref{eq:LflavorIR} could be interesting channels to pursue. In fact, it has been recently shown in Ref.~\cite{Rodriguez:2021urv} that LHCb can have a sensitivity at the level of $\text{Br} \sim 3\times 10^{-5}-10^{-4}$ for decays of the type $\Lambda_b^0 \to K^+\, \pi^-\, \bar{\psi}$.
    
    \item $B_s^\star \to B\to \psi \,\mathcal{B} $ Searches: 
    Very recently, the LHCb experiment has been able to set the constraint $\text{Br}\left(B^+ \to K^+\mu^-\tau^+\right)<3.9\times 10^{-5}$~\cite{Aaij:2020mqb} at 90\% CL by using $B^+$ mesons arising from the resonant $B$ meson decay $B_{s2}^{0\star} \to B^+ K^-$. In~\cite{Aaij:2020mqb}, the $\tau^+$ four-momentum is reconstructed from the kinematics of the rest of the particles and is thus effectively treated as a missing particle. Therefore, it is potentially feasible that LHCb could also search for $B_s^\star \to B\to \psi \,\mathcal{B} $ processes following a similar approach and achieve sensitivities comparable to that of $B^+ \to K^+\mu^-\tau^+$~\cite{Aaij:2020mqb}.
    
    \item $B_s^0\to \psi \, \mathcal{B}$ with Oscillations:
    Measuring the mass of $\psi$ in hadron colliders seems challenging. However, Ref.~\cite{Poluektov:2019trg} has pointed out that one could gain a handle on the mass of $\psi$ by studying the oscillation pattern of $B_s^0\to \psi \, \mathcal{B}$ decays. Given that this approach needs to trigger on the $B_s^0\to \psi \, \mathcal{B}$ decay explicitly, it necessarily requires searches such as the ones described in the two previous paragraphs to find the decay first.

\end{itemize}

To conclude, although missing energy searches at the LHC seem at first complicated, there are interesting avenues that are currently being explored and that may shed light on the new decay modes required for the $B$-Mesogenesis. Of course, all of the above considerations could also apply to potential future hadron collider experiments, see e.g.~\cite{Chobanova:2020vmx}.

\subsection{Relating Exclusive to Inclusive Decays}
\label{sec:excl_vs_inclu}

While the observable directly linked to baryogenesis is the inclusive $B\to \psi \, \mathcal{B}\,\mathcal{M}$ branching ratio, the experimental searches discussed above are best suited to test exclusive final states without additional mesons, $B\to \psi \, \mathcal{B}$.
In fact, the presence of additional hadronic states accompanying the final-state baryon can significantly modify the expected sensitivities in Eqs.~\eqref{eq:BR_Babar_sensi} and~\eqref{eq:BR_BelleII_sensi}.
It is therefore crucial to estimate the relative size of the exclusive modes that contribute to a given inclusive channel.

\begin{figure*}[t]
\centering
\begin{tabular}{cc}
		\label{fig:Br_ratio_Bplus1}
		\includegraphics[width=0.48\textwidth]{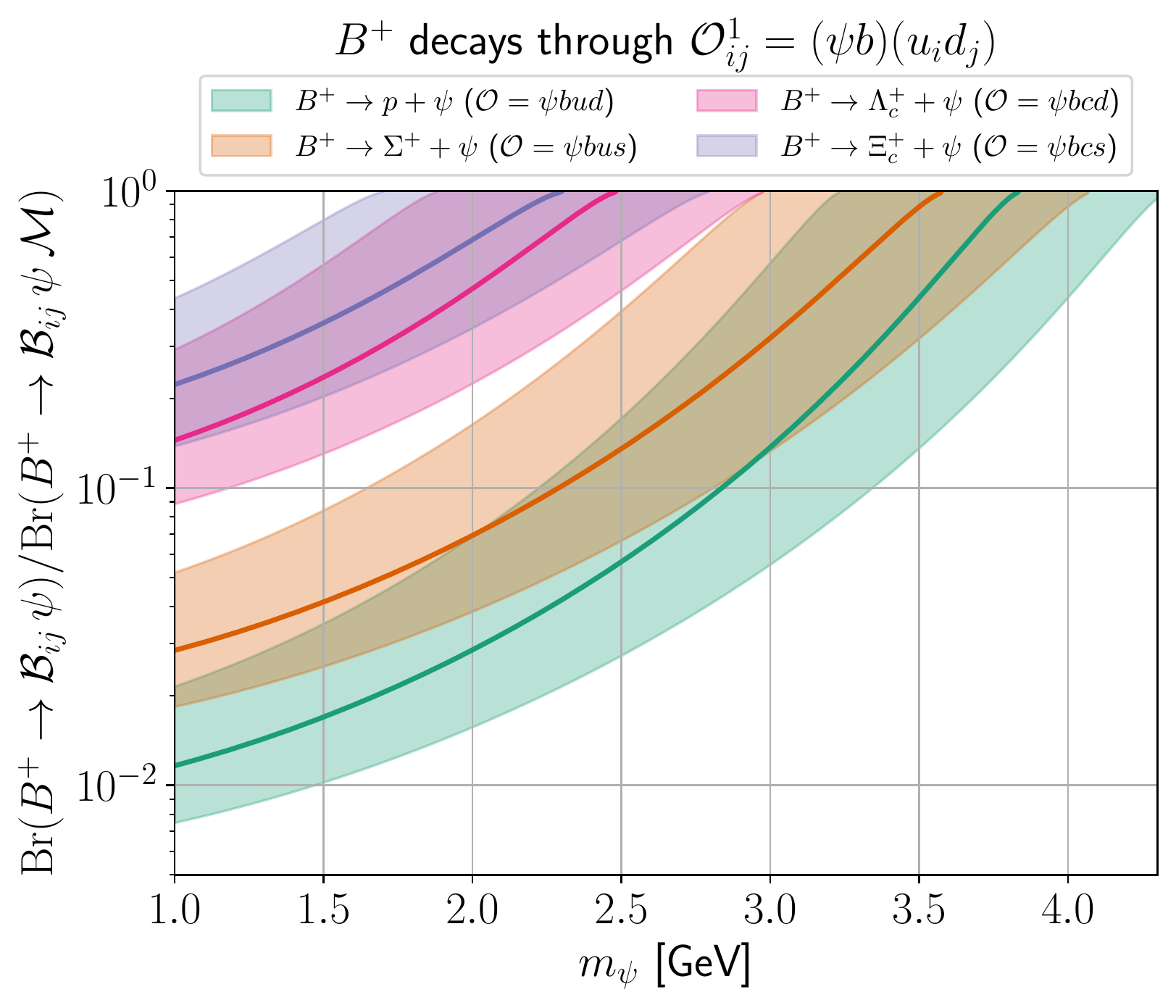}
&
		\label{fig:Br_ratio_Bplus2}
		\includegraphics[width=0.48\textwidth]{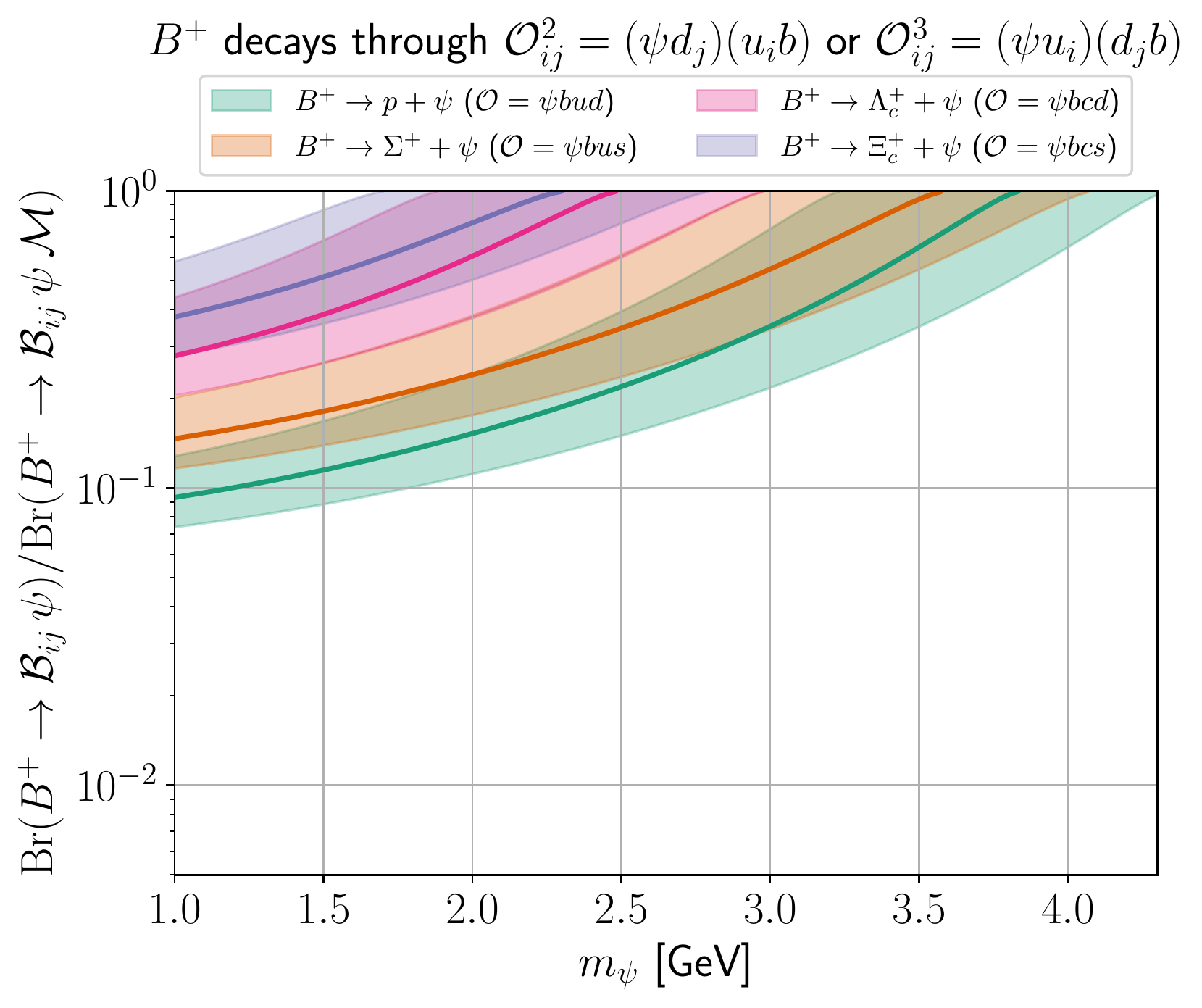}
\end{tabular}
\vspace{-0.4cm}
\caption{Fraction of $B^+\to \psi\,\mathcal{B}\,\mathcal{M}$ decays that are not expected to contain hadrons other than $\mathcal{B}$ in the final state, as a function of the mass of the dark fermion $\psi$.
Different colors correspond to decays induced by the different operators listed in Table~\ref{tab:hadronmasses}.
Each panel corresponds to a different kinematic structure of the effective four-fermion operator as listed in Table~\ref{tab:mchibounds}.
The width of the band represents an estimation of the uncertainty in our computations, and is obtained by varying the $b$-quark mass used in the calculation between $\bar{m}_b(\mu=\bar{m}_b)$, $m_b^{\rm pole}$ (solid line), and $m_{B^0_d}$.
}
\label{fig:Br_ratio_Bplus}
\end{figure*}

The task of obtaining form factors for the exclusive branching ratios induced by the operators listed in Table~\ref{tab:hadronmasses} requires sophisticated methods that can deal with the non-perturbative hadronization process.
In the simplest case of $B\rightarrow \psi\,\mathcal{B}$, there is only one form factor which is a function of the recoil energy $q^2=m_{\psi}^2$.
Form factors are often obtained using data-driven methods, which are, however, not helpful in this case given that baryon number-violating $B$ decays have never been observed.
As an alternative, one could hope to obtain a purely theoretical estimate using the vacuum insertion approximation in the lines of~\cite{Gaillard:1974nj}.
However, the presence of a spectator quark makes this technique challenging to apply to our decays.
Furthermore, the vacuum insertion approximation requires that all the states involved share the same quantum numbers as the vacuum and can therefore only be applied to neutral decays.
More sophisticated possibilities would be to perform a lattice QCD calculation, or use a QCD sum rule, which are, however, beyond the scope of this paper.

In this humbling situation, the only way forward is to resort to phase-space considerations in order to obtain an estimate.
The kinematic arguments that we use were first employed by Bigi in~\cite{Bigi:1981bb} to predict a sizeable decay rate of $B$ mesons to baryons within the SM. 
In this prescription, the probability for the final state of the $B$ decay induced by a $\bar{b}\rightarrow u_i d_j \psi$ transition to contain a single baryonic state is calculated as the fraction of phase space in which the invariant mass $M_{u_id_j}$ of the diquark system does not exceed a certain energy scale $\Lambda$, which we take to be the mass of the corresponding baryon.
In the remaining region of phase space, $M_{u_id_j} \gg \Lambda$ and we would expect that resonant baryons are produced, which naturally leads to additional hadrons (mostly pions) in the final state.
Although simple in nature, the predictions made using this approach~\cite{Bigi:1981bb,Dunietz:1996es,Dunietz:1998uz,Bigi:1995jr} were in all cases subsequently validated by experiment.
For example, in~\cite{Dunietz:1998uz} these considerations were used to understand the softness of the $B \to \Lambda_c +X$ spectrum, and in~\cite{Bigi:1981bb} it was predicted that $\text{Br}\left(B\to \mathcal{B}+ X\right) \simeq 5-10\%$ which agrees well with the observed inclusive measurements, see Eq.~\eqref{eq:BR_inclu_p}. 

\begin{figure*}[t]
\centering
\begin{tabular}{cc}
		\label{fig:Br_ratio_Bd1}
		\includegraphics[width=0.48\textwidth]{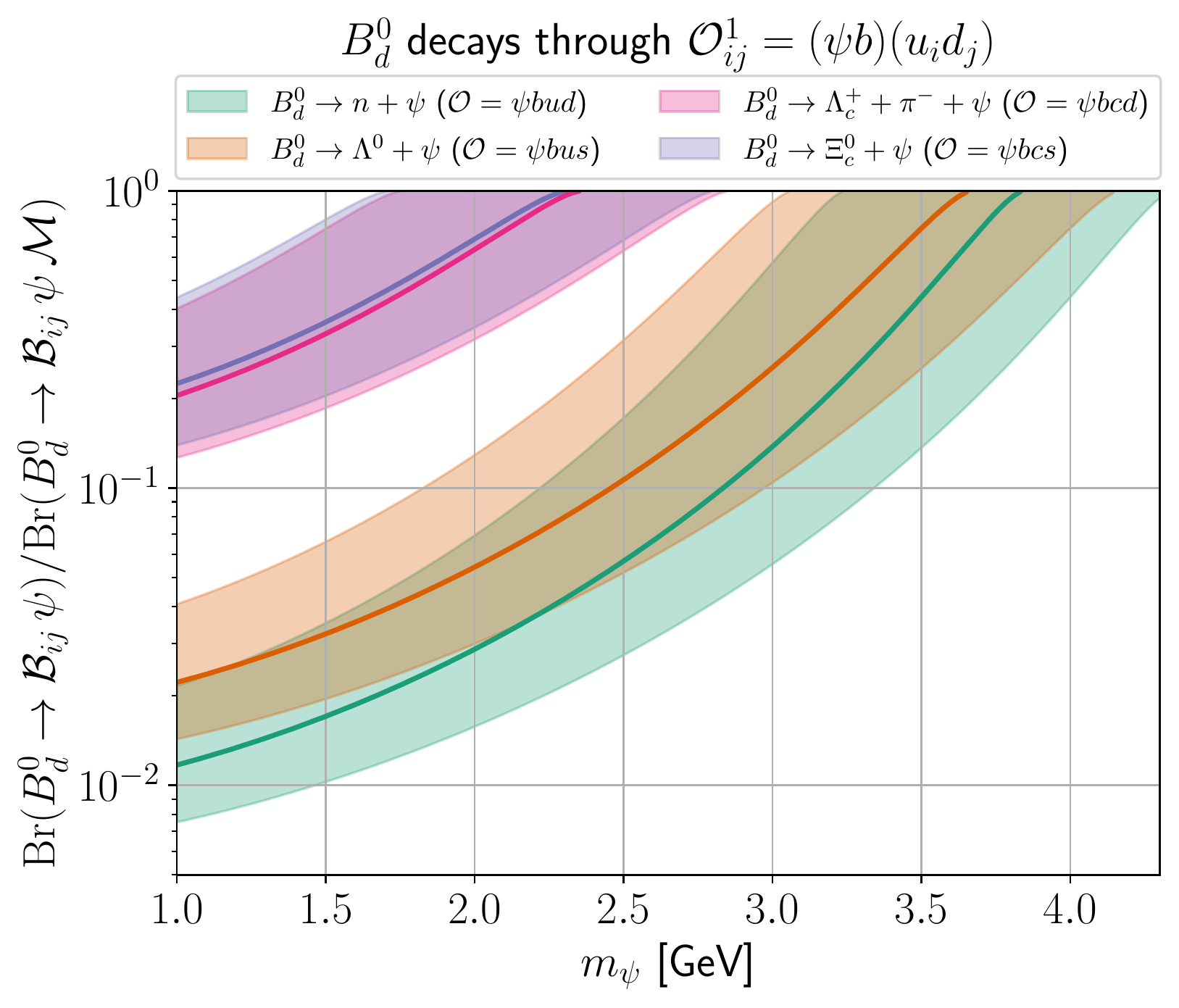}
&
		\label{fig:Br_ratio_Bd2}
		\includegraphics[width=0.48\textwidth]{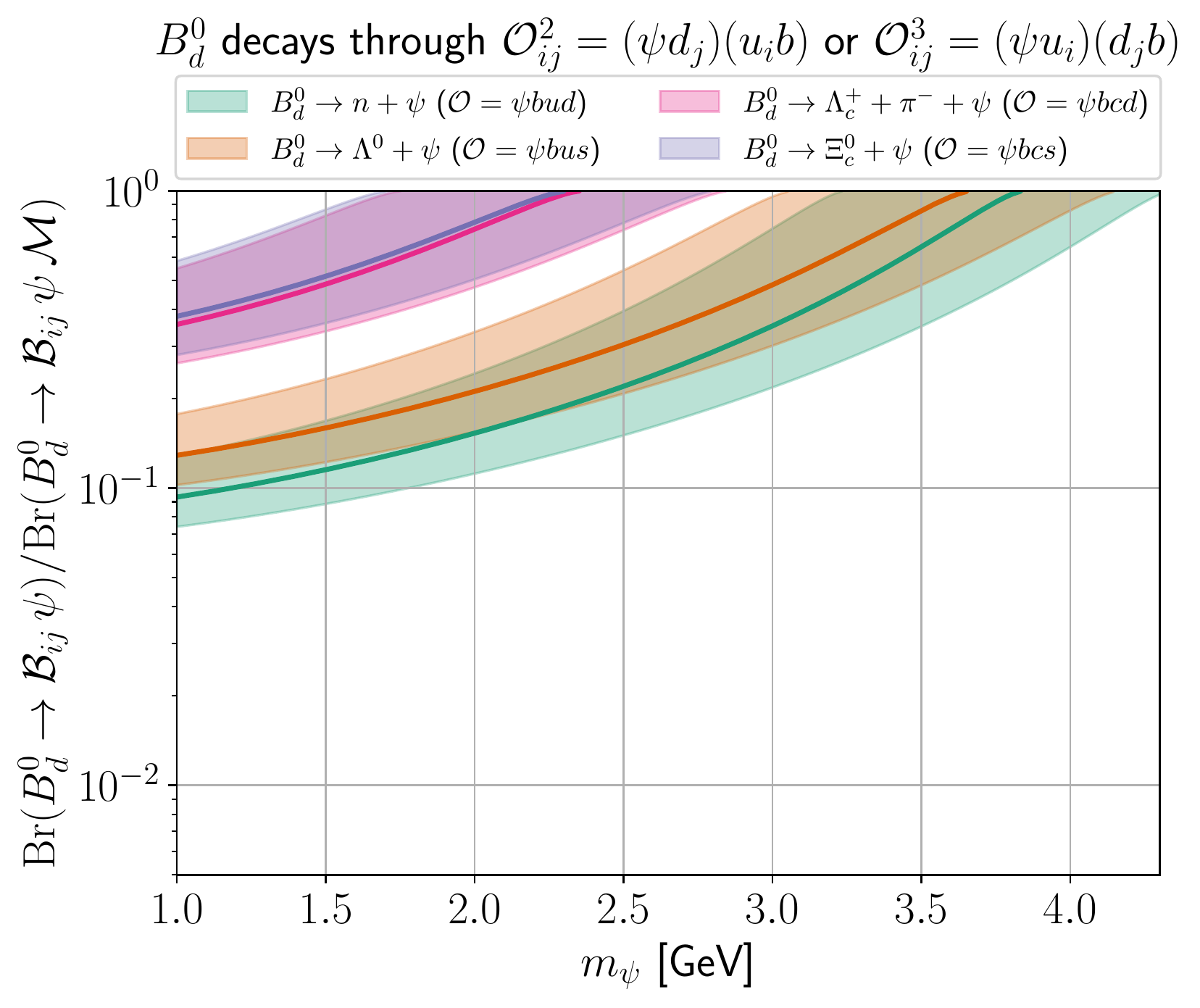}
\end{tabular}
\vspace{-0.4cm}
\caption{Same as Fig.~\ref{fig:Br_ratio_Bplus} but for $B^0_d\to \psi\,\mathcal{B}\,\mathcal{M}$ decays.
}
\label{fig:Br_ratio_Bd}
\end{figure*}

We choose to follow this avenue and study the kinematics of the $B\to \psi +\mathcal{B}+\mathcal{M}$ decays of interest.
Defining the phase-space constrained width of the parton-level decay $\bar{b}\rightarrow u_id_j\psi$ as
\begin{equation}\label{eq:phase_space}
    \gamma(\Lambda) \equiv \int_{(m_{u_i}+m_{d_j})^2}^{\Lambda^2} \frac{\partial\Gamma}{\partial M^2_{u_id_j}}\,\mathop{\diff M^2_{u_id_j}}\, ,
\end{equation}
the desired ratio of phase-space integrals corresponds to
\begin{equation}\label{eq:Br_ratio_phase_space}
    \frac{\mathrm{Br}(B\rightarrow\mathcal{B}_{ij}+\psi)}{\mathrm{Br}(B\rightarrow \mathcal{B}_{ij} + \psi + \mathcal{M})} \simeq \frac{\gamma(m_{\mathcal{B}_{ij}})}{\gamma(m_b-m_\psi)}\, ,
\end{equation}
where $m_{\mathcal{B}_{ij}}$ denotes the mass of the lightest baryon with matching flavor content as listed in Table~\ref{tab:hadronmasses}.
The ratio above critically depends on the mass of the dark Dirac antibaryon $\psi$, which is unknown but bound to lie in the $0.94 \GeV < m_\psi < 4.34 \GeV$ window (see Eq.~\eqref{eq:psirange}).
The resulting estimation for the expected fraction of decays containing only a baryon and $\psi$ in the final state is shown in Figs.~\ref{fig:Br_ratio_Bplus} and~\ref{fig:Br_ratio_Bd} as a function of $m_\psi$.
For brevity, here we only display the results for charged $B$ meson and neutral $B_d$ decays, as they were found to be the most promising experimental targets in Sec.~\ref{sec:BR_Bfactories}.
The results for $B_s^0$ mesons as well as $\Lambda_b$ baryons are presented in Appendix~\ref{sec:Exclusive_vs_Inclusive}.

For each operator $\mathcal{O}_{ij}=\psi bu_id_j$, the phase-space integration depends on the matrix element obtained from the effective Lagrangian~\eqref{eq:b_decay1}.
Different combinations of the quarks in the dimension-6 operators in Eq.~\eqref{eq:LflavorIR} lead to different contractions of external momenta.
Given this dependence on the kinematic structure of the matrix element, we choose to separate the results of different quark combinations in Figs.~\ref{fig:Br_ratio_Bplus} and~\ref{fig:Br_ratio_Bd}.
In these figures, the left panel corresponds to the ``type-1'' operator $\mathcal{O}^1_{ij} = (\psi b)(u_id_j)$, while the right one corresponds to the ``type-2'' and ``type-3'' cases $\mathcal{O}^2_{ij} = (\psi d_j)(u_ib)$ and $\mathcal{O}^3_{ij} = (\psi u_i)(d_jb)$, for which the phase-space integration is very similar.
Note that the type-2 and type-3 combinations always yields a larger phase-space ratio than the type-1 one.
This means that it is easier to probe the inclusive branching ratio $B\to \psi \,\mathcal{B}\,\mathcal{M}$ by measuring the exclusive channel $B\to \psi \,\mathcal{B}$ if the effective operators are of the former types.

An important question is related to the value of $m_b$ that should be used when evaluating the phase-space integral in Eq.~\eqref{eq:phase_space}.
Arguments can be made in favour of using the pole mass $m_b^{\rm pole}=4.78\,\mathrm{GeV}$~\cite{pdg} or the $\overline{\rm MS}$ mass at the corresponding energy scale $\bar{m}_b(\mu=\bar{m}_b)=4.18\,\mathrm{GeV}$~\cite{pdg}, or even the mass of the decaying $B$ meson against which the diquark system is recoiling. 
In order to be conservative regarding this choice, we decide to use this indetermination as one measure of the uncertainty in our calculation.
We choose the intermediate value $m_b^{\rm pole}$ as our benchmark, corresponding to the solid lines in Figs.~\ref{fig:Br_ratio_Bplus} and~\ref{fig:Br_ratio_Bd}, while $\bar{m}_b(\mu=\bar{m}_b)$ and $m_B$ respectively delineate the upper and lower edges of the shaded bands in those figures.
As can be seen in the figures, this amounts to a factor of $\sim 2$ uncertainty in our predictions for the inclusive vs. exclusive rates, which is reasonable given the purely kinematic nature of our arguments.

From this analysis and as can be seen in Figs.~\ref{fig:Br_ratio_Bplus} and~\ref{fig:Br_ratio_Bd}, we learn that the exclusive modes that do not contain any extra pions in the final state are expected to constitute a $1-100\%$ fraction of the inclusive width of $B$ mesons, where the larger numbers correspond to heavier $\psi$-particles, which restrict the available phase-space.
Although this estimate is clearly rough and a dedicated calculation using lattice QCD or QCD sum rules is highly desirable, we expect it to be a good order-of-magnitude indicator of the behaviour of the actual form factor.
As a consequence, the searches proposed in Sec.~\ref{sec:BR_Bfactories} should aim to test exclusive $B\to \psi \, \mathcal{B}$ modes down to a branching fraction of $\sim 10^{-6}-10^{-5}$ in order to completely probe the parameter space that allows for successful $B$-Mesogenesis, see Eq.~\eqref{eq:BR_constraint_ALL_meas}.
It is also worth noting that a search for $B\to \psi \, \mathcal{B}^\star$ with a sensitivity of $\text{Br}\sim 10^{-5}-10^{-4}$ would yield complementary information to $B\to \psi \, \mathcal{B}$ searches.
The reason is that given that $\text{Br}\left(B\to \psi \, \mathcal{B} \right) +\text{Br}\left(B\to \psi \, \mathcal{B}^\star \right) \simeq \br $ and that $\text{Br}\left(B\to \psi \, \mathcal{B} \right)$ is small for light $\psi$ masses, one expect $\text{Br}\left(B\to \psi \, \mathcal{B}^\star \right)$ to be large in this regime.
We believe that this serves as further motivation to perform searches for these exotic $B$-meson decays at LHCb in addition to BaBar, Belle and Belle II, taking advantage of the channels containing excited baryons as described in Sec.~\ref{sec:BR_LHC}.
LHCb also offers the possibility to search for exotic $b$-baryon decays such as $\Lambda_b\rightarrow\bar{\psi}\mathcal{M}$.
Fig.~\ref{fig:Br_ratio_Lambdab} shows that a large fraction of these decays are expected to be into final states with multiple mesons, which is to be expected given the large phase available in these decays if $\psi$ is not too heavy.
This information should help design appropriate search strategies when targeting these decays in order to test $B$-Mesogenesis.

\section{Color-Triplet Scalar}\label{sec:TripletScalar}
\begin{figure*}[t]
\centering
\includegraphics[width=0.95\textwidth]{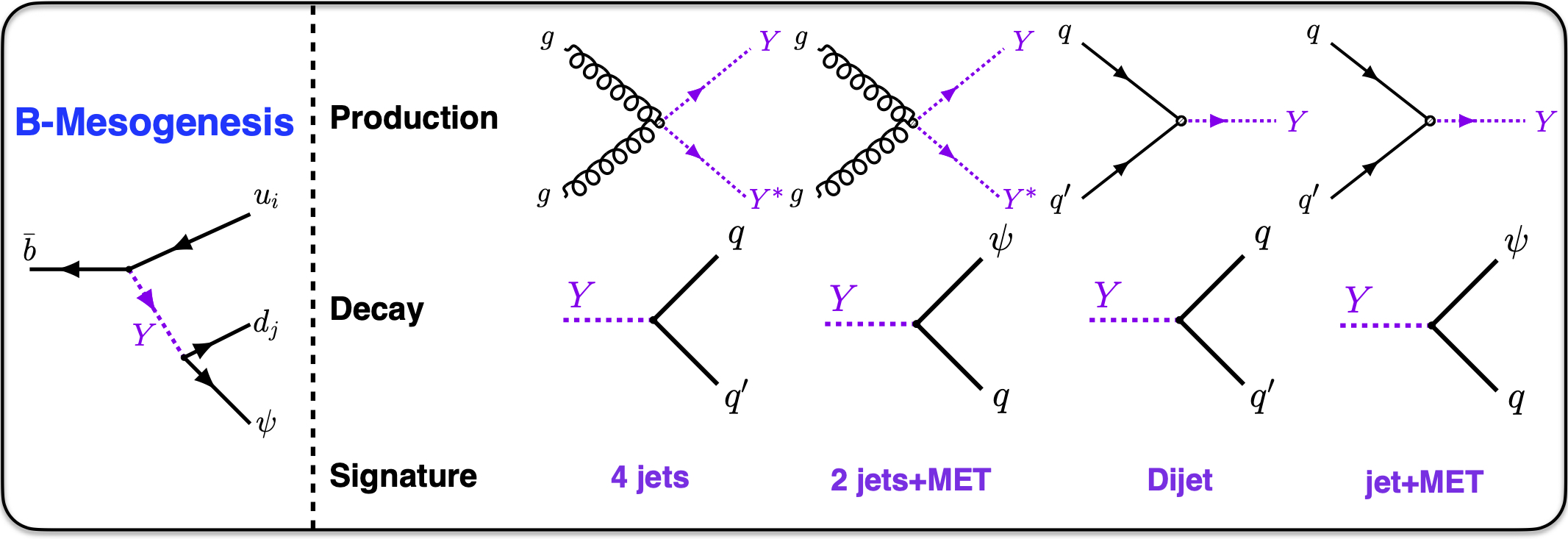}
\vspace{-0.3cm}
\caption{Summary of the most important production and decay modes of a color-triplet scalar at the LHC, together with their associated signatures. Some of the couplings involved in these processes can be directly identified with the ones that mediate baryogenesis and dark matter production.}
\label{fig:Y_cartoon}
\end{figure*}

The four-fermion operator in Eq.~\eqref{eq:LflavorIR}, which triggers the new decay mode of the $B$ mesons necessary for baryogenesis, can arise in a UV model with a color-triplet scalar mediator with baryon number $-2/3$. We denote this scalar mediator by $Y$.
It must be a $SU(2)_L$ singlet and carry hypercharge $-1/3$ or $+2/3$, just like a right-handed $d$- or $u$-type squark.
While the discussion in~\cite{Elor:2018twp} focused on the $Y\sim (3,\,1,\,-1/3)$ option, here we also consider the choice of possible charge assignment $Y\sim (3,\,1,\,2/3)$.
Although the results are qualitatively similar for both scenarios, current experimental constraints are less stringent for some flavorful variations of the hypercharge $2/3$ version.
As we will see, this has important consequences for determining which of the operators in Table~\ref{tab:hadronmasses} are best suited for $B$-Mesogenesis.

The most general renormalizable Lagrangian that can be written for a (hypercharge $-1/3$ or $2/3$) color-triplet scalar interacting with quarks and the SM singlet baryon $\psi$ is
\begin{subequations}\label{eq:LflavorUV}
\begin{align}
\!\! \mathcal{L}_{-1/3} &= - \sum_{i,\,j} y_{u_id_j} Y^\star \bar{u}_{iR} d_{jR}^c - \sum_k y_{\psi d_k} Y d_{kR}^c \bar{\psi} + \text{h.c.} \,, \label{eq:LflavorUV_13}\\
\!\! \mathcal{L}_{2/3} &= - \sum_{i,\,j} y_{d_id_j} Y^\star \bar{d}_{iR} d_{jR}^c - \sum_k y_{\psi u_k} Y  u_{kR}^c \bar{\psi} + \text{h.c.} \,, \label{eq:LflavorUV_23}
\end{align}
\end{subequations}
where the $y$'s are coupling constants, the sum is performed over all up and down type quarks, and we are working in the quark mass basis (i.e. where the Higgs-quark Yukawa matrix is diagonal).
The color indices in the diquark operators are contracted in a totally antisymmetric way, so that $y_{d_i d_j}$ must be an antisymmetric matrix with only 3 relevant entries.
Note that all quarks here are right handed and $Y$ carries baryon number $-2/3$ so that Eq.~\eqref{eq:LflavorIR} is a \emph{baryon number conserving} Lagrangian. The interactions of $Y$ are reminiscent of those of squarks in R-parity violating supersymmetric scenarios, see~\cite{Alonso-Alvarez:2019fym} for the details of such a realization.

In this section we turn our attention to the phenomenology associated with this color-triplet scalar. First, in Sec.~\ref{sec:Ytrip_baryo} we discuss the requirements on $Y$ such that the requisite $\br$ needed for baryogenesis is achieved. As we shall see, this requires the triplet scalar to have a TeV-scale mass. Next in Sec.~\ref{sec:LHC_colored_triplet_scalar}, we study constraints arising from LHC searches for colored scalars on the couplings that contribute to $\br$, which can therefore indirectly test $B$-Mesogenesis. In Sec.~\ref{sec:Ytrip_Flavor}, we also present a study of flavor mixing constraints due to observables in the neutral $B$, $D$, and $K$ systems that shape the flavor structure of the $y$ coupling matrices but that do not necessarily test the mechanism indirectly. In particular, in Sec.~\ref{sec:Decays} we show that sizeable new tree-level $b$ decays can arise in this setup. Finally, in Sec.~\ref{sec:ModelBuilding_triplet} we present a combined discussion of the searches for this color-triplet scalar in light of its role as mediator in the generation of the baryon asymmetry and the dark matter of the Universe.

\subsection{Requirements for \texorpdfstring{$B$}{B}-Mesogenesis}\label{sec:Ytrip_baryo}
Before diving into the experimental phenomenology of the triplet scalar $Y$, we first delineate the parametric requirements of the Lagrangian in Eq.~\eqref{eq:LflavorUV} to successfully generate the new $B$ meson decay.
The rate for a new decay mode of the b-quark into lighter quarks forming a baryon $\mathcal{B}$ and $\psi$ can be estimated as~\cite{Elor:2018twp}\footnote{This estimate should be accurate up to a $20\%$ QCD correction, compare the first number of the first and last rows of Table 1 of~\cite{Krinner:2013cja}.}
\begin{align}
\br \simeq 10^{-3} \left(\frac{\Delta m}{3~\rm GeV}\right)^4  \left(\frac{1.5~\rm TeV}{{M_Y}} \frac{\sqrt{y^2}}{0.53} \right)^4 \,,
\label{eq:bdecayrate}
\end{align}
where $\Delta m = m_B - m_\psi - m_{\mathcal{B}}- m_{\mathcal{M}}$ is the mass difference between the initial and final states and $y^2 \equiv y_{u_{i} d_{j}}\,y_{\psi d_k}$ or $y^2 \equiv y_{d_{i} d_{j}}\,y_{\psi u_k}$ represents the product of two of the couplings in the interaction Lagrangians in Eq.~\eqref{eq:LflavorUV}, provided a $b$-quark is involved and depending on the hypercharge of $Y$.
The benchmark value for $M_Y$ is chosen to be consistent with LHC bounds on colored scalars, as is discussed in the following Sec.~\ref{sec:LHC_colored_triplet_scalar}.
Given the results shown in Fig.~\ref{fig:semileptonic_asymmetries}, successful baryogenesis requires $\br > 10^{-4}$.
This implies that the size of the couplings is bound to be
\begin{align}\label{eq:Baryogenesis_range}
 \sqrt{y^2} \,\,>\,\, 0.3 \frac{M_Y}{1.5\,\text{TeV}}  \frac{3\,\text{GeV}}{\Delta m} \left( \frac{\br}{10^{-4}} \right)^{\tfrac{1}{4}},
\end{align}
which in turn implies rather large coupling constants for $M_Y \gtrsim 1.5\,\text{TeV}$.
Unitarity of the processes $q\bar{q}\to q\bar{q}$ and $q\psi \to q\psi $ require $\sqrt{y^2} < \sqrt{4\pi}$.
This means that if $Y$ is to trigger the new decay mode of the $b$ quark as needed for baryogenesis and dark matter production in the early Universe, then $Y$ cannot be too heavy.
More concretely, we can put upper bounds on its mass depending on the branching ratio assumed:
\begin{subequations}\label{eq:MYupperlimits}
\begin{align}
      \!\!\!\! M_Y &< 5\,(2.5)\,\text{TeV} && \mathrm{for}\;\;\br = 10^{-2}\,, \\
       \!\!\!\! M_Y &< 9\,(4.5)\,\text{TeV} && \mathrm{for}\;\;\br = 10^{-3}\,,\\
       \!\!\!\! M_Y &< 16\,(8)\,\text{TeV} && \mathrm{for}\;\;\br = 10^{-4}\,.
\end{align} 
\end{subequations}
The numbers with and without parentheses correspond to setting $\Delta m = 1.5\,\text{GeV} $ and $\Delta m = 3\,\text{GeV} $, respectively.
These are approximately the maximum mass differences for decays with and without a $c$-quark in the final state baryon.
Comparing this with Fig.~\ref{fig:semileptonic_asymmetries}, we conclude that the color-triplet scalar may be within the reach of direct searches at the LHC. This has important implications for upcoming ATLAS and CMS searches as detailed below.

\begin{figure*}[t]
\centering
\begin{tabular}{cc}
		\label{fig:Y_Baryo1}
		\includegraphics[width=0.48\textwidth]{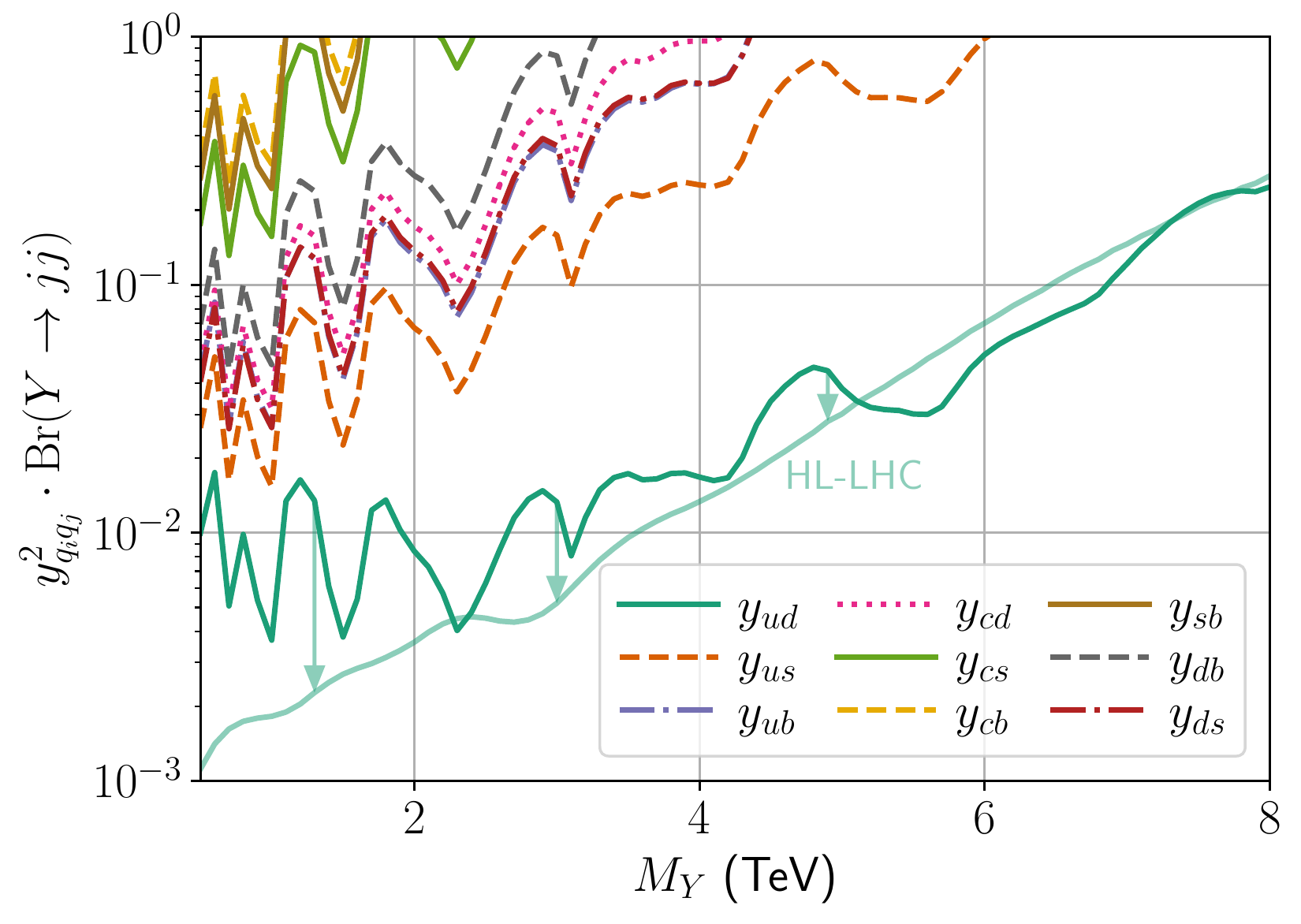}
&
		\label{fig:Y_Baryo2}
		\includegraphics[width=0.48\textwidth]{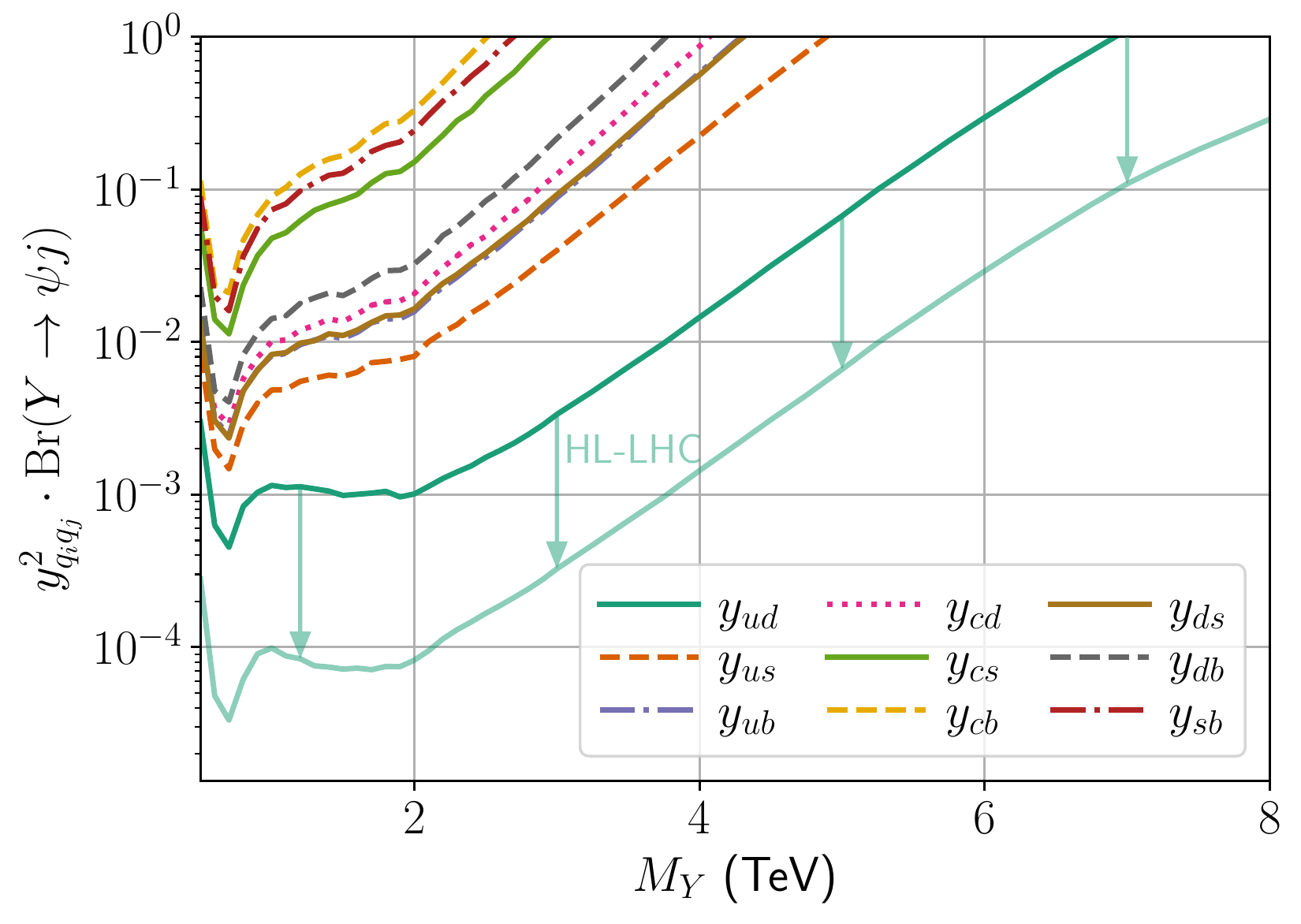}
\end{tabular}
\vspace{-0.4cm}
\caption{\textit{Left panel}: Constraints on $y_{q_iq_j}^2\, \text{Br}(Y\to j \,j)$ from dijet searches. Bounds are based on a recast of the CMS search of~\cite{Sirunyan:2018xlo} that used $36\,\text{fb}^{-1}$ of data at 13 TeV. We have assumed $\Gamma_Y/M_Y = 0.04$ and, as discussed in~\cite{Sirunyan:2018xlo}, an acceptance of $A=0.56$. \textit{Right panel}: Constraints on $y_{q_iq_j}^2\, \text{Br}(Y\to j \,\psi)$ from jet plus missing energy signals. Bounds have been recasted from the analysis of ATLAS~\cite{Aaboud:2017phn} that used $36\,\text{fb}^{-1}$ of data at 13 TeV. We have assumed $\Gamma_Y/M_Y = 0.06$ and performed a fit without correlating errors to the exclusive signal regions of~\cite{Aaboud:2017phn}.}
\label{fig:Y_LHC}
\end{figure*}
\subsection{LHC Searches for Color-Triplet Scalars}\label{sec:LHC_colored_triplet_scalar}
A color-triplet scalar baryon with a TeV-scale mass has the potential to be copiously produced at the LHC.
Depending on the production and decay channels, one expects various different signatures as highlighted in Fig.~\ref{fig:Y_cartoon}. Firstly, an ATLAS~\cite{Aaboud:2017nmi} search for resonant 4-jets with 36.7 $\text{fb}^{-1}$ of data shows that pair-produced triplet scalars that decay into quarks should have $M_Y > 0.5\,\text{TeV}$. In addition, ATLAS~\cite{Aad:2020aze} and CMS~\cite{Sirunyan:2019ctn} SUSY searches for pair-produced squarks decaying into a neutralino and a quark rule out the existence of this kind of strongly interacting bosons in the mass region below $1.2\,\text{TeV}$ with the current 139 $\text{fb}^{-1}$ of data.
For larger masses, pair production becomes kinematically suppressed and the bounds weaken drastically.
However, resonant single $Y$ production is still possible above this threshold via the process $q_{i}q_j\rightarrow Y$, enabled by the first couplings in Eq.~\eqref{eq:LflavorUV}.
The production cross-section varies significantly depending on the exact combination of flavors in each $y_{u_i d_j}$ or $y_{d_i d_j}$ coupling due to the proton PDFs, with a difference of about two orders of magnitude between the most favorable combination $ud$ and the least favorable one $cb$ (we do not consider contribution from top quarks, as they are too heavy to participate in the $b$-quark decay).
While the bounds on pair-produced triplet scalars have been thoroughly inspected, single production has been less explored (see, however,~\cite{Dobrescu:2019nys,Pascual-Dias:2020hxo} for recent studies of other heavy colored resonances) and we therefore analyze it in detail below.

Once produced, the triplet scalar can decay via any of the two types of Yukawa couplings in Eq.~\eqref{eq:LflavorUV}, either into two quarks or into a $\psi$-quark pair.
For $M_Y \gg m_q,\,m_\psi$, the corresponding partial decay rates are
\begin{align} \label{eq:PartialWidths}
\!\!\! \Gamma(Y\to \bar{q_i} \bar{q_j}) &= 2 \frac{y_{q_i q_j}^2}{16\pi} M_Y \,,\,\,\,\,
\Gamma(Y\to \psi q_i) =  \frac{y_{\psi q_i}^2}{16\pi} M_Y \,.
\end{align}
These two decay channels lead to very different signatures at the LHC detectors.
The $qq$ final state can be targeted using dijet searches.
In order to obtain a limit, we perform a recast of the CMS analysis presented in~\cite{Sirunyan:2018xlo}, which uses $36\,\text{fb}^{-1}$ of data at $13$~TeV.
For that, we first implement our particles and interactions in \texttt{FeynRules}~\cite{Alloul:2013bka} and then use \texttt{MadGraph5\_aMC@NLO}~\cite{Alwall:2014hca} in order to calculate the lowest order (LO) $Y$ production cross section as a function of $y_{q_i q_ j}$ and $M_Y$. Then, we compare the LO prediction with the constraints on production cross section limits times the dijet branching fraction in Fig. 12 of~\cite{Sirunyan:2018xlo} by taking the acceptance $A=0.57$ as relevant for the isotropic decays that we consider~\cite{Sirunyan:2018xlo}. 

Finally, using the narrow width approximation and following the previously described procedure, we can translate the constraints from~\cite{Sirunyan:2018xlo} into a bound on the product $y_{q_iq_j}^2\, \text{Br}(Y\to j \,j)$.
The corresponding results are shown in the left panel of Fig.~\ref{fig:Y_LHC} as a function of $M_Y$ and for a fixed value of $\Gamma_Y/M_Y = 0.04$.
We note that the bounds in Fig.~\ref{fig:Y_LHC} can only be applied to $y < 1$ in order to maintain the validity of the narrow width approximation. For larger couplings, the $Y$ particle has a wider width which in turn would lead to more relaxed bounds on $y_{ud}^2\, \text{Br}(Y\to j \,j)$, see~\cite{Sirunyan:2018xlo}. In addition, we also display in this figure a forecast for the reach of the high luminosity (HL) LHC running at $\sqrt{s} = 14\,\text{TeV}$ with $3\,\text{ab}^{-1}$ of data, which has been obtained using the results of Sec.~6.4 of~\cite{CidVidal:2018eel}. 
Note that one does not expect a large improvement for the masses shown in Fig.~\ref{fig:Y_LHC} as a result of the large SM backgrounds, though HL measurements could be sensitive to $M_Y \sim 10\,\text{TeV}$ scalars. 

We now turn to the $\psi q$ final state which produces a jet and missing energy in the detector.
In this case, we perform a recast of the ATLAS search~\cite{Aaboud:2017phn}, again based on $36\,\text{fb}^{-1}$ of data at $13$~TeV.
To do this, we make use of a procedure similar to the one described in the previous paragraph, but for this channel we employ a publicly available \texttt{MADANALYSIS5}~\cite{Conte:2012fm,Conte:2014zja,Conte:2018vmg} implementation~\cite{recast_jetMET} of the relevant analysis~\cite{Aaboud:2017phn}. We use \texttt{PYTHIA8}~\cite{Sjostrand:2014zea} to model the partonic showering and hadronization of the jets and use the \texttt{DELPHES3}~\cite{deFavereau:2013fsa} program to simulate the ATLAS detector response following~\cite{recast_jetMET}.
As before, we use the narrow width approximation to translate the limit on the number of observed events for each signal region in~\cite{Aaboud:2017phn} to find a limit on $y_{q_iq_j}^2\, \text{Br}(Y\to j \,\psi)$; this is shown in the right panel of Fig.~\ref{fig:Y_LHC}.
The resulting limit displayed there is the 
combination of the individual limits obtained for each of the 10 different exclusive signal regions in the analysis.
In addition to the current bounds, we show the maximal reach of the HL-LHC with a total of $3\,\text{ab}^{-1}$ of data for the exemplary case of the $y_{ud}$ coupling (all other limits scale analogously).
This forecast is obtained by assuming that the uncertainties are and will remain dominated by systematics.
This allows one to simply scale the uncertainties on the SM background given in~\cite{Aaboud:2017phn} with the inverse square root of the luminosity.

\begin{figure}[!t]
\centering
\includegraphics[width=\linewidth]{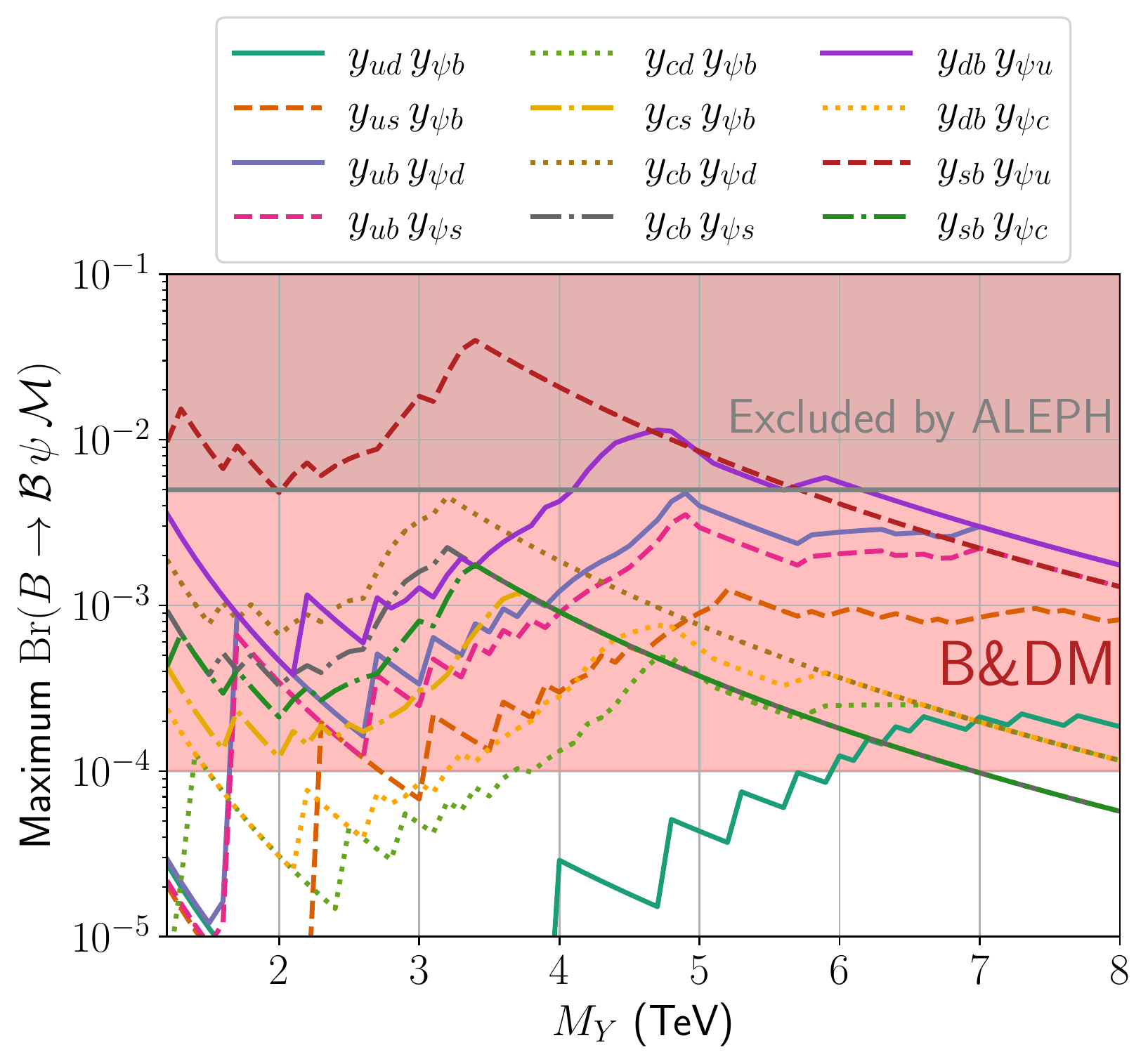}
\vspace{-0.3cm}
\caption{Maximum possible exotic branching ratio of $B$ meson decays mediated by a hypercharge $-1/3$ or $2/3$ color-triplet scalar, in view of the LHC constraints displayed in Fig.~\ref{fig:Y_LHC}.
The value of $\br$ is obtained using Eq.~\eqref{eq:bdecayrate} with a fixed value of $m_\psi=1.5\,\mathrm{GeV}$ (larger branching ratios are possible for a lighter $\psi$).
The red band highlights the values of $\br$ where $B$-Mesogenesis can proceed, while the gray line shows the lower limit of the ALEPH bound on $\br$ as found in Sec.~\ref{sec:BR_current_LEP}.
}
\label{fig:LHC_BR}
\end{figure}

The findings of both analyses described above can be combined using Eq.~\eqref{eq:PartialWidths}, which allows us to compute the branching ratio relevant for each search in terms of the $y$ couplings.
The resulting constraints are compiled in Figs.~\ref{fig:LHC_combined_plots} and~\ref{fig:LHC_combined_plots_23} in App.~\ref{sec:LHC_plots}. The information showcased in those figures can be further used to obtain an upper limit on the product $[\sqrt{y^2} {\text{TeV}}/{M_Y}]^4$ and therefore to directly constrain $\br$ through Eq.~\eqref{eq:bdecayrate}.
We show the obtained constraints in Table~\ref{tab:mchibounds} and Fig.~\ref{fig:LHC_BR} for all the possible flavor combinations corresponding to different products of $y$ couplings, for both hypercharge $-1/3$ and $2/3$ triplet scalars.
For all the possible effective operators $\mathcal{O}^k_{u_id_j}$, the first column of Table~\ref{tab:mchibounds} gives the maximum possible value of the $[\sqrt{y^2} {\text{TeV}}/{M_Y}]^4$ product, while for the latter columns we use Eq.~\eqref{eq:bdecayrate} to obtain the maximum dark fermion mass that allows for $\br$ at the $10^{-4}$, $10^{-3}$, and $10^{-2}$ levels.
In order to showcase the dependence of the constraints on the mass of the triplet scalar $Y$, in Fig.~\ref{fig:LHC_BR} we display the maximum possible $\br$ as a function of $M_Y$, for a fixed dark fermion mass $m_\psi = 1.5\,\mathrm{GeV}$, and once more for all the possible flavor combinations. The red bands displays the region that allows for $B$-Mesogenesis with new-physics enhanced semileptonic asymmetries. Fig.~\ref{fig:LHC_BR} shows that triplet scalars with masses in the $\sim 3-5\,\mathrm{TeV}$ range produce the largest exotic $B$ branching ratios allowed by the dijet and jet+MET constraints on the triplet scalar mediator.
It also shows the strong dependence of the limits on the flavor structure of the $Y$ couplings, with diquark couplings to heavy flavors being much less constrained than the ones to light quarks.

\vspace{0.2cm}

In light of Fig.~\ref{fig:LHC_BR}, we conclude that despite the stringent ATLAS and CMS bounds, it is still possible for a color-triplet scalar to mediate the exotic decay $B\rightarrow\mathcal{B}\,\psi\,\mathcal{M}$ with a branching fraction above the $10^{-4}$ level, no matter what the flavor combination of its couplings in Eq.~\eqref{eq:LflavorUV} is responsible for it.
This means that $B$-Mesogenesis can proceed through any of the operators listed in Table~\ref{tab:hadronmasses}, provided that the semileptonic asymmetries in the $B^0_s$ and/or $B^0_d$ system are enhanced by new physics to a sufficient extent.
That said, if the enhancement of $A_{\rm SL}^{d,s}$ is relatively modest and a branching fraction $\mathcal{O}(10^{-3})$ is required, then colored scalar searches at the LHC could be instrumental in determining which of the operators in Table~\ref{tab:mchibounds} may give rise to the exotic $B$ meson decay. Excitingly, our projections using the final luminosity of $3\,\mathrm{ab}^{-1}$ show that the HL-LHC has great potential to discover a color-triplet scalar with the couplings required for $B$-Mesogenesis via searches for jet plus missing energy, as can be appreciated in the right panel of Fig.~\ref{fig:Y_LHC}.

\begin{table}[!t]
\vspace{0.3cm}
\renewcommand{\arraystretch}{1.2}
  \setlength{\arrayrulewidth}{.25mm}
\centering
\small
\setlength{\tabcolsep}{0.2 em}

\begin{tabular}{ c || c | c | c | c  }
    \hline\hline
             \multirow{3}{*}{Operator} 			& Max: &\multicolumn{3}{c}{Upper limit on $m_\psi$} \\  \cline{3-5} 
          						&  \multirow{2}{*}{$\left[\frac{\sqrt{y^2}}{0.53} \frac{1.5\,\text{TeV}}{M_Y}\right]^4$}  &\multicolumn{3}{c}{Inclusive $\br$} \\  \cline{3-5} 
			&	&   $\,\,\,\,10^{-4}\,\,\,\,$  &  $\,\,\,\,10^{-3}\,\,\,\,$ &   $\,\,\,\,10^{-2}\,\,\,\,$       \\    \hline\hline 
    $\mathcal{O}_{ud}^1 = (\psi \, b)\, (u\, d)$   &   0.3	 &   2.0	   &   	-	&  -   	 \\ 
    $\mathcal{O}_{ud}^2 = (\psi \, d)\, (u\, b)$   &   6.7	 &   3.2	   &   	2.3	&   -		   	 \\ 
    $\mathcal{O}_{ud}^3 = (\psi \, u)\, (d\, b)$   &   16.2	 &   3.4	   & 2.8  		&   1.6 	   	 \\  \hline 
     $\mathcal{O}_{us}^1 = (\psi \, b)\, (u\, s)$   &   2.4	 &   2.7	   & 1.6  		&    -	   	 \\ 
    $\mathcal{O}_{us}^2 = (\psi \, s)\, (u\, b)$    &   6.7	&   3.2	   & 2.3  		&  -  		   	 \\ 
    $\mathcal{O}_{us}^3 = (\psi \, u)\, (s\, b)$    &   75.8	&   3.5	   & 3.0  		&   2.2 	   	 \\  \hline 
    $\mathcal{O}_{cd}^1 = (\psi \, b)\, (c\, d)$    &   10.4	&   2.0	   & 1.2  		&    -		   	 \\ 
    $\mathcal{O}_{cd}^2 = (\psi \, d)\, (c\, b)$   &   96.6	 &   2.4	   & 1.9  		&   1.2 		   	 \\ 
    $\mathcal{O}_{cd}^3 = (\psi \, c)\, (d\, b)$    &   16.2	&   2.1	   & 1.4  		&   - 		   	 \\  \hline 
    $\mathcal{O}_{cs}^1 = (\psi \, b)\, (c\, s)$    &   50.9	&   2.0	   & 1.5  		&   - 		   	 \\ 
    $\mathcal{O}_{cs}^2 = (\psi \, s)\, (c\, b)$    &   96.6	&   2.4	   &  1.9 		&   1.2 		   	 \\ 
    $\mathcal{O}_{cs}^3 = (\psi \, c)\, (s\, b)$    &   75.8	&   2.1	   &  1.7 		&    -		   	 \\  \hline\hline
\end{tabular}
\caption{Maximum possible value of $[\sqrt{y^2} {\text{TeV}}/{M_Y}]^4$ as constrained by a combination of constraints on color-triplet scalars from ATLAS and CMS on resonant high $p_T$ jets and jets plus missing energy. The columns on the right hand side show the heaviest expected possible dark fermion mass in the $\bar{b}\to u_i d_j \psi$ decay as a function of the operator that triggers it given the constraint on $[\sqrt{y^2} {\text{TeV}}/{M_Y}]^4$, see Eq.~\eqref{eq:bdecayrate}.}
\label{tab:mchibounds}
\end{table}

\subsection{Flavor Mixing Constraints}\label{sec:Ytrip_Flavor}

A color-triplet scalar with inter-generation couplings as in~\eqref{eq:LflavorUV} can induce $\Delta F = 2$ processes and therefore contribute to mass differences and CP violating parameters in the neutral $B$, $D$ and $K$ meson systems, see e.g.~\cite{Davidson:1993qk,Giudice:2011ak,Agrawal:2014aoa,Fajfer:2020tqf}.
The SM prediction for the mass differences matches observations except for the $K$ and $D$ meson systems, in which the SM prediction cannot be reliably calculated from first principles.
In this situation, any new physics contribution to $\Delta M_K$ and $\Delta M_D$ can only be bounded to be smaller than the corresponding experimentally measured value.
CP violation is particularly constraining in the Kaon sector, given that $\epsilon_K$ is experimentally known down to the half percent level~\cite{pdg}, while recent theoretical evaluations~\cite{Brod:2019rzc} have achieved a $\sim 10\%$ uncertainty in the SM prediction.

Triplet-scalar contributions to neutral meson mixing arise from box diagrams, which are only non-vanishing if at least two different $y_{q_iq_j}$ or $y_{\psi q_i}$ couplings are 
nonzero.
It is important to stress that such coupling combinations are not the same ones that enter Eq.~\eqref{eq:bdecayrate} and therefore flavor mixing constraints do not directly constrain $B$-Mesogenesis.
The relevant diagrams and corresponding amplitudes are given in Appendix~\ref{sec:MesonMixing}.
Comparing the calculated values of $\Delta M$ and $\epsilon_K$ with the most recent experimental measurements~\cite{pdg} and SM predictions~\cite{Lenz:2019lvd,Brod:2019rzc} allows us to obtain constraints on the couplings of the triplet scalar.

We show first the results for a hypercharge $-1/3$ triplet scalar. The limits for the $Y$ interactions involving two SM quarks are
\begin{subequations}\label{eq:flavor_constraints_13}
\begin{alignat}{2}
|y_{u_id} y_{u_is}^\star| &\,<\, 4\!\times\!10^{-2}\left({M_Y}/{1.5\text{TeV}}\right) \,, &&\,\, \Delta M_K \\
|y_{ud} y_{us}^\star| &\,<\, 1\!\times\!10^{-3}\left({M_Y}/{1.5\text{TeV}}\right) \,, &&\,\,\epsilon_K \\
|y_{cd} y_{cs}^\star| &\,<\, 8\!\times\!10^{-4}\left({M_Y}/{1.5\text{TeV}}\right) \,, &&\,\,\epsilon_K \\
|y_{td} y_{ts}^\star| &\,<\, 3\!\times\!10^{-4}\left({M_Y}/{1.5\text{TeV}}\right) \,, &&\,\,\epsilon_K \\
|y_{cd_j} y_{ud_j}^\star|  &\,<\, 2\!\times\!10^{-2}\left({M_Y}/{1.5\text{TeV}}\right) \,, &&\,\,\Delta M_{D} \\
|y_{u_ib} y_{u_id}^\star|  &\,<\, (2-4)\!\times\!10^{-2}\left(\frac{M_Y}{1.5\text{TeV}}\right) \,, &&\,\,\Delta M_{B_d} \\
|y_{u_ib} y_{u_is}^\star|  &\,<\, (1-2)\!\times\!10^{-1}\left(\frac{M_Y}{1.5\text{TeV}}\right) \,,\; &&\,\,\Delta M_{B_s} \label{eq:mixing_Bs_13}
\end{alignat}
while the ones for the $Y$-$\psi$-SM quark couplings read
\begin{alignat}{2}
\! |y_{\psi d} y_{\psi s}^\star| &\,<\, 4\!\times\!10^{-2}\left({M_Y}/{1.5\text{TeV}}\right) \,, &&\,\,\Delta M_K \\
\! |y_{\psi d} y_{\psi s}^\star| &\,<\, 1\!\times\!10^{-3}\left({M_Y}/{1.5\text{TeV}}\right) \,, &&\,\,\epsilon_K \\
\! |y_{\psi b} y_{\psi d}^\star|  &\,<\, (2-4)\!\times\!10^{-2}\left({M_Y}/{1.5\text{TeV}}\right) \,, &&\,\,\Delta M_{B_d} \\
\! |y_{\psi b} y_{\psi s}^\star|  &\,<\, (1-2)\!\times\!10^{-1}\left({M_Y}/{1.5\text{TeV}}\right) \,, &&\,\,\Delta M_{B_s}.
\end{alignat}
\end{subequations}

The limits for the hypercharge $2/3$ triplet scalar are obtained analogously. For the $Y$ interactions involving two SM quarks, we have
\begin{subequations}\label{eq:flavor_constraints_23}
\begin{alignat}{2}
|y_{db} y_{sb}^\star| &\,<\, 4\!\times\!10^{-2}\left({M_Y}/{1.5\text{TeV}}\right) \,, &&\,\, \Delta M_K \\
|y_{db} y_{sb}^\star| &\,<\, 1\!\times\!10^{-3}\left({M_Y}/{1.5\text{TeV}}\right) \,, &&\,\,\epsilon_K \\
|y_{ds} y_{sb}^\star|  &\,<\, (2-4)\!\times\!10^{-2}\left({M_Y}/{1.5\text{TeV}}\right) \,, &&\,\,\Delta M_{B_d} \\
|y_{db} y_{sd}^\star|  &\,<\, (1-2)\!\times\!10^{-1}\left({M_Y}/{1.5\text{TeV}}\right) \,,\; &&\,\,\Delta M_{B_s} \label{eq:mixing_Bs_23}
\end{alignat}
while for the $Y$-$\psi$-SM quark couplings, we find
\begin{alignat}{2}
\! |y_{\psi u} y_{\psi c}^\star| &\,<\, 2\!\times\!10^{-2}\left({M_Y}/{1.5\text{TeV}}\right) \,, &&\quad\Delta M_D\,.
\end{alignat}
\end{subequations}
To be conservative, all the above constraints are given at $95\%$~CL and assume that there is no cancellation between the contributions from different combinations of couplings.
Note that there is no bound on $y_{\psi d_i}$ or $y_{d_i d_j}$ couplings from $D^0$ mesons because these interactions only involve down-type quarks.
Conversely, bounds for $y_{\psi u_i}$ arise only from $D^0$ mesons.
Although we quote the bounds in terms of the absolute value of the couplings, the dependence on the phase is important, especially for CP-violating parameters.
In particular, the bound from $\Delta M_K$ only applies to the real part of the corresponding coupling product, while the one from $\epsilon_K$ constrains the imaginary one.
For the $B$ mesons, the intervals quoted cover all the possible phases, ranging from aligned to anti-aligned with the SM contribution.
The only case where the flavor of the internal quark in the box diagrams is relevant is $\epsilon_K$, due to the fact that the dominant contribution in this case comes from the right diagram in Fig.~\ref{fig:box_diagrams}, which depends on CKM factors.

In light of these results and given the size of the couplings in Eq.~\eqref{eq:Baryogenesis_range} required for baryogenesis, we conclude that the color-triplet scalar should preferentially couple to one particular flavor combination.
In other words, if the coupling responsible for generating a large $\br$ is $y_{q_iq_j}\sim\mathcal{O}(1)$ for some particular flavor combination, all the other couplings $y_{q_kq_l}$ should be suppressed at the $\sim 10^{-1}-10^{-2}$ level (or up to $\sim 10^{-4}$ depending on the complex phase and the exact flavor structure).

We end this section by noting that the study presented here of the flavor-mixing effects of the color-triplet scalar $Y$ can be improved in a number of ways.
First, we have neglected any renormalization group running from the scale $\mu\sim M_Y$ at which the box diagrams in Fig.~\ref{fig:box_diagrams} are computed down to the relevant hadronic scale $\mu\sim\mathrm{GeV}$.
Furthermore, for simplicity we have disregarded CP violation in the $B$ and $D$ systems.
Although these CPV observables are generally less constraining than $\Delta M$, they are relevant when considering the full dependence of the constraints on the complex phases of the $Y$ couplings.
Of particular interest is the fact that modifications of the width mixing $\Gamma_{12}^q$ are possible as is discussed in the next section.
These caveats call for a more exhaustive study of this topic, ideally by means of a global fit including all CP-conserving and -violating observables along the lines of~\cite{Bona:2007vi}.

\subsection{New Tree-Level \texorpdfstring{$b$}{b} Decays}\label{sec:Decays}

The operators in Lagrangian~\eqref{eq:LflavorUV} can induce three qualitatively distinct tree-level decays of a $b$ quark.
All of them are mediated by the triplet scalar $Y$, which can be integrated out to obtain different effective $4$-fermion operators depending on the hyperharge of $Y$ and the flavor variation. For hypercharge $-1/3$, we have
\begin{subequations}
\begin{align}
\!\!\!\!\mathcal{L}_{b\rightarrow \bar{u}_i \bar{d}_j \bar{\psi}} &= \frac{1}{M_Y^2} \, \epsilon_{\alpha\beta\gamma} \, \Bigl[ y^*_{u_i b} y^*_{\psi d_j} \, \left( b_R^\alpha u_{iR}^\beta \right) \Bigl( d_{jR}^\gamma \psi_R \Bigr) \nonumber \\ 
& \hspace{1.3cm} + y^*_{u_i d_j} y^*_{\psi b} \, \left( d_{jR}^\alpha u_{iR}^\beta \right) \Bigl( b_R^\gamma \psi_R \Bigr) \Bigr] \,, \label{eq:b_decay1} 
\end{align}
\vspace{-0.4cm}
\begin{align}
\!\!\!\!\! \mathcal{L}_{b\rightarrow u_i \bar{u}_j d_k} &= \half \frac{1}{M_Y^2} \, y_{u_j b}^* y_{u_i d_k} \, \Bigl[ \left( \bar{u}_{iR}^\alpha \gamma^\mu b_R^\beta \right) \left( \bar{d}_{kR}^\beta \gamma_\mu u^\alpha_{jR} \right) \nonumber \\ 
& \hspace{1.5cm} - \Bigl( \bar{u}_{iR}^\alpha \gamma^\mu b_R^\alpha \Bigr) \left( \bar{d}_{kR}^\beta \gamma_\mu u^\beta_{jR} \right) \Bigr] \,, \label{eq:b_decay2} 
\end{align}
\vspace{-0.4cm}
\begin{align}
\!\!\!\!\! \mathcal{L}_{b\rightarrow \psi \bar{\psi} d_j} &= -  \frac{1}{2 M_Y^2} \, y^*_{\psi b} y_{\psi d_j} \, \Bigl( \bar{\psi}_R \gamma^\mu b_R^\alpha \Bigr) \Bigl( \bar{d}_{jR}^\alpha \gamma_\mu \psi_R \Bigr) \,, \label{eq:b_decay3}
\end{align}
\end{subequations}
while for hypercharge $2/3$, we get
\begin{subequations}
\begin{align}
\!\!\!\!\mathcal{L}_{b\rightarrow \bar{u}_i \bar{d}_j \bar{\psi}} &= \frac{1}{M_Y^2} \, \epsilon_{\alpha\beta\gamma} \, y^*_{d_j b} y^*_{\psi u_i} \, \left( b_R^\alpha d_{jR}^\beta \right) \Bigl( u_{iR}^\gamma \psi_R \Bigr) \,, \label{eq:b_decay1b} 
\end{align}
\vspace{-0.4cm}
\begin{align}
\!\!\!\! \mathcal{L}_{b\rightarrow d_i \bar{d}_j d_k} &= \half \frac{1}{M_Y^2} \, y_{d_j b}^* y_{d_i d_k} \, \Bigl[ \left( \bar{d}_{iR}^\alpha \gamma^\mu b_R^\beta \right) \left( \bar{d}_{kR}^\beta \gamma_\mu d^\alpha_{jR} \right) \nonumber \\ 
& \hspace{1.5cm} - \Bigl( \bar{d}_{iR}^\alpha \gamma^\mu b_R^\alpha \Bigr) \left( \bar{d}_{kR}^\beta \gamma_\mu d^\beta_{jR} \right) \Bigr] \,. \label{eq:b_decay2b} 
\end{align}
\end{subequations}
Here, $\alpha$, $\beta$, and $\gamma$ are color indices while $u_i$ and $d_j$ represent any up- or down-type quark such that the corresponding decay is kinematically allowed.
We have performed Fierz rearrangements~\cite{Nieves:2003in} to bring the operators into the above form.

The decay channels~\eqref{eq:b_decay1} and~\eqref{eq:b_decay1b} are the ones responsible for baryogenesis and are studied in detail in Sec.~\ref{sec:BR_Bfactories}.
The remaining three are not strictly required for the baryogenesis mechanism, but appear if we consider a generic flavor structure for the couplings of the color-triplet scalar $Y$.
While the channels in~\eqref{eq:b_decay2} and~\eqref{eq:b_decay2b} are the ``right-handed versions" of flavor-changing charged current and flavor-changing neutral current SM decays, Eq.~\eqref{eq:b_decay3} constitutes a radically novel decay channel for the $b$ quark.

Decays of neutral $B^0_{d,s}$ mesons that are triggered by~\eqref{eq:b_decay2},~\eqref{eq:b_decay3}, or~\eqref{eq:b_decay2b} can produce final states that are common to both mesons and antimesons.
As a consequence, these tree-level $b$ decays contribute to the width mixing $\Gamma_{12}^{d,s}$, and can thus enhance the semileptonic asymmetries above the small values predicted in the SM.
As discussed in Sec.~\ref{sec:CP_violation}, these tree-level decays do not contribute to CP violation in interference and are therefore largely unconstrained.

In order to highlight the relevance of this phenomenology, we consider the particular decay $b\rightarrow c\bar{c}s$, as a complete model-independent analysis of potential new physics contributions to this channel was performed in~\cite{Jager:2019bgk}.
In the notation of~\cite{Jager:2019bgk}, our operators~\eqref{eq:b_decay2} can be mapped to $Q_1^{c\prime}$ and $Q_2^{c\prime}$.
The strongest constraints on these operators are found to be due to their loop-level contributions to the semileptonic decays of $B$ mesons (more specifically, to the operator $Q'_{9V}$ as described in~\cite{Aebischer:2019mlg}).
From the bounds on the Wilson coefficients displayed in Fig.~$3$ of~\cite{Jager:2019bgk}, we can obtain the $1\sigma$ limit
\begin{equation}
\left| y_{cb} y_{cs}^* \right| \lesssim 7\times 10^{-2} \left( {M_Y}/{1.5\TeV} \right)^2 ,
\end{equation}
which is somewhat stronger than the corresponding bound from $\Delta M_{B^0_s}$ in Eq.~\eqref{eq:mixing_Bs_13}. 
This showcases the fact that a full study of the impact of $Y$ in the neutral meson mixing system including modifications in $\Gamma_{12}^q$ is highly desirable. 
As interesting as this may be, the matching of the operators~\eqref{eq:b_decay2}-\eqref{eq:b_decay3} to observables of the $B^0_q$ system would involve hadronic calculations that are out of the scope of this paper.

In addition, the decays of the form $b\rightarrow \psi\bar{\psi}s$ and $b\rightarrow \psi\bar{\psi}d$, which involve invisible dark sector states, can in principle significantly modify $\Gamma_{12}^{d,s}$ and thus $\phi_{d,s}$.
That said, these decays containing missing energy are constrained by the LEP search~\cite{Barate:2000rc} discussed in Sec.~\ref{sec:BR_current_LEP}.
Although a dedicated recast would be required to obtain precise numerical values, the bound on the similar process $\mathrm{Br}(b\rightarrow s \nu \bar{\nu}) < 6.4\times 10^{-4}$ at $90\%$ CL found in~\cite{Barate:2000rc} constitutes a conservative comparison.

Overall, we conclude that our present setup has the potential to significantly enhance $\Gamma_{12}^q$ and thus the semileptonic asymmetries beyond their SM values through the introduction of new tree-level $b$-quark decay channels.
Although precise numerical evaluations are needed in order to make a more quantitative statement, these modifications may be the source of the extra CP violation required by $B$-Mesogenesis to explain the baryon asymmetry of the Universe.
This interesting possibility would make the framework presented here completely self-contained.

\subsection{Implications for Searches and Models}\label{sec:ModelBuilding_triplet}

We now combine the requirements found for successful baryogenesis and dark matter production with the multiple experimental constraints on color-triplet scalars discussed above. This provides a global perspective on the implications for \textit{
(i)} direct searches of $B\to\psi\,\mathcal{B}\,\mathcal{M}$ at $B$ factories and colliders, and \textit{(ii)} model-building efforts to reproduce the flavor structure of the triplet scalar interactions required by $B$-Mesogenesis.

\begin{figure}[t!]
\centering
\hspace{-0.4cm}\includegraphics[width=0.50\textwidth]{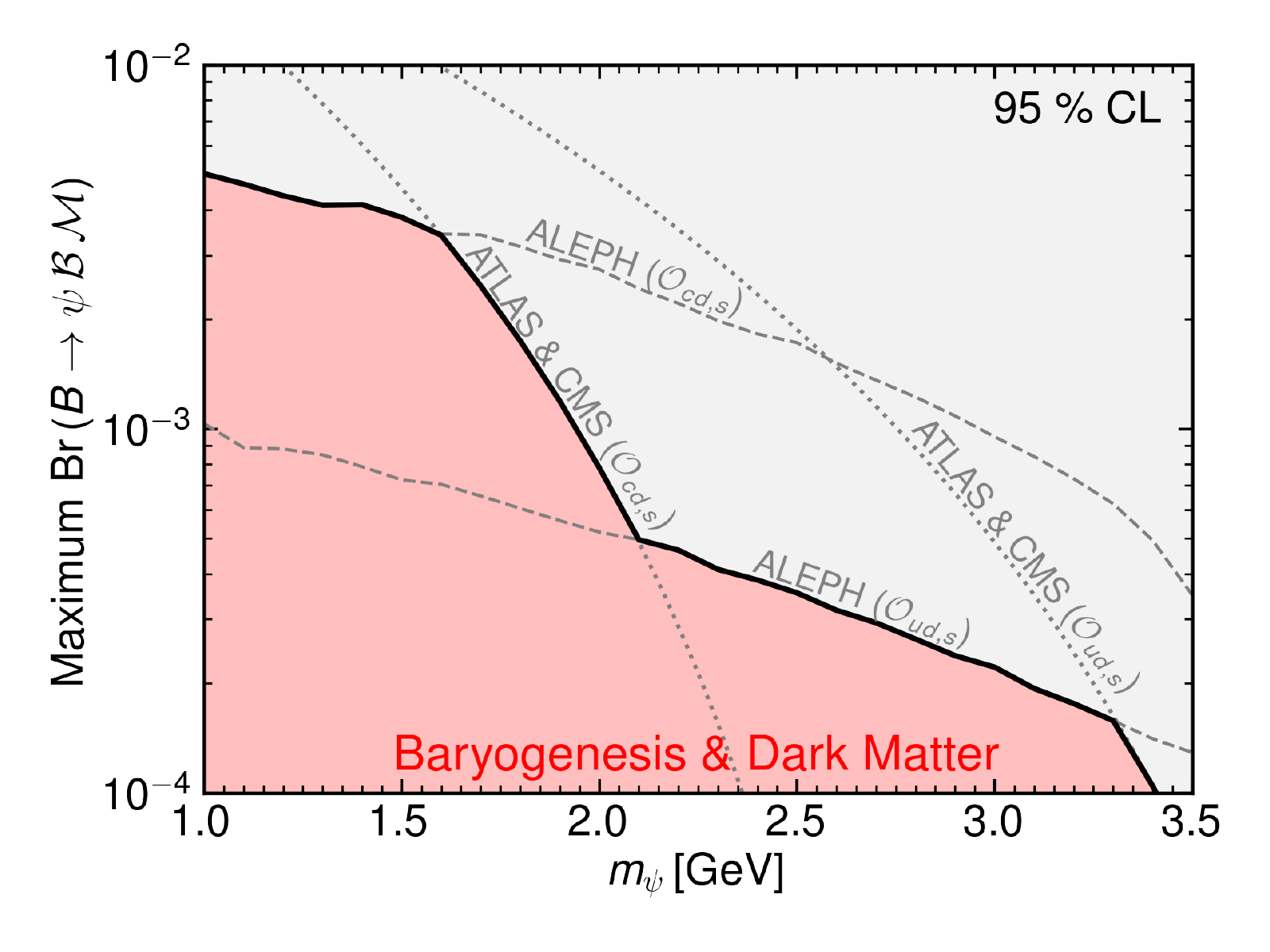}
\vspace{-0.65cm}
\caption{Maximum inclusive $\br$ as a function of $m_\psi$ as directly constrained by an ALEPH search~\cite{Barate:2000rc} for $b$ decays with large missing energy (see Fig.~\ref{fig:LEP_constraints}), and indirectly by ATLAS \& CMS searches for TeV-scale color-triplet scalars. In red we highlight the region of parameter space compatible with baryogenesis and dark matter production. Each line represents the constraints on each of the four possible $b$ decay operators in Table~\ref{tab:hadronmasses}.}
\label{fig:BR_mpsi}
\end{figure}

In Fig.~\ref{fig:BR_mpsi} we highlight the implications of current searches for TeV-scale color-triplet scalars at ATLAS and CMS on the possible size of $\br$. The dotted lines in Fig.~\ref{fig:BR_mpsi}  represent the region of parameter space excluded by the lack of BSM signals of high $p_T$ dijets and jets plus missing energy. From the results, it is clear that the  impact of these indirect constraints is very relevant for $B$-Mesogenesis. In particular, for operators containing a charm quark, the indirect constraint from ATLAS and CMS is substantially more stringent than the direct one set by ALEPH for $m_\psi> 1.7\,\text{GeV}$. In addition, achieving successful baryogenesis in light of these constraints requires $m_\psi \lesssim 3.5\,\text{GeV}$. This can be understood  as due to the strong dependence of $\br$ on $\Delta m$, see Eq.~\eqref{eq:bdecayrate}. Globally, Fig.~\ref{fig:BR_mpsi} showcases the relevant size and kinematic properties of each decay mode that should be targeted with direct searches at BaBar, Belle, Belle-II, and LHCb.
Finally, we note that, barring a detection, these bounds will be significantly improved by the High-Luminosity upgrade of the LHC, as shown in Fig.~\ref{fig:Y_LHC}.

\begin{figure}[t]
\centering
\vspace{0.25in}
\begin{tabular}{c}
		\label{fig:Matrices_1}
		\includegraphics[width=0.45\textwidth]{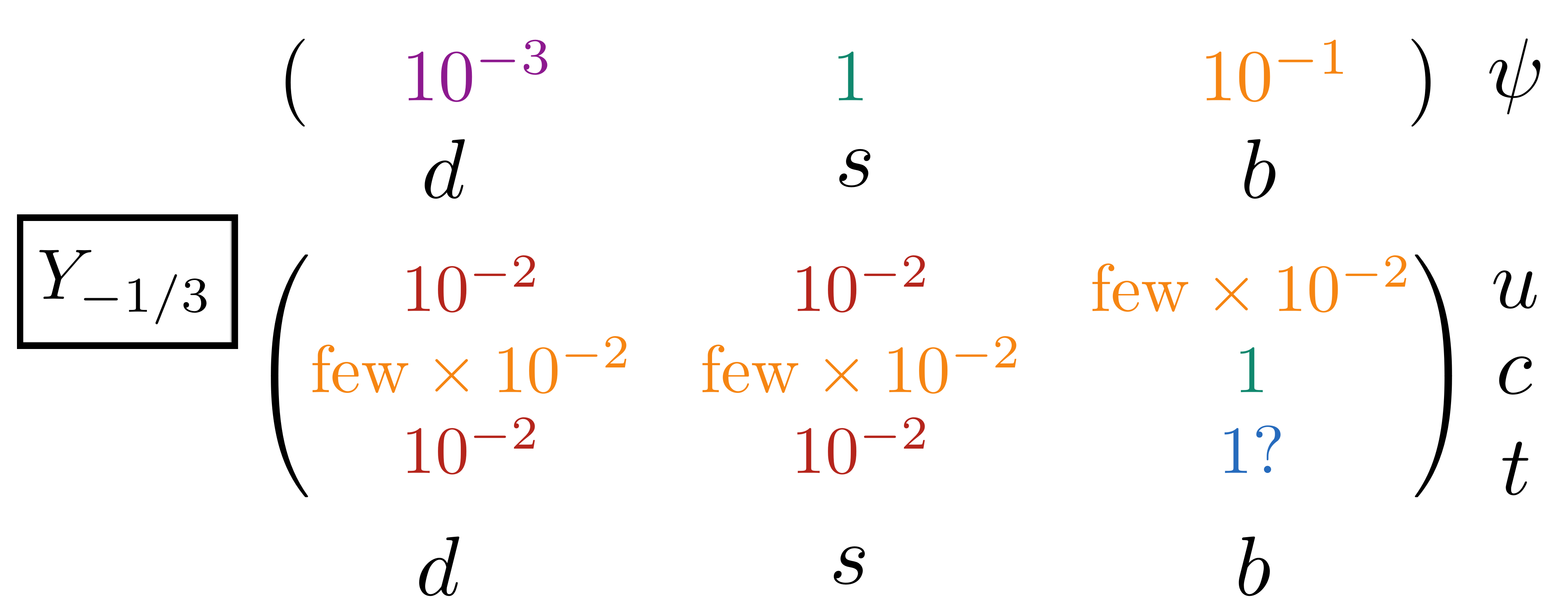}
\\
\hline
		\label{fig:Matrices_2}
		\includegraphics[width=0.45\textwidth]{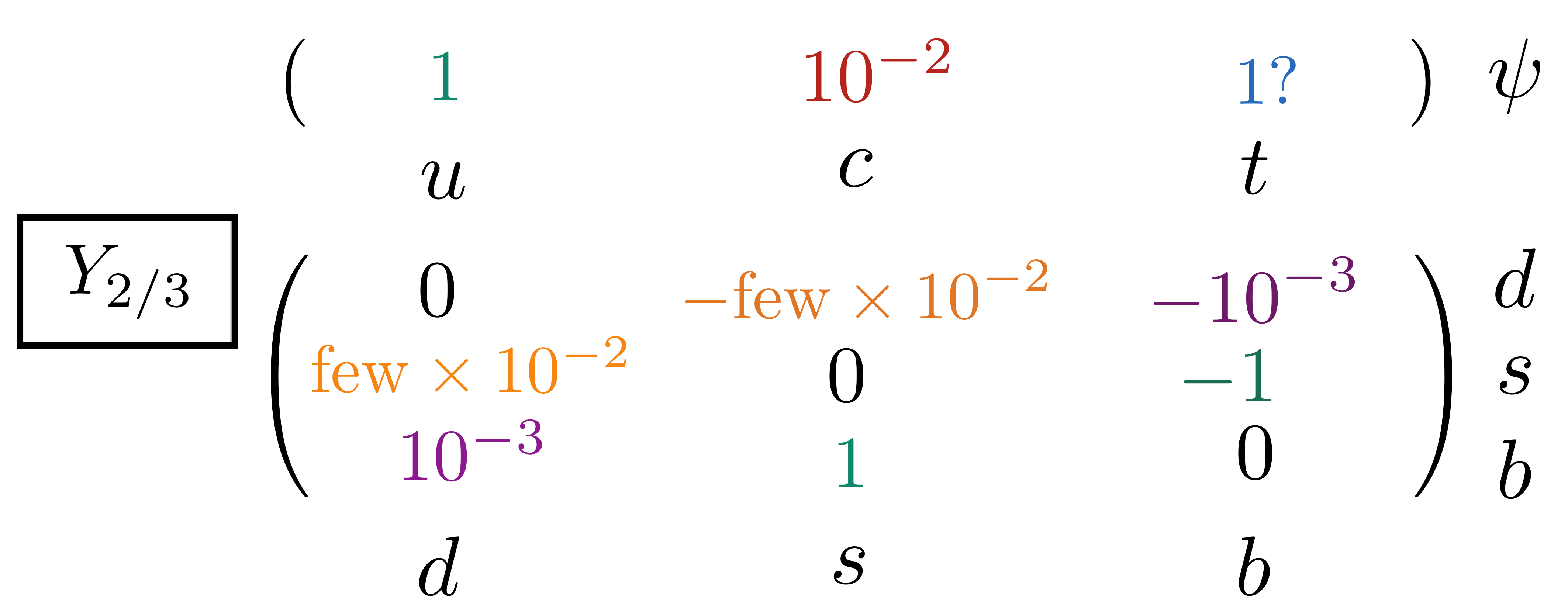}
\end{tabular}
\vspace{-0.2cm}
\caption{Two examples of flavor structure of the couplings of the color-triplet scalar $Y$ that can produce the observed baryon asymmetry and the dark matter abundance while satisfying all current constraints.
The heavy scalar mass here is fixed to $M_Y=1.5$~TeV, but masses up to $20$~TeV are possible with the subsequent relaxation of experimental constraints. 
The couplings controlling the $B\to \psi\,\mathcal{B}\,\mathcal{M}$ decay as relevant for baryogenesis are shown in green. Note that $y_{bt}$ and $y_{\psi t}$ are largely unconstrained.
}
\label{fig:Matrices}
\end{figure}

We now turn our attention to the implications for model building and the LHC. As shown in Figs.~\ref{fig:LHC_BR} and~\ref{fig:BR_mpsi}, any of the operators in Eq.~\eqref{eq:LflavorIR} could be responsible for baryogenesis. Depending upon the exact $\br$ needed for baryogenesis, the color-triplet scalar mediator $Y$ could be much heavier and potentially out of reach of direct LHC searches. Even in such a scenario, indirect effects on flavor observables such as neutral meson mixing can still shed some light on the flavor structure of the $Y$ couplings.
The important observation is that flavor observables constrain the product of two different couplings as can be seen in Eqs.~\eqref{eq:flavor_constraints_13} and~\eqref{eq:flavor_constraints_23}, while baryogenesis only requires a single coupling to be $\mathcal{O}(1)$.
Therefore, flavor bounds can be evaded if, for instance, the couplings of $Y$ have a hierarchical structure. In Fig.~\ref{fig:Matrices} we show two examples of such structures, one for the case of $Y$ carrying hypercharge $-1/3$ and another one for hypercharge $2/3$. It would be interesting to explore how flavor structures like the ones in Fig.~\ref{fig:Matrices} may arise in a UV complete model\footnote{We refer to~\cite{Alonso-Alvarez:2019fym} for a scenario where $Y$ is a supersymmetric squark with hypercharge $-1/3$.}. We leave this interesting model-building task to future work.  

\section{Discussion: Collider Complementarity}\label{sec:global}

In the previous sections we have systematically explored the various collider implications of $B$-Mesogenesis. In this one, we take a global view and discuss the correlations between signals relevant for testing this baryogenesis and dark matter production mechanism. Given our discussion, we conclude that a combination of measurements at BaBar, Belle, Belle II, LHCb, ATLAS and CMS would be capable of testing the entire parameter space of $B$-Mesogenesis on a timescale of $4-5$ years.

\begin{figure*}[t!]
\centering
\hspace{-0.4cm}\includegraphics[width=0.70\textwidth]{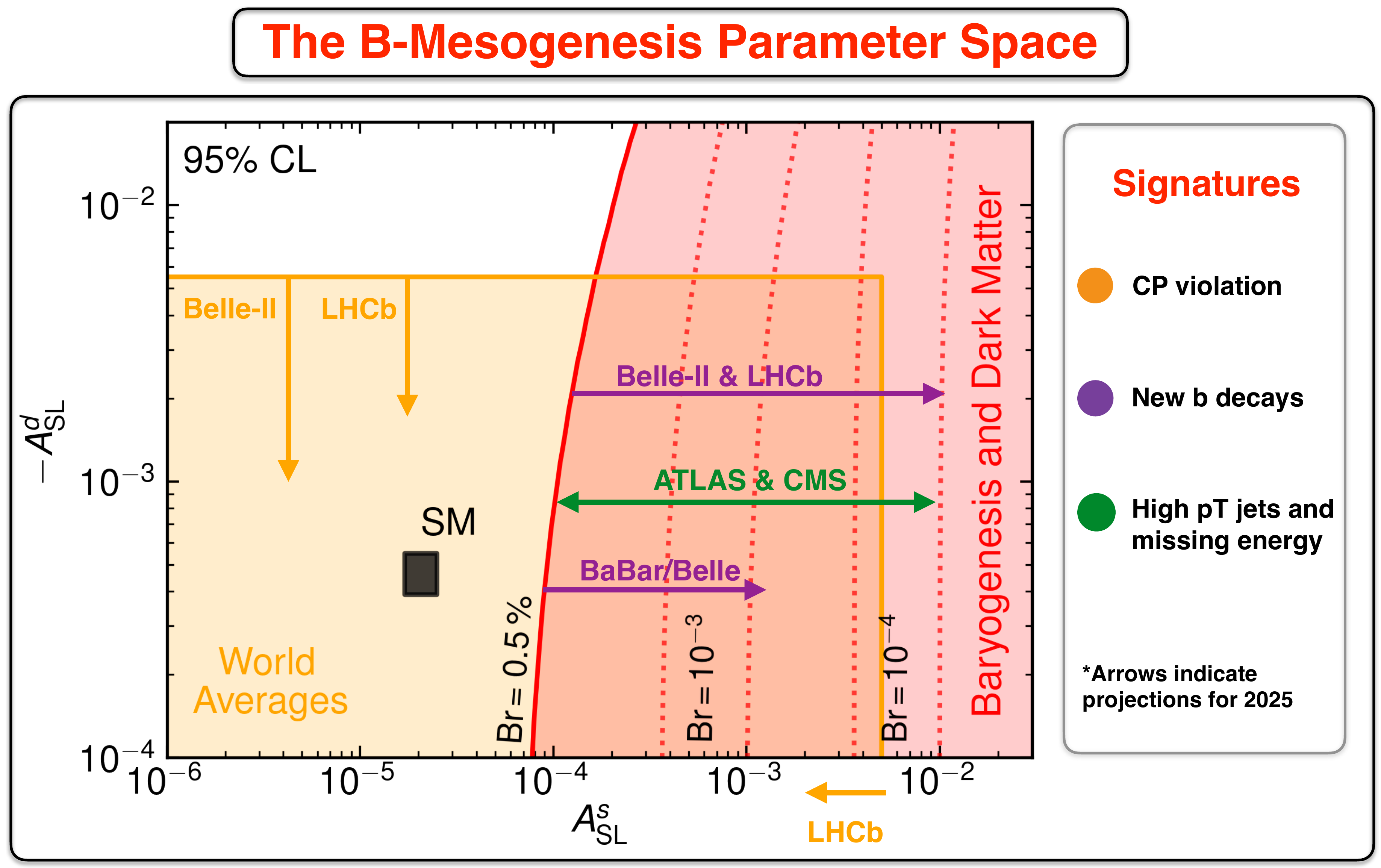} 
\caption{Parameter space for successful baryogenesis and dark matter generation within $B$-Mesogenesis in the $A_{\rm SL}^s$-$A_{\rm SL}^d$ plane. Similarly to Fig.~\ref{fig:semileptonic_asymmetries}, the red bands indicate the minimum $\br$ needed to obtain the observed baryon asymmetry of the Universe. In black we highlight the SM prediction for the semileptonic asymmetries, while in orange we show current experimental constraints. The arrows indicate the upcoming sensitivity from collider experiments on these quantities. The purple arrows highlight the reach on $\br$ from analyses at BaBar or Belle, and from Belle-II and LHCb. With a green arrow we indicate the region of parameter space for which ATLAS and CMS could be sensitive to the colored triplet scalar needed to trigger the new $B\to \psi\, \mathcal{B}\,\mathcal{M}$ decay. We see that the upcoming measurements and searches at Belle-II and LHCb have the potential to test the entire $B$-Mesogenesis parameter space. }
\label{fig:ParameterSpaceFuture}
\end{figure*}

Fig.~\ref{fig:ParameterSpaceFuture} summarizes the prospects for the aforementioned experiments to test $B$-Mesogenesis. As usual, we highlight in red the entire viable parameter space for which the observed baryon and dark matter abundances of the Universe are reproduced. This roughly corresponds to
\begin{align}\label{eq:parameterspace_nutshell}
    \br \times \left(25 \, A_{\rm SL}^s + A_{\rm SL}^d \right) > 10^{-5}\,.
\end{align}
Note that the product of these quantities is critically positive since we live in a Universe dominated by matter rather than antimatter.

Guided by Eq.~\eqref{eq:parameterspace_nutshell}, one can appreciate that there is a strong complementarity between the various current collider experiments that
\begin{itemize}
    \item Directly search for $B\to \psi\, \mathcal{B}\,\mathcal{M}$ decays, or
    \item constrain the CP violation in the $B_q^0$ system, or
    \item indirectly constrain $\br$ by searches for the heavy colored scalar needed to trigger the $B\to \psi\, \mathcal{B}\,\mathcal{M}$ decay.
\end{itemize}

At present, we know that $\br < 0.5\%$ at 95\% CL as a result of an old ALEPH search, see Sec.~\ref{sec:BR_current}. This constraint on $\br$ in turn tells us that at least one semileptonic asymmetry should be positive and larger than $10^{-4}$. At present, the semileptonic asymmetries are measured to be $A_{\rm SL}^d = (-2.1 \pm 1.7)\times 10^{-3}$ and $A_{\rm SL}^s =(-0.6 \pm 2.8)\times 10^{-3}$, meaning that current measurements have error bars which are a factor of $20-30$ larger than the minimum value required by successful baryogenesis. That said, and as discussed in Sec.~\ref{sec:CP_violation}, Belle II and LHCb expect to improve the precision on these quantities by a factor of $3-4$ on a 5 year timescale. The impact of this improvement on the parameter space is highlighted in orange in Fig.~\ref{fig:ParameterSpaceFuture}. We can clearly appreciate that the improved measurements of $A_{\rm SL}^s$ by LHCb will narrow down the minimum required value of $\br$. 
A potential measurement of a nonzero positive value for the semileptonic asymmetry would not only represent a clear signal of new physics, but also a strong indication in favour of $B$-Mesogenesis. 

The most distinctive signal of $B$-Mesogenesis is the presence of a new decay mode of $B$ mesons into a dark sector antibaryon (missing energy in the detector), a visible baryon and any number of light mesons, $B\to \psi\, \mathcal{B}\,\mathcal{M}$.
As we have discussed in Sec.~\ref{sec:BR}, such processes are already constrained in an inclusive way by LEP at the $\br < 0.5\%$ level.
We stress that while $B$-Mesogenesis is only sensitive to the inclusive $\br$ rate, searches for exclusive channels like $B\rightarrow \psi + \mathcal{B}$ are typically easier to carry out in collider experiments.
Indeed, LEP constrains these exclusive modes down to $\text{Br}(B\to \psi\, \mathcal{B})\lesssim 10^{-4}-10^{-3}$ at 95\% CL. 
It is therefore crucial to know the relative exclusive vs.\! inclusive branching ratios, which for the simplest case containing no additional mesons in the final state we estimate to be in the $1-10\%$ range following our discussion in Sec.~\ref{sec:excl_vs_inclu}.

In light of the current constraints on the semileptonic asymmetries, we know that $\br > 10^{-4}$ is required for successful baryogenesis. 
With this target in mind, the sensitivity of past and current collider experiments to these decays is already promising. In particular, a direct search using old BaBar or Belle data should improve upon existing constraints by more than an order of magnitude, reaching $\text{Br}(B\to \psi \,\mathcal{B})\sim 3\times 10^{-5}$ for some of the flavor combinations. Regarding upcoming experiments, Belle-II is expected to improve this figure by another order of magnitude and constrain exclusive branching fractions down to $3\times 10^{-6}$. Given this sensitivity, searches at  Belle-II for these decays together with improved measurements of the semileptonic asymmetries could be enough to probe the entirety of the parameter space of $B$-Mesogenesis. 

Additionally, LHCb can also search for decays of the type $B\to \psi\, \mathcal{B}^\star$, where $B^\star$ is a resonant baryon. The sensitivity of LHCb with $15\,\text{fb}^{-1}$ for decays of this type can reach the $\text{Br} \sim 10^{-7}$ level~\cite{Rodriguez:2021urv}. From the theoretical point of view, it is not straightforward to translate a limit on $\text{Br}(B\to \psi \mathcal{B}^\star)$ to $\br$. However, these modes should represent some fraction of the inclusive decay and probing $\text{Br}(B\to \psi \mathcal{B}^\star)$ down to the $10^{-6}-10^{-7}$ level would represent an important test of $B$-Mesogenesis. 

Given all of the above sensitivity expectations, we conclude that Belle-II and LHCb have the capabilities of performing a direct search for the decays predicted by $B$-Mesogenesis in order to test the its entire parameter space. The reach of the individual searches are indicated with purple arrows in Fig.~\ref{fig:ParameterSpaceFuture}. 

An alternative avenue to test $B$-Mesogenesis is to directly search for the colored particle that mediates the $B$ meson decay mode into a visible baryon and a dark antibaryon. This new mediator cannot be too heavy and could therefore be produced and searched for at the LHC. In fact, as we have shown in Sec.~\ref{sec:TripletScalar}, ATLAS and CMS searches for high $p_T$ jets, and jets plus missing energy already place relevant constraints on the couplings of this mediator to quarks and dark sector states. Given that $\br$ is completely determined by these couplings and the mass of $Y$ (see Eq.~\eqref{eq:bdecayrate}), one can use these bounds to place indirect constraints on $\br$. At present, these constraints are particularly relevant for decay operators involving a charm quark, as Fig.~\ref{fig:BR_mpsi} shows, but with the planned increase in luminosity, the LHC has great potential to discover (or exclude) this mediator.

In summary, we have argued that there is an intimate interplay  between \emph{(i)} direct searches for $B\to \psi\, \mathcal{B}\,\mathcal{M}$ decays at BaBar, Belle, Belle II and LHCb, \emph{(ii)} the various possible tests of CP violation in the $B_q^0$ system at Belle II, LHCb, ATLAS and CMS, and \emph{(iii)} the searches for color-triplet scalars at ATLAS and CMS in order to test $B$-Mesogenesis. Not only could it be possible to detect multiple signatures of this scenario at different collider experiments, but, most importantly, a combination of measurements has the potential to conclusively confirm whether or not this mechanism is responsible for the origin of the baryon asymmetry and the dark matter of the Universe.

\section{Conclusions}\label{sec:conclusions}

Within the framework of baryogenesis and dark matter from $B$ mesons---$B$ Mesogenesis~\cite{Elor:2018twp}, the baryon asymmetry of the Universe is directly proportional to two experimental observables: \textit{i)} the CP asymmetries in neutral $B_q^0$ decays, and \textit{ii)} the branching fraction of a new decay mode of $B$ mesons into dark matter (missing energy) and a visible baryon, $\br$ (recall Eq.~\eqref{eq:basypara}). In this work, we have presented a detailed study of experimental tests that could confirm or refute this mechanism---see Figs.~\ref{fig:Summary} and~\ref{fig:ParameterSpaceFuture} for a summary of these possible searches and their implications.
We have found that $B$-Mesogenesis should be fully testable at current and upcoming collider experiments.  

In order to asses all possible theoretical uncertainties, we have first revisited the calculation of the evolution of $B$ meson oscillations in the early Universe using a density matrix approach (see Appendix~\ref{sec:decoherence} for details). The corresponding predictions of the mechanism have then been contrasted against current CP violating measurements in the $B_q^0$ system in Sec.~\ref{sec:CP_violation}. In Sec.~\ref{sec:BR} we have discussed current constraints and the expected sensitivity of possible searches of the $B\rightarrow\psi\,\mathcal{B}\,\mathcal{M}$ decay at recent and current collider experiments. We have also presented a rough phase-space calculation that suggests that the ratio between the rate of the minimal exclusive channel $B\to \psi \,\mathcal{B}$ and that of the aforementioned inclusive one should be $\gtrsim 1\,\%$, see Fig.~\ref{fig:Br_ratio_Bplus}. In Sec.~\ref{sec:TripletScalar} we have considered the collider implications of the new TeV-scale color-triplet scalar $Y$, needed to mediate the $B\to \psi \,\mathcal{B}\,\mathcal{M}$ decay. Finally, in Sec.~\ref{sec:global} we have presented a global view of the experimental landscape with which the implications of $B$-Mesogenesis for current collider experiments have been assessed in an all-encompassing way.

\subsection{Summary}
\label{sec:summaryofresults}

\noindent The main results obtained in this work are the following:
\begin{itemize}[leftmargin=0.7cm,itemsep=0.4pt]
\item Current measurements of the semileptonic asymmetries in $B_q^0$ decays imply that baryogenesis and dark matter generation in the early Universe requires (see Fig.~\ref{fig:semileptonic_asymmetries})
\begin{align}\label{eq:PredictionCPaverage}
\br \gtrsim 10^{-4}\,,
\end{align}
where $\br$ is the inclusive branching fraction of a $B$ meson decaying into a visible baryon $\mathcal{B}$, a dark sector antibaryon $\psi$, and any number of light mesons collectively denoted by $\mathcal{M}$. 
\item The most stringent current constraint on these new decay modes arises from an old search at ALEPH for $b$ decay events with large missing energy~\cite{Barate:2000rc}. Our recast of this search yields constraints at the level of
\begin{align}
\,\,\,\,\,\,  \br &\lesssim 10^{-4}-0.5\%\,\,\text{(ALEPH)}\,, \\
\text{Br}(B\to \psi\,  \mathcal{B}) &\lesssim 10^{-4}-10^{-3}\,\,\text{(ALEPH)}\,.
\end{align}
The weakest constraints are for $m_\psi \sim 1\,\text{GeV}$, while the more stringent limit applies for heavier $\psi$ states, see Fig.~\ref{fig:LEP_constraints} for the precise dependence of the limits on $m_\psi$. 

\item Given the above constraints on $\br$, we conclude that at least one of the semileptonic asymmetries should be positive and have a magnitude
\begin{align}
A_{\rm SL}^q > 10^{-4}
\end{align}
in order to generate a sufficient baryon asymmetry.

\item We estimate that $B$ factories should have a sensitivity to exclusive $B$ meson decays of (see Sec.~\ref{sec:BR_Bfactories})
\vspace{-0.4cm}
\begin{subequations}\label{eq:Bfactories_sensitivitytoBR}
\begin{align}
    \qquad \text{Br}\left(B\to \psi + \mathcal{B} \right) &\sim 3\times 10^{-5}\,\,\text{(BaBar-Belle)}\,, \\ 
     \qquad     \text{Br}\left(B\to \psi + \mathcal{B} \right) &\sim 3\times 10^{-6}\,\,\text{(Belle II)}\,, 
\end{align}
\end{subequations}
which are up to two orders of magnitude better than existing limits.

\item Given our (rather primitive) phase space calculation of the $B\to \psi \, \mathcal{B}$ decay, we expect the ratio between exclusive and inclusive decays to be (see Fig.~\ref{fig:Br_ratio_Bplus})
\begin{align}
\frac{\text{Br}\left(B\to \psi \, \mathcal{B} \right)}{\br} \gtrsim (1-10)\,\%\,. 
\end{align}
This, together with Eq.~\eqref{eq:PredictionCPaverage} and Eq.~\eqref{eq:Bfactories_sensitivitytoBR}, supports our expectation that direct searches at $B$ factories have a great potential to test wide regions (if not all) of parameter space of $B$-Mesogenesis.

\item The LHC has the potential to search for $B\to \psi \,\mathcal{B}$, $B\to \psi \,\mathcal{B}^\star$, and $\mathcal{B}_b\to \bar{\psi} \,\mathcal{M}$ decays. In fact, a recent analysis shows that LHCb could reach sensitivities at the level of $\text{Br}\sim 10^{-7}-10^{-5}$~\cite{Rodriguez:2021urv}. We have discussed the implications of irreducible backgrounds for these searches in Sec.~\ref{sec:BR_LHC}, which need to be carefully analyzed to reach these target sensitivities.

\item The color-triplet scalar $Y$ needed to mediate the $B\to \psi \,\mathcal{B}\,\mathcal{M}$ decay has the potential to be discovered at the LHC by ATLAS and CMS. Current dijet and jet+MET searches already represent a powerful indirect constraint on the parameter space of the mechanism as shown in Fig.~\ref{fig:BR_mpsi}. In fact, we expect that upcoming squark searches at the HL-LHC will provide an independent probe of relevant and yet uncharted parameter space of the mechanism, as argued in Sec.~\ref{sec:ModelBuilding_triplet}.

\item While meson-mixing constraints do not directly constrain the mechanism, they do shape the $Y$-quark-quark and $Y$-$\psi$-quark coupling matrices. The fact that $A_{\rm SL}^q > 10^{-4}$ is needed for successful baryogenesis implies that there should be new physics contributions to the phases of $M_{12}^q$ or $\Gamma_{12}^q$ in the $B_q$ meson system, which could be probed via meson-mixing observables.
In fact, the new tree-level $b$ decays enabled by the triplet-scalar mediator could be directly responsible for such contributions, making the mechanism completely self-contained.

\item Our analysis of meson-mixing constraints in Sec.~\ref{sec:MesonMixing} shows that only one of the four operators highlighted in Sec.~\ref{sec:BaryogenesisnandDM} can be sizeable and thus responsible for baryogenesis. Namely, we do not expect any scenarios in which two operators are simultaneously large. However, it is in general not possible to favor any particular operator above the others provided that $\br$ and $m_\psi$ lie within the allowed regions depicted in Fig.~\ref{fig:BR_mpsi}. As a consequence, searches for all possible flavor variations of the $B\rightarrow \psi\mathcal{BM}$ are necessary in order to fully test the mechanism.
 
\item In Sec.~\ref{sec:global} we have presented a global analysis of the signals of $B$-Mesogenesis that highlights the remarkable complementarity between recent, current, and upcoming collider experiments in testing the parameter space of $B$-Mesogenesis. In particular, we have shown that it is possible that \emph{simultaneously} \textit{i)} BaBar, Belle and Belle II could discover the new decay mode $B\to \psi\,\mathcal{B}$, \textit{ii)} LHCb, ATLAS \& CMS could measure the relevant CP violation in the $B_s^0-\bar{B}_{s}^0$ system, and \textit{iii)} ATLAS and CMS could discover a TeV scale color-triplet scalar responsible for the $B\to \psi\,\mathcal{B}\,\mathcal{M}$ decay.
\end{itemize}

In summary, we have presented an exhaustive and global study of the experimental signals of baryogenesis and dark matter from $B$ mesons. Importantly, if the mechanism of~\cite{Elor:2018twp} is at play in the early Universe, there should be clear signals at $B$ factories of $B$ mesons decaying into a baryon and missing energy at rates that are within the reach of current data from BaBar and Belle and certainly from upcoming data at Belle II. In addition, upcoming measurements of the CP violation in the $B_q^0-\bar{B}_q^0$ system will play a key role in constraining the parameter space of the mechanism. This is relevant for measurements at Belle II, LHCb, ATLAS and CMS. Complementarily, dijet and jet+MET searches of TeV-scale color-triplet scalars can indirectly constrain the mechanism.

\subsection{Outlook: Future Directions}
\label{sec:outlook}

In the course of our analysis, the following additional experimental, phenomenological, and theoretical studies that remain to be explored have been identified: 

\begin{itemize}[leftmargin=0.4cm,itemsep=0.4pt]

\item Theory:
\begin{itemize}[leftmargin=0.4cm,itemsep=0.4pt]

\item \textit{UV models:} 
Across the entire paper, we have employed a minimal implementation of the particle content necessary for the mechanism and have remained agnostic about possible UV completions that can accommodate $B$-Mesogenesis. Needless to say, one would expect additional signals to arise after an exhaustive UV completion is constructed. For example, in a Supersymmetric version of the mechanism~\cite{Alonso-Alvarez:2019fym}, one finds new signals involving auxiliary long-lived states and further implications for flavor and neutrino physics. It is therefore interesting to pursue the construction of UV models of this kind further, as they may help to not only shed light on the origin of baryogenesis and dark matter, but also more generally on the nature of the physics beyond the Standard Model. \\

\item \textit{QCD calculations:} 
 In Sec.~\ref{sec:excl_vs_inclu}, we have presented a very rough calculation that has allowed us to quantify the relative size of exclusive and inclusive $B\to \psi\,\mathcal{B} \,\mathcal{M}$ decay modes. Unfortunately and due to the purely kinematic nature of our argumentation, our predictions in Figs.~\ref{fig:Br_ratio_Bplus} and~\ref{fig:Br_ratio_Bd} could have uncertainties of up to an order of magnitude. We would like to remark that a lattice or a sum-rule calculation of the exclusive decay rates required for baryogenesis and dark matter production would be highly desirable to provide more accurate estimates. \\ 
 Similarly, we have calculated the inclusive missing energy spectrum of $B$ mesons decays at the parton level in order to contrast it with a search performed at ALEPH~\cite{Barate:2000rc}. This is clearly a rough estimation of the actual inclusive spectrum, and a robust constraint can only be obtained after properly accounting for the relevant QCD corrections and for the effect of the $b$ quark momentum inside the $B$ meson. 

\end{itemize}

\item Phenomenology:
\begin{itemize}[leftmargin=0.4cm,itemsep=0.4pt]
\item \textit{Flavor Structures:} 
In Sec.~\ref{sec:ModelBuilding_triplet}, we have assessed what structures of the $Y$-quark-quark and $Y$-$\psi$-quark coupling matrices are needed in order to accommodate baryogenesis while satisfying all known collider and meson-mixing constraints on color-triplet scalars. Although we have not aimed to provide a theoretical explanation of the particular shapes of the resulting coupling matrices, we believe that such a study would be very relevant as it could yield indirect constraints on viable UV completions of the mechanism. Finally, given that the models aimed to explain the anomalous measurements in $B\to K \ell \ell $ decays~\cite{Aebischer:2019mlg} modify the mixing pattern of $B_s^0-\bar{B}_s^0$ oscillations~\cite{DiLuzio:2017fdq} and thus lead to $A_{\rm SL}^s \neq A_{\rm SL}^s|_{\rm SM}$, it would be very interesting to explore connections between the mechanism of~\cite{Elor:2018twp} and these current anomalies in $b\to s \ell \ell$ transitions.

\item \textit{CP violation in $\Gamma_{12}^q$:}
The new tree-level $b$ decays present in our model generically contribute to the width mixing in $B^0_{d,s}$ mesons.
We have not analyzed the sizes and phases of couplings of the triplet scalar that could result in an enhancement of $A_{\rm SL}^{d}$ and $A_{\rm SL}^s$ with respect to their SM values, as such calculation requires non-trivial QCD matching and a careful analysis of many tree-level decays, which is beyond the scope of this work.
Given that current global fits including some of the operators~\cite{Lenz:2019lvd,Lenz:2020efu} present in our model highlight that these modes can lead to large values of $A_{\rm SL}^q$, we believe that it would be particularly interesting to systematically consider the possible modifications to $\Gamma_{12}^q$ from new tree-level decays.

\item \textit{LHC simulations:} 
In Sec.~\ref{sec:LHC_colored_triplet_scalar} we have presented an analysis of dijet and jet+MET signals involving color-triplet scalars at the LHC. It would be interesting to re-evaluate these analyses at next-to-leading order in QCD and foregoing the narrow width approximation, in a similar fashion to~\cite{Albert:2017onk,Arina:2020tuw} but in the context of a singly-produced color-triplet scalar.\\
\end{itemize}

\item  Experiment:

\begin{itemize}[leftmargin=0.4cm,itemsep=0.4pt]
\item \textit{LHC searches:} 
In Sec.~\ref{sec:BR_LHC}, we have commented on the opportunities and difficulties when searching for the new decay modes of $B$ mesons and b-flavored baryons at the LHC. Although recent ideas to target these decays have been put forward~\cite{Rodriguez:2021urv,Poluektov:2019trg,Stone:2014mza}, evaluating the full impact of direct LHC searches on the parameter space of $B$-Mesogenesis remains an exciting avenue to explore.

\item \textit{Searches at BaBar, Belle and Belle II:} 
In Sec.~\ref{sec:BR_Bfactories} we have estimated that upcoming measurements at $B$ factories have the potential to test wide regions of parameter space for baryogenesis and dark matter from $B$ mesons as originally proposed in~\cite{Elor:2018twp}.
In fact, searches are currently ongoing for $B_d^0\to \psi \,\Lambda$ decays~\cite{privateBaBarBR,privateBelleBR} at BaBar, Belle and Belle II.
We look forward to the results of upcoming searches which we expect will significantly test $B$-Mesogenesis.
\end{itemize}
\end{itemize}

\newpage
Uncovering the mechanism responsible for generating an asymmetry of matter over antimatter in the early Universe is key to explaining our very existence.  That such a process may also be at the origin of dark matter, thereby explaining the nature of the component that dominates the matter density of the Universe, is particularly appealing. $B$-Mesogenesis is such a mechanism,  and in this work we have shown it to be fully \emph{testable} at current hadron colliders and  $B$ factories.
This scenario resurrects the dream that discovering baryogenesis can be a physics goal of $B$ factories and hadron colliders. In fact, we are looking forward to forthcoming results from searches and studies already underway at BaBar, Belle, Belle II and LHCb. By providing a roadmap towards the discovery of baryogenesis and dark matter from $B$ mesons, it is our hope that many more experimental programs will be set forth to discover or refute this mechanism.

\vspace{0.2cm}
\begin{center}
    \textbf{Acknowledgments}
\end{center}
\vspace{0.2cm}
We are forever thankful to Ann Nelson, our collaborator, mentor, and friend who pioneered the ideas that this paper is based on. We would also like to thank David McKeen for collaboration during the early stages of this work. We are grateful to Zoltan Ligeti, Jorge Martin-Camalich and Dean Robinson for useful discussions regarding the use of the vacuum insertion approximation for our decays, to Alexander Lenz for useful discussions regarding tree-level decays effects in $\Gamma_{12}^q$,  to Ville Vaskonen for helpful discussions on neutron star bounds, to Christos Hadjivasiliou, Jan Strube, and the whole Belle/Belle II team at PNNL for useful discussions regarding missing energy searches at $B$ factories, to Xabier Cid Vidal for helpful discussions on the LHCb capabilities to missing energy final states, and to Veronika Chobanova and Diego Mart\'inez Santos for useful discussions regarding $\phi_s^{c\bar{c}s}$ measurements and analyses. Finally, we are grateful to Veronika Chobanova, Xabier Cid Vidal, Bertrand Echenard,  Diego Mart\'inez Santos and Jan Strube for their very useful comments and suggestions on a preliminary version of this manuscript. GA acknowledges support by a ``la Caixa" postgraduate fellowship from the Fundaci\'on ``la Caixa", by a McGill Trottier Chair Astrophysics Postdoctoral Fellowship, and by NSERC (Natural Sciences and Engineering Research Council, Canada). GE is supported by the U.S. Department of Energy, under grant number DE-SC0011637. GE thanks the Berkeley Center for Theoretical Physics and Lawrence Berkeley National Laboratory for their hospitality during the completion of this work. ME is supported by a Fellowship of the Alexander von Humboldt Foundation.

\newpage
\section{Appendices}

\subsection{\texorpdfstring{$B$}{B} Meson Oscillations in the Early Universe}
\label{sec:decoherence}
In this appendix we describe the set of Boltzmann equations that characterize the evolution of the population of $B$ mesons and their decay products in the early Universe in order to establish the benchmarks for the experimental searches proposed in the main body of this work.
Although our discussion is based on that of~\cite{Elor:2018twp} and~\cite{Nelson:2019fln}, we have updated the numerical results and identified and quantified the main sources of uncertainty in the calculation.
Compared to~\cite{Elor:2018twp} and as was done in~\cite{Nelson:2019fln}, we use a density-matrix approach to track the evolution of the number densities, thus consistently taking into account decoherence effects in the neutral $B$ meson oscillations.
It is important to note that the results originally obtained in~\cite{Nelson:2019fln} show some discrepancies with the one in~\cite{Elor:2018twp} in the limit of vanishing decoherence, where both approaches should agree.
Crucially and thanks to our updated numerical treatment, we resolve this discrepancy and fully agree with~\cite{Elor:2018twp} in this limiting case.
We therefore strongly advise to use the numerical results presented in this appendix rather than the ones in~\cite{Elor:2018twp,Nelson:2019fln}.

At times well before the relevant dynamics take place (i.e. at times much shorter than $1/\Gamma_\Phi$), the Universe is assumed to be dominated by an admixture of radiation and non-relativistic $\Phi$ particles.
The Hubble parameter can thus be expressed as
\begin{equation}
H^2 = \frac{8\pi}{3} \frac{\rho_\mathrm{r} + \rho_\Phi}{\Mpl^2}\,,
\end{equation}
in terms of the energy densities, $\rho_\Phi = M_\Phi n_\Phi$, and $\rho_\mathrm{r} = (\pi^2/30)g_\star(T)T^4$, where $g_\star(T)$ are the number of relativistic degrees of freedom as a function of temperature which we take from~\cite{Laine:2015kra}. 

The evolution of the energy and number densities of radiation and the decaying $\Phi$ particles is given by the following Boltzmann equations
\begin{align}\label{eq:Boltzmann_rad_Phi}
\frac{\diff n_\Phi}{\diff t} + 3Hn_\Phi = -\Gamma_\Phi n_\Phi \, ,\\
\frac{\diff \rho_\mathrm{r}}{\diff t} + 4H\rho_\mathrm{r} = \Gamma_\Phi M_\Phi n_\Phi \, .
\end{align}
Note that although the decay of $\Phi$ is assumed to be predominantly into non-relativistic $b$ quarks, these hadronize and decay into radiation in a timescale much shorter than that of the Universe expansion, $H^{-1}$.
This justifies neglecting the brief existence of $B$ mesons in~\eqref{eq:Boltzmann_rad_Phi} to only include the final relativistic products of the decay chain of $\Phi$.

Instead of using the lifetime of $\Phi$ as a parameter, we use the ``reheat temperature" to which it leads, i.e. $3H(T_\mathrm{R}) \equiv \Gamma_\Phi$. Physically, this temperature must be below the hadronization scale $\Lambda_{\mathrm{QCD}}\sim 200\MeV$ (otherwise $B$ mesons do not hadronize) but above $\sim 4\MeV$~\cite{deSalas:2015glj,Hasegawa:2019jsa} so that the decays of $\Phi$ do not spoil the successful predictions of Big Bang Nucleosynthesis.
In this way, the $b$ quarks produced via the decay of $\Phi$ hadronize into $B^\pm$, $B^0_d$, $B^0_s$ mesons and $b$-baryons.
Among those, the neutral $B$ mesons undergo CP violating oscillations and subsequently decay.
If the oscillation process is coherent, i.e. it is not damped by interactions with the plasma, then the resulting chain of events can lead to the generation of a net particle-antiparticle asymmetry in the Universe.

The most accurate framework to study these processes in the early Universe is the density-matrix formalism. This formalism has been applied to study similar set ups in~\cite{Kneller:2004jz,Cirelli:2011ac,Tulin:2012re,Ipek:2016bpf} and was specifically described in~\cite{Nelson:2019fln} for $B$-Mesogenesis. We therefore follow closely the notations and equations~\cite{Nelson:2019fln}. In terms of the density matrix
\begin{equation}
n = 
\begin{pmatrix}
n_{BB} & n_{B\bar{B}} \\
n_{\bar{B}B} & n_{\bar{B}\bar{B}}
\end{pmatrix},
\end{equation}
the Boltzmann equations that govern the oscillations can be compactly written as
\begin{align}\label{eq:Boltzmann_density_matrix}
\frac{\diff n}{\diff t} + 3Hn = &-i(\mathcal{H}n - n\mathcal{H}^\dagger) - \half \Gammasc \left[ O_{-},\,\left[ O_{-}, n \right] \right] \nonumber \\ 
&+ \half \Gamma_\Phi n_\Phi \mathrm{Br}_{\Phi\rightarrow B} O_{+}\, .
\end{align}
Here, $O_{\pm}\equiv \mathrm{diag}(1,{\pm}1)$ discriminates between flavor-blind and flavor-sensitive interactions.
The first term describes oscillations and decays in terms of the $\Delta F=2$ effective Hamiltonian $\mathcal{H} = M - i \Gamma / 2$.
The second one describes elastic scatterings with the electrons and positrons in the plasma, which occur due to the non-trivial charge distribution in the neutral mesons.
These interactions are flavor-sensitive and therefore act to decohere the $B^0-\bar{B}^0$ oscillations via the quantum Zeno effect~\cite{Misra:1976by}.
We use the estimation of the scattering rate computed in~\cite{Elor:2018twp},
\begin{equation}\label{eq:eB_scattering_rate}
\Gammasc \simeq 10^{-11}\GeV \left( \frac{T}{20\MeV} \right)^5 \left( \frac{\braket{r^2_{B}}}{0.187\,\text{fm}^2} \right)^2,
\end{equation}
that depends on the charge radius of neutral $B$ mesons which can only be estimated theoretically and for which we have taken the value obtained in~\cite{Hwang:2001th}.

The particle-antiparticle asymmetry $\Delta_B = n_{BB}-n_{\bar{B}\bar{B}}$ that is generated through the coherent oscillations is translated into a baryon-antibaryon asymmetry through the new $B$ meson decay into a baryon ($\mathcal{B}$) and a dark antibaryon ($\psi$): $B\to \psi \, \mathcal{B}\,\mathcal{M}$.
In this manner, the visible positive baryon asymmetry is compensated by a negative one that is stored in the dark matter.
Thus, dark matter is antibaryonic in our set up. 
The baryon asymmetry in the visible sector, which we denote by $n_{\mathcal{B}}$, is therefore sourced by the $B$ meson asymmetry $\Delta_B$.
The time evolution of $n_{\mathcal{B}}$ is given by
\begin{equation}\label{eq:Boltzmann_asymmetry}
\frac{\diff n_{\mathcal{B}}}{\diff t} + 3Hn_{\mathcal{B}} = \mathrm{Br}_{B\rightarrow\mathcal{B}}\Gamma_B \Delta_B \, ,
\end{equation}
from which the baryon-to-entropy ratio
\begin{equation}
Y_\mathcal{B} = \frac{n_{\mathcal{B}}}{\tfrac{2\pi^2}{45}g_{\star, S}(T) T^3}\,,
\end{equation}
can be obtained.

\begin{figure}[t!]
\centering
\includegraphics[width=0.99\linewidth]{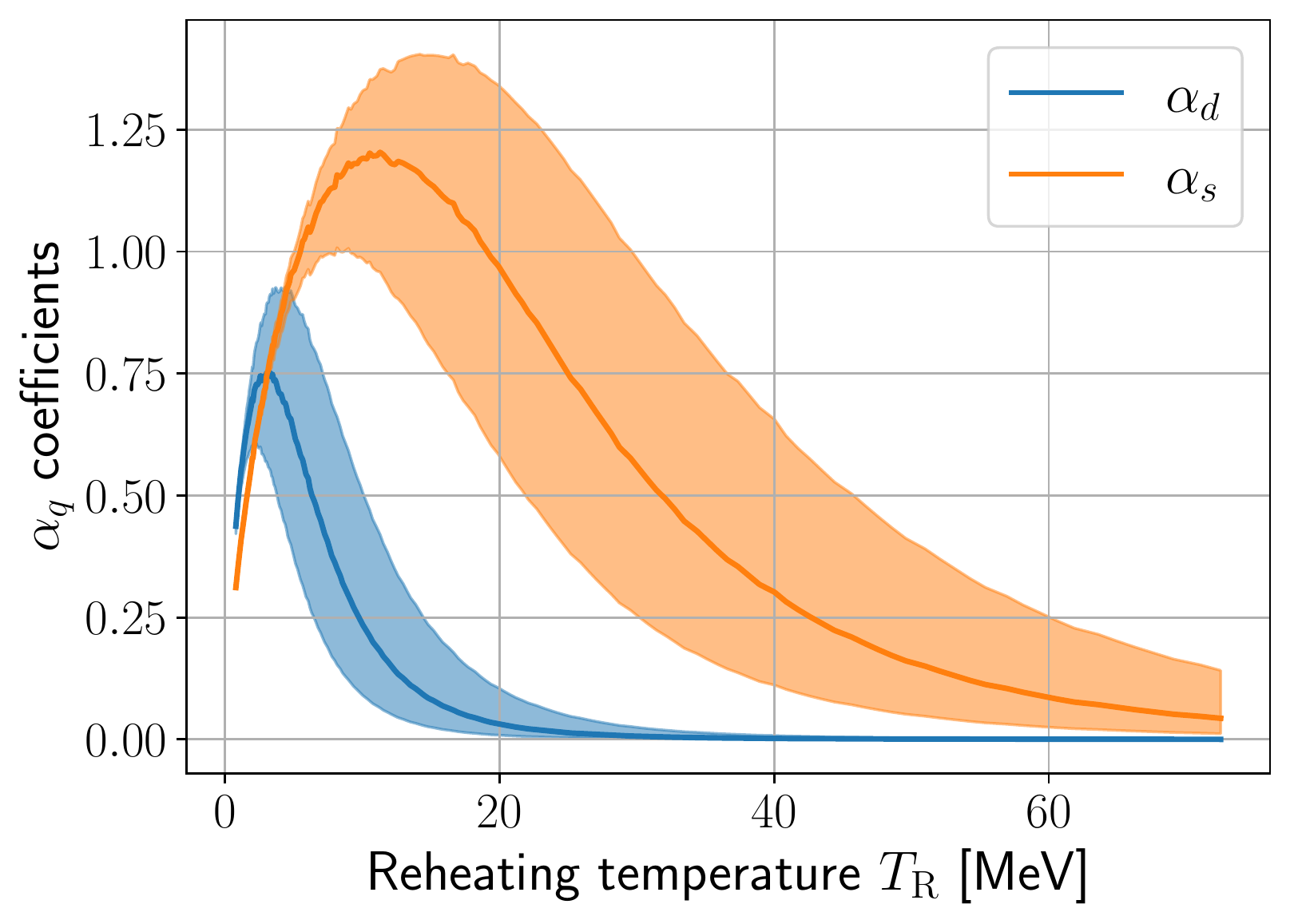} 
\vspace{-0.2cm}
\caption{Functions describing the dependence of the generated baryon asymmetry on the reheat temperature, as parametrized in~\eqref{eq:basypara} and~\eqref{eq:alpha_ds_parametrization}. The severe suppression of the baryon yield at high temperatures is caused by the decoherence of the $B^0-\bar{B}^0$ oscillations due to interactions with electrons and positrons in the plasma. The shaded regions represent the uncertainty in our calculation arising from the lack of precise knowledge of the charge radius of the neutral $B$ mesons.}
\label{fig:alpha_ds}
\end{figure}

Owing to the fact that both the oscillation frequency and the lifetime of neutral $B$ mesons are much smaller than $H^{-1}$ at the relevant temperatures, the system described by~\eqref{eq:Boltzmann_density_matrix} reaches equilibrium in a very short timescale.
This allows us to solve~\eqref{eq:Boltzmann_density_matrix} algebraically for the fixed point of $\Delta_B$ as a function of time, once~\eqref{eq:Boltzmann_rad_Phi} is solved numerically. With this, one can numerically integrate~\eqref{eq:Boltzmann_asymmetry}, including the contributions from both $B^0_d$ and $B^0_s$, to obtain the baryon asymmetry.

The resulting asymmetry can be parametrized as in~\eqref{eq:basypara} in terms of the functions $\alpha_{d,s}(T_R,M_\Phi)$, which encompass the dependence of the result on the cosmological parameters.
In turn, these functions can be written as
\begin{subequations}\label{eq:alpha_ds_parametrization}
\begin{align}\label{eq:alpha_ds_parametrization1}
\!\!\!\! \alpha_d(T_R,M_\Phi) &= \left( \frac{11\GeV}{M_\Phi} \right) \left( \frac{\mathrm{Br}_{\Phi\rightarrow B^0_d/\bar{B}^0_d}}{0.4} \right) c_d(T_R) \, , \\
\label{eq:alpha_ds_parametrization2}
\!\!\!\! \alpha_s(T_R,M_\Phi) &= \left( \frac{11\GeV}{M_\Phi} \right) \left( \frac{\mathrm{Br}_{\Phi\rightarrow B^0_s/\bar{B}^0_s}}{0.1} \right) c_s(T_R) \, ,
\end{align}
\end{subequations}
where the functions $c_{s,d}$ enclose the dependence on the reheat temperature and have to be obtained numerically.
The reheat temperature is ultimately controlled by the lifetime of $\Phi$, and it determines the precise moment when the out-of-equilibrium population of $B$ mesons is injected.
Keeping this in mind, the functional form of $\alpha_{s,d}(T_R)$, which is displayed in Fig.~\ref{fig:alpha_ds}, can be understood as follows.

If the decay of $\Phi$ happens at low temperatures, the rate at which electrons and positrons scatter with the neutral $B$ mesons, given in~\eqref{eq:eB_scattering_rate}, is small and the oscillations can proceed unimpeded.
In this limit, the asymmetry scales proportionally to the number of $B$ mesons produced in the decay of $\Phi$ relative to the number of photons, which grows as $\propto T_R$.

At higher reheat temperatures, scatterings with $e^{\pm}$ become more frequent and decohere the system once their rate overcomes the frequency at which the neutral mesons oscillate.
This results in a sharp decline in the generated asymmetry, which happens earlier for $B^0_d$ mesons given that their oscillation frequency $\Delta M_d = 0.5065\pm 0.0019\,\mathrm{ps}^{-1}$ is smaller than that of $B^0_s$ mesons, $\Delta M_d = 17.757\pm 0.021\,\mathrm{ps}^{-1}$.

The overall conclusion is that the generation of the asymmetry is maximally efficient at $T_R\sim 5\MeV$ for $B^0_d$ mesons and $T_R\sim 10\MeV$ for $B^0_s$ mesons.
Given that current experimental data favor a negative value of the semileptonic asymmetry $A_{\rm SL}^d$ of $B^0_d$ mesons, these are expected to contribute negatively to the baryon asymmetry.
Under this assumption, the production of the observed positive baryon asymmetry is driven by $B^0_s$ mesons and occurs most favorably at temperatures $T_R\sim 10-20\MeV$, depending on the exact magnitude of $A_{\rm SL}^s$ and $A_{\rm SL}^d$, which are currently not well-determined experimentally.

The uncertainty of the numerical calculation of $c_{d,s}$ is shown as shaded regions in Fig.~\ref{fig:alpha_ds} and is dominated by the lack of precise knowledge of the charge radius of neutral $B$ mesons.
A larger (smaller) value of the charge radius results in more (less) efficient scatterings with the $e^\pm$ in the plasma and reduces (increases) the generated asymmetry.
For our benchmark calculation we use the value $\braket{r^2_B} = -0.187\,\mathrm{fm}^2$, which was estimated in~\cite{Hwang:2001th}.
However, given the dispersion in its determination by different authors~\cite{Hwang:2001th,Becirevic:2009ya,Das:2016rio}, we choose to conservatively allow for a factor of $2$ uncertainty in this parameter, which leads to the upper and lower bands for $c_{d,s}$ depicted in Fig.~\ref{fig:alpha_ds}.

\begin{table}[t]
\label{Table:fragmentation_ratios}
\renewcommand{\arraystretch}{1.4}
  \setlength{\arrayrulewidth}{.18mm}
\centering
\setlength{\tabcolsep}{0.18 em}
\begin{tabular}{ |c | c | c | c |}
    \hline
     & $Z$ decays & Tevatron & $\Upsilon (5S)$ \\
     \hline
     $f_u$ & 0.408 & 0.344 & 0.379 \\
     \hline
     $f_d$ & 0.408 & 0.344 & 0.379 \\
     \hline
     $f_s$ & 0.100 & 0.115 & 0.199 \\
     \hline
     $f_{\rm baryon}$ & 0.084 & 0.198 & 0.045 \\
    \hline
\end{tabular}
\caption{Experimentally measured values for the fragmentation ratios of $b$ quarks into $B^\pm$,  $B^0_d$, $B^0_s$, and $b$-baryons as compiled by~\cite{pdg}. In addition, ATLAS has measured a value $f_s/f_d = 0.252$~\cite{Aad:2015cda} in $pp$ collisions at $7$~TeV.}
\label{tab:fragmentation_ratios}
\end{table}

Aside from this, the functions $\alpha_{d,s}$ depend on the mass of $\Phi$, which has to be $M_\Phi>2m_{b}$, and on its branching fraction to $B^0_d$ and $B^0_s$.
Assuming for simplicity that the decay of $\Phi$ always produces $b$ quarks, the important quantity is the fragmentation ratio of $b$ quarks into the different $b$-hadrons.
These fractions are usually denoted $f_u$ (for $B^\pm$), $f_d$ (for $B^0_d$), $f_s$ (for $B^0_s$), and $f_{\rm baryon}$ (for $b$-baryons).
These ratios are challenging to compute from first principles and have experimentally been found to be very process dependent, as can be seen in Table~\ref{tab:fragmentation_ratios}.
Clearly, the fragmentation ratios for $\Phi$ decays are completely unknown.
For concreteness, in Fig.~\ref{fig:alpha_ds} we choose a benchmark ratio $4:4:1:1$,  consistent with measurements from $Z$ decays into $b{\bar{b}}$~\cite{pdg}.
However, the actual value for $\Phi$ decays could significantly deviate from this benchmark as, for example, the $B^0_d$ to $B^0_s$ ratio is measured to be closer to $3:1$ in $p\bar{p}$ collisions~\cite{pdg}.
Fortunately, it is straightforward to use Eqs.~\eqref{eq:alpha_ds_parametrization} to scale our results for different values of these ratios.
For the predictions presented in the main text, we choose to account for the uncertainty in the fragmentation fractions by allowing the ratio $f_s/f_d$ to vary over the range of experimentally measured values given in Table~\ref{tab:fragmentation_ratios}, which amounts to $f_s/f_d\in [0.22,0.37]$.

\subsection{Missing energy spectrum from $b$-decays at ALEPH}\label{sec:ALEPH_recast}
This appendix contains supplementary material to our recast of the ALEPH search for events with large missing energy at LEP~\cite{Barate:2000rc}. In particular, we provide the relevant formulas needed to calculate the missing energy spectrum of the decays predicted in $B$-Mesogenesis as produced at the $Z$ peak in $Z\to b\bar{b}$ processes. We first comment on the relationship between the center of mass energy ($\sqrt{s} = M_Z$) and the energy of the decaying $b$-quark as encoded in the b-quark fragmentation function. Then we discuss how to calculate the missing energy spectrum in exclusive 2-body $B$ decays and for inclusive decays, $b\to  \psi u_i d_j $. We finish by showing some representative results for each of the cases.

\begin{center}
    \textbf{\small{The b-fragmentation function at LEP}}
\end{center}

A relevant quantity to calculate the missing energy spectrum at ALEPH is the fragmentation function of $b$ hadrons at the $Z$ peak. This essentially represents the fraction of the center-of-mass energy that the decaying $b$ quark carries, $x_b = E_b/(\sqrt{s}/2)$. The distribution of $x_b$ values is dependent upon the hadronization of the $b$-quark and is typically characterized using theoretical models which are then fitted to b-quark decay data. In~\cite{Barate:2000rc}, the Peterson model~\cite{Peterson:1982ak} is used, which gives
\begin{align}
   \frac{d N_b }{d x_b}  =  \frac{N}{x_b}\left[1-\frac{1}{x_b} - \frac{\epsilon_b}{1-x_b}\right]^{-2},
\end{align}
where $N$ is a normalization constant, and $\epsilon_b$ must be obtained from a fit to kinematic decay data and is found to be $\epsilon_b = 0.0032\pm 0.0017$~\cite{Buskulic:1994dx}. We note that the used fragmentation function does lead to a relevant $\sim 20-40\%$ effect on the resulting missing energy spectrum at high energy as relevant for the ALEPH analysis. In order to derive our constraints, and as is done in~\cite{Barate:2000rc}, we use $\epsilon_b = 0.0032+0.0017$ in order to remain conservative.

\begin{center}
    \textbf{\small{Exclusive Missing Energy Spectrum:}\\ \small{2-body hadronic decays}}
\end{center}

We study the missing energy spectrum in exclusive two-body decays of the type $B \to \psi\,\mathcal{B}$. For these decays, the kinematics is fixed in the $B$ meson rest frame, and the energy of the dark antibaryon $\psi$ simply reads
\begin{align}
E_\psi^{\rm B-frame} = \frac{m_B^2 +m_\psi^2- m_\mathcal{B}^2}{2 m_B}\,.
\end{align}
This energy can be easily boosted to the laboratory frame, 
\begin{align}\label{eq:boost_LEP}
E_{\rm Miss}^{\rm LEP} = \gamma \left(E_\psi^{\rm B-frame} -\beta p_\psi^{\rm B-frame}  \cos \theta  \right),
\end{align}
where $p_\psi = \sqrt{E_\psi^2-m_\psi^2}$, $\beta = \sqrt{\gamma^2-1}/\gamma $ and $\gamma = E_B/m_b$, and where $E_B = x_b \sqrt{s}/2$. In addition, $\theta$ is a random angle that characterizes the direction of the emitted $\psi$ with respect to the direction of flight of the $B$ meson. By choosing $\sqrt{s} = M_Z = 91.2\,\text{GeV}$ and generating random samples of $\theta$, the missing energy spectrum at LEP can be easily simulated.

\begin{figure*}[t]
\centering
\begin{tabular}{cc}
		\label{fig:Histo2body}
		\includegraphics[width=0.48\textwidth]{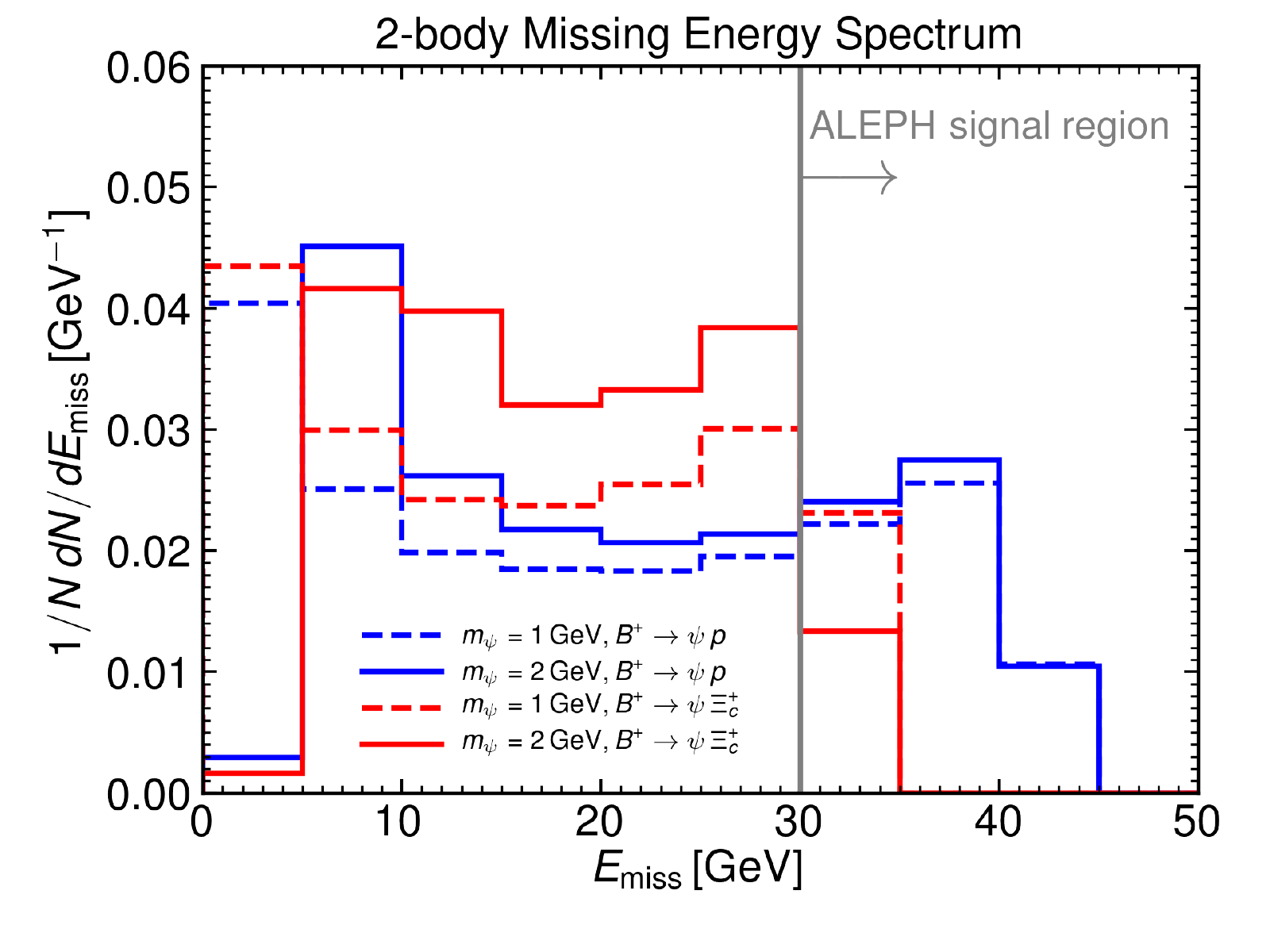}
&
		\label{fig:Histo3body}
		\includegraphics[width=0.48\textwidth]{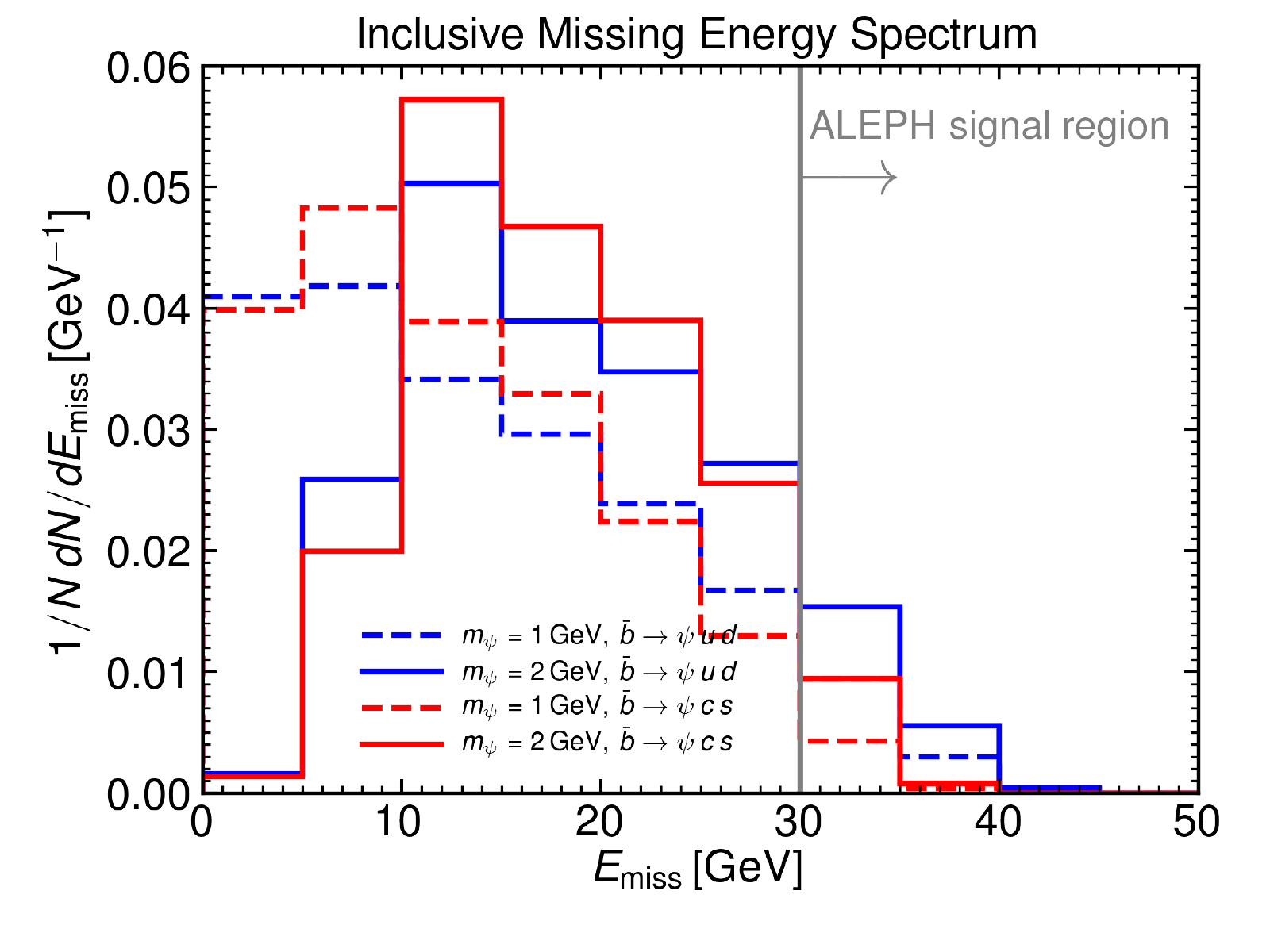}
\end{tabular}
\vspace{-0.4cm}
\caption{Missing energy spectrum from the decays predicted by $B$-Mesogenesis at $\sqrt{s} = M_Z$ arising from $Z\to \bar{b}b$ processes. \textit{Left panel:} exclusive two body decays, $B\to \psi \, \mathcal{B}$. \textit{Right panel:} inclusive decays (calculated here at the parton level as a 3-body decay without any QCD corrections). We also highlight the signal region targeted in the ALEPH~\cite{Barate:2000rc} search. Note that the region with $E_{\rm miss} < 30\,\text{GeV}$ suffers from the presence of substantially large backgrounds.  
}
\label{fig:MissingEnergy}
\end{figure*}

\begin{center}
    \textbf{\small{Inclusive Missing Energy Spectrum:}\\ \small{3-body parton decays}}
\end{center}

We also calculate the missing energy of inclusive decays at the parton level, $\bar{b}\rightarrow \psi u d$, where $u$ and $d$ stand for quarks of any generation. In the $b$-frame the differential decay rate can be written as
\begin{align}\label{eq:Dgamma_dMud}
    d\Gamma = \frac{1}{(2\pi)^3} \frac{1}{8m_b} |\mathcal{M}|^2 dE_u dE_\psi\,.
\end{align}
Given the purely right-handed nature of the triplet scalar couplings, the matrix elements can be expressed as
\begin{align}
	\mathcal{O} = (\psi b)(ud) \quad &\rightarrow \quad |\mathcal{M}|^2 \propto (p_b\cdot p_\psi) (p_u\cdot p_d),\\
	\mathcal{O} = (\psi d)(ub) \quad &\rightarrow \quad |\mathcal{M}|^2 \propto (p_b\cdot p_u) (p_\psi\cdot p_d),\\
	\mathcal{O} = (\psi u)(db) \quad &\rightarrow \quad |\mathcal{M}|^2 \propto (p_b\cdot p_d) (p_\psi\cdot p_u)\,.
\end{align}
Taking into account the relevant 3-body decay phase space (see e.g. the PDG review for kinematics) one can rewrite these momentum products in terms of $E_u$ and $E_\psi$ and integrate over $E_u$. The formulas that we obtain for the differential decay rate $d\Gamma/d E_\psi$ are not particularly insightful and we do not explicitly reproduce them here. In any case, this yields $d\Gamma/d E_\psi$, the differential missing energy spectrum in the b-quark rest frame as a function of the $\psi$ mass for each type of operator. 

The last ingredient in order to calculate the missing energy spectrum is to know the kinematic regions for the decay. Clearly, these correspond to $E_\psi^{\rm min} = m_\psi$ ($\psi$ at rest) and $E_\psi^{\rm max} = (m_b^2+m_\psi^2-(m_u+m_d)^2)/2m_b$ ($ud$ diquark at rest). Finally, the shift to the LAB frame is done by performing boost similar to that described in Eq.~\eqref{eq:boost_LEP}.
For our numerical evaluation, we use the following values for the quark masses:
\begin{align}
&m_b^{\rm pole} = 4.8\,{\rm GeV},\quad
m_c = 1.275\,{\rm GeV}\,,\quad
m_s = 0.093\,{\rm GeV}\,,\quad \nonumber \\
&m_d = 0.0047\,{\rm GeV}\,,\quad
m_u = 0.0022\,{\rm GeV}\,.
\end{align}

\begin{center}
    \textbf{\small{Simulation}}
\end{center}

We generate 20000 random events for every value of $m_\psi$ for each of the 4 different operators. The 3 random variables are 
\begin{enumerate}
    \item $x_b$, which is generated with the Peterson probability distribution and ranging between $0 \leq x_b\leq 1$.
    \item $\theta$, which is generated randomly over an uniform distribution in the range $0\leq \theta \leq \pi $.
    \item For 3-body decays, $E_{\psi}$, which is generated randomly given the probability distribution dictated by $d\Gamma/dE_{\psi}$ and in the region $m_\psi \leq E_{\psi} \leq (m_b^2+m_\psi^2-(m_u+m_d)^2)/2m_b$. 
    \end{enumerate}

The missing energy spectrum from exclusive ($B\to \psi\,\mathcal{B}$) and inclusive ($\bar{b}\to \psi\,u_i\,d_j$) decays is shown in Fig.~\ref{fig:MissingEnergy}. In the left panel, we can appreciate that the exclusive missing-energy spectrum significantly depends upon the $\psi$ mass and upon the mass of the resulting baryon in the final state. Heavier baryons in the final state naturally lead to a smaller missing energy. In the right panel of Fig.~\ref{fig:MissingEnergy}, we show the missing energy spectrum for inclusive decays for operators of the type $(\psi b)(ud)$ (similar results are obtained for the others). In this case, the number of events with large missing energy is substantially smaller than for exclusive 2-body decays. This is simply because in the 3-body decay the $\psi$ particle carries less energy than in the exclusive decays in the b rest frame. It is important to highlight that QCD and heavy quark corrections to the missing energy spectrum are expected to lead to an even smaller fraction of events in this window---in a similar way to what happens with $b\to s \bar{\nu}\nu$ decays, see the erratum of~\cite{Grossman:1995gt} for a discussion. 

\subsection{Exclusive vs. Inclusive decays: \texorpdfstring{$B^0_s$}{B0s} and \texorpdfstring{$\Lambda_b$}{Lambdab}}\label{sec:Exclusive_vs_Inclusive}
\begin{figure*}[t]
\centering
\begin{tabular}{cc}
		\label{fig:Br_ratio_Bs1}
		\includegraphics[width=0.48\textwidth]{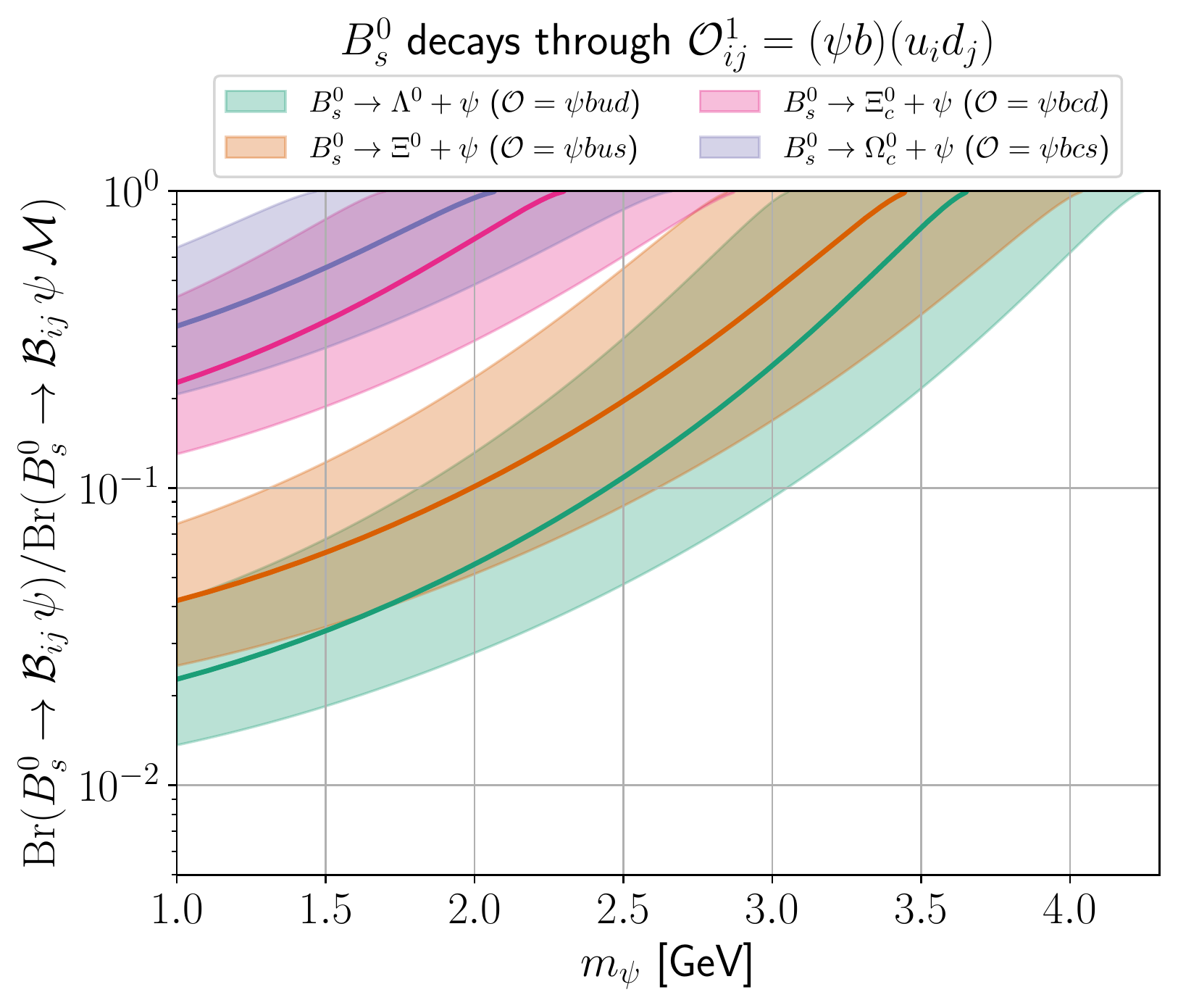}
&
		\label{fig:Br_ratio_Bs2}
		\includegraphics[width=0.48\textwidth]{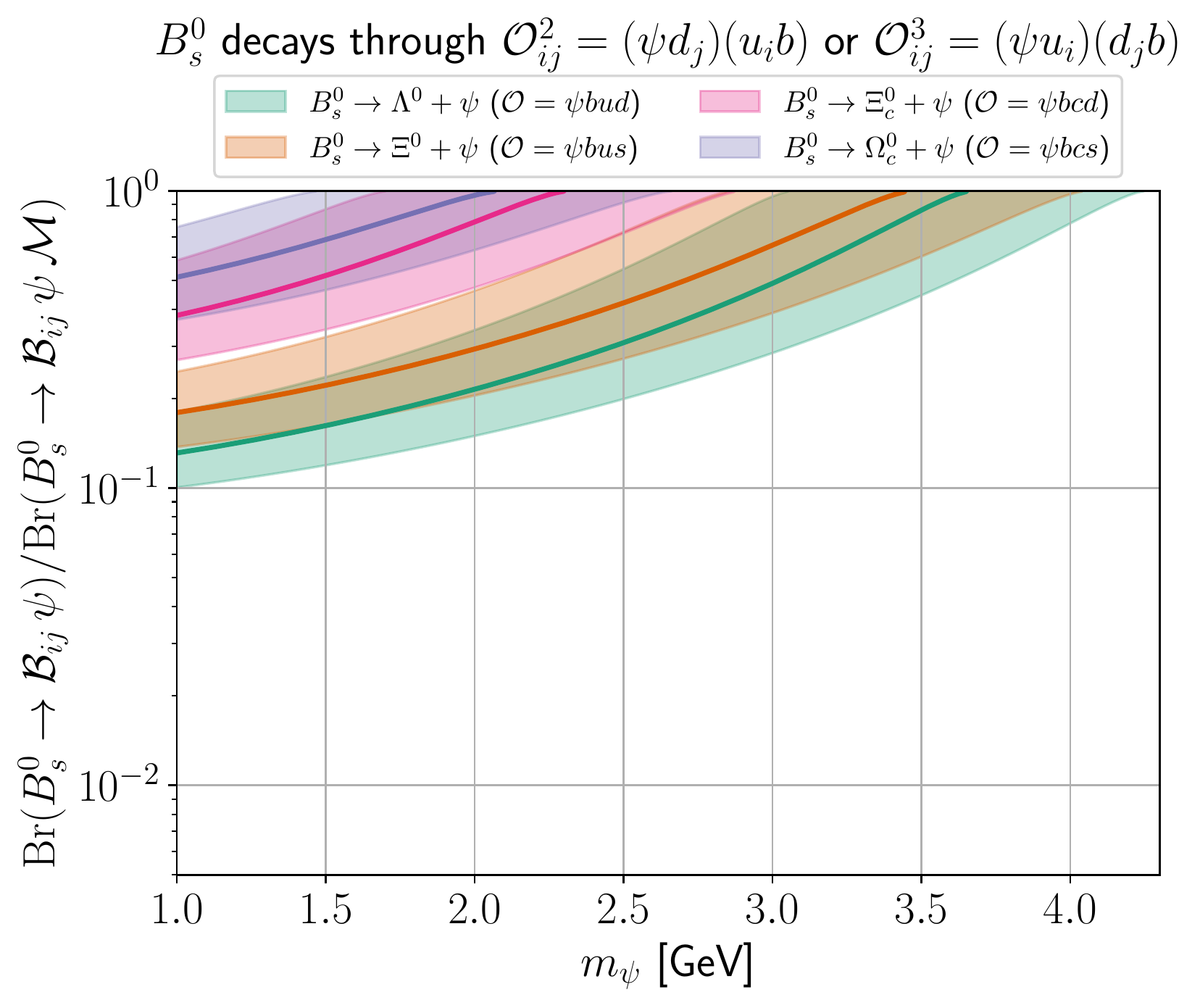}
\end{tabular}
\vspace{-0.4cm}
\caption{Same as Fig.~\ref{fig:Br_ratio_Bplus} but for $B^0_s\to \psi\,\mathcal{B}\,\mathcal{M}$ decays.
}
\label{fig:Br_ratio_Bs}
\end{figure*}

\begin{figure*}[t]
\centering
\begin{tabular}{cc}
		\label{fig:Br_ratio_Lambdab1}
		\includegraphics[width=0.48\textwidth]{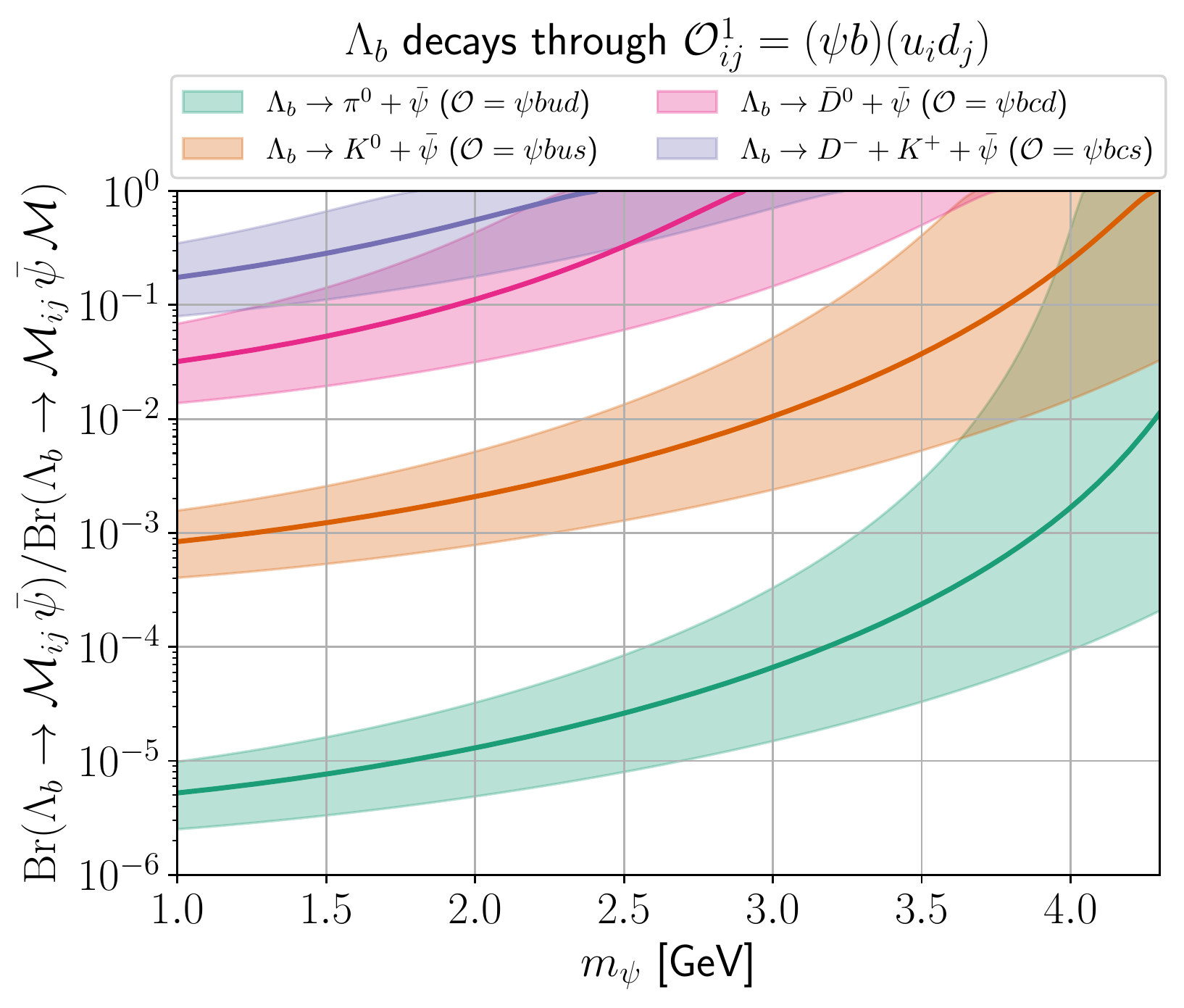}
&
		\label{fig:Br_ratio_Lambdab2}
		\includegraphics[width=0.48\textwidth]{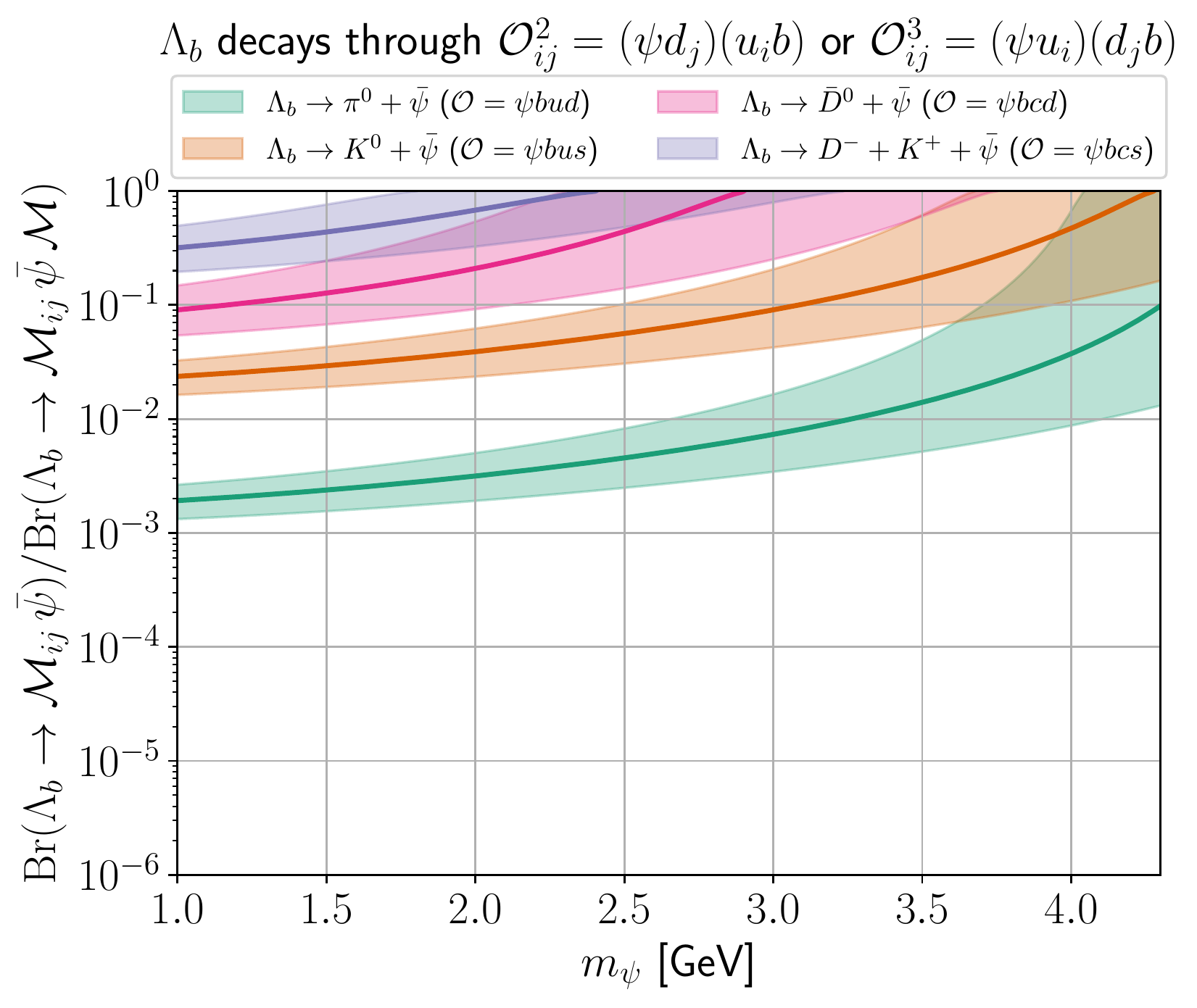}
\end{tabular}
\vspace{-0.4cm}
\caption{Same as Fig.~\ref{fig:Br_ratio_Bplus} but for $\Lambda_b\to \psi\,\mathcal{M}$ decays.
}
\label{fig:Br_ratio_Lambdab}
\end{figure*}

Figs.~\ref{fig:Br_ratio_Bs} and~\ref{fig:Br_ratio_Lambdab} show the fraction of $B^0_s$ and $\Lambda_b$ decays that are not expected to contain hadrons other than the ones listed in Table~\ref{tab:hadronmasses} in the final state, as a function of the mass of the dark fermion $\psi$. These fractions are obtained based on the phase-space arguments detailed in Sec.~\ref{sec:excl_vs_inclu}, and in particular using Eq.~\eqref{eq:Br_ratio_phase_space} with the appropriate flavor content for each case. From the right panel of Fig.~\ref{fig:MissingEnergy} we can appreciate that the missing energy spectrum is not very large in the signal region, $E_{\rm miss}>30\,\text{GeV}$, with only a fraction of $\sim 10-15\,\%$ of the events in such kinematic region. This is simply because the 3-body kinematics imply typically lead to  less energetic $\psi$ particles in the b-quark rest frame (as compared to the exclusive 2-body decay).


\begin{figure*}[t]
\centering
\begin{tabular}{c c}
\feynmandiagram[scale=1.6,transform shape, horizontal=a to b] { 
i1 [particle={\tiny \(q_1\)}]
-- a
-- [edge label={\tiny \( q_{\scaleto{j}{3pt}}/\psi\)}] b
-- f1 [particle={\tiny\( q_2 \)}],
i2 [particle={\tiny\( \overline q_2\)}]
-- c
-- [edge label'={\tiny \( q_{\scaleto{i}{3pt}}/\psi \)}] d
-- f2 [particle={\tiny\( \overline q_1 \)}],
{ [same layer] a -- [scalar, edge label'={\tiny \( Y \)}] c },
{ [same layer] b -- [scalar, edge label={\tiny \( Y \)}] d},
};
&
\feynmandiagram[scale=1.6,transform shape, horizontal=a to b] { 
i1 [particle={\tiny \( q_1\)}]
-- a
-- [edge label={\tiny \( q_{\scaleto{j}{3pt}}\)}] b
-- f1 [particle={\tiny\( \overline q_1 \)}],
i2 [particle={\tiny\( \overline q_2 \)}]
-- c
-- [edge label'={\tiny \( q_{\scaleto{i}{3pt}} \)}] d
-- f2 [particle={\tiny\( q_2 \)}],
{ [same layer] a -- [scalar, edge label'={\tiny \( Y \)}] c },
{ [same layer] b -- [boson, edge label={\tiny \( W,H \)}] d},
};
\end{tabular}
\vspace{-0.4cm}
\caption{\footnotesize{Box diagrams involving the heavy colored scalar $Y$ and contributing to the neutral meson oscillations.
The one involving the dark baryon $\psi$ is only possible for down-type external quarks.}}
\label{fig:box_diagrams}
\end{figure*}
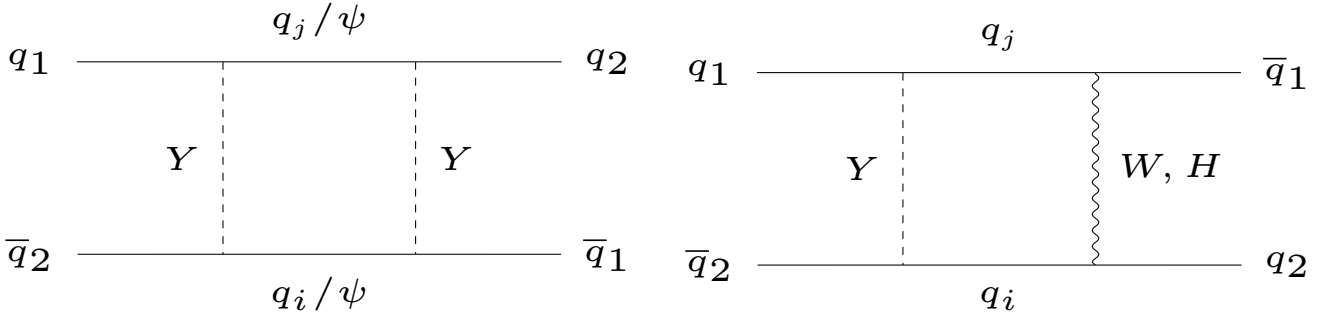
%

\newpage
\subsection{LHC Bounds on Color-Triplet Scalars}\label{sec:LHC_plots}

\begin{figure*}[p]
\centering
\setlength{\tabcolsep}{10pt}
\renewcommand{\arraystretch}{1}
\begin{tabular}{ccc}
		\label{fig:yud_1_5TeV}
		\includegraphics[width=0.26\textwidth]{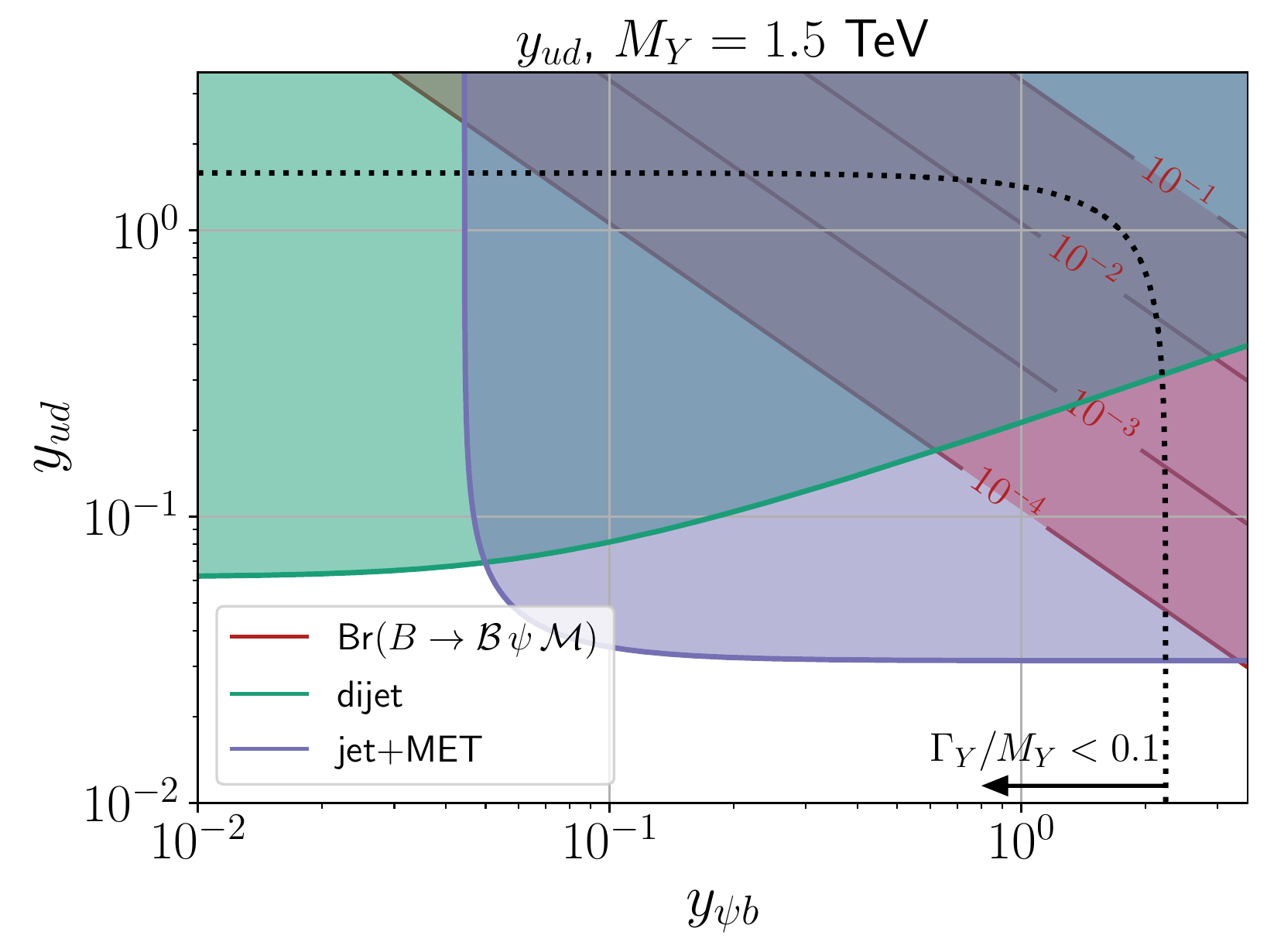}
&
		\label{fig:yud_3TeV}
		\includegraphics[width=0.26\textwidth]{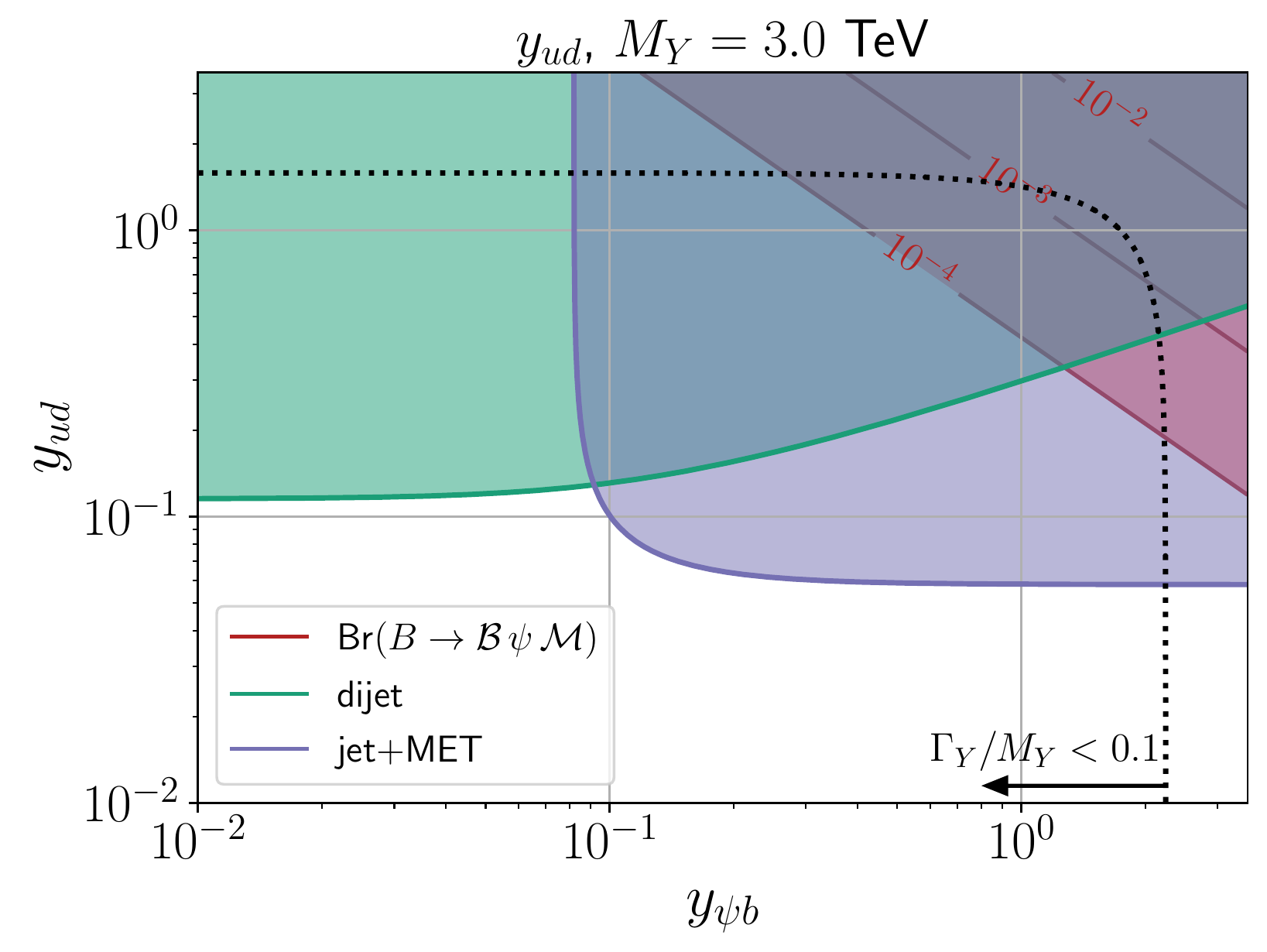}
&
		\label{fig:yud_5TeV}
		\includegraphics[width=0.26\textwidth]{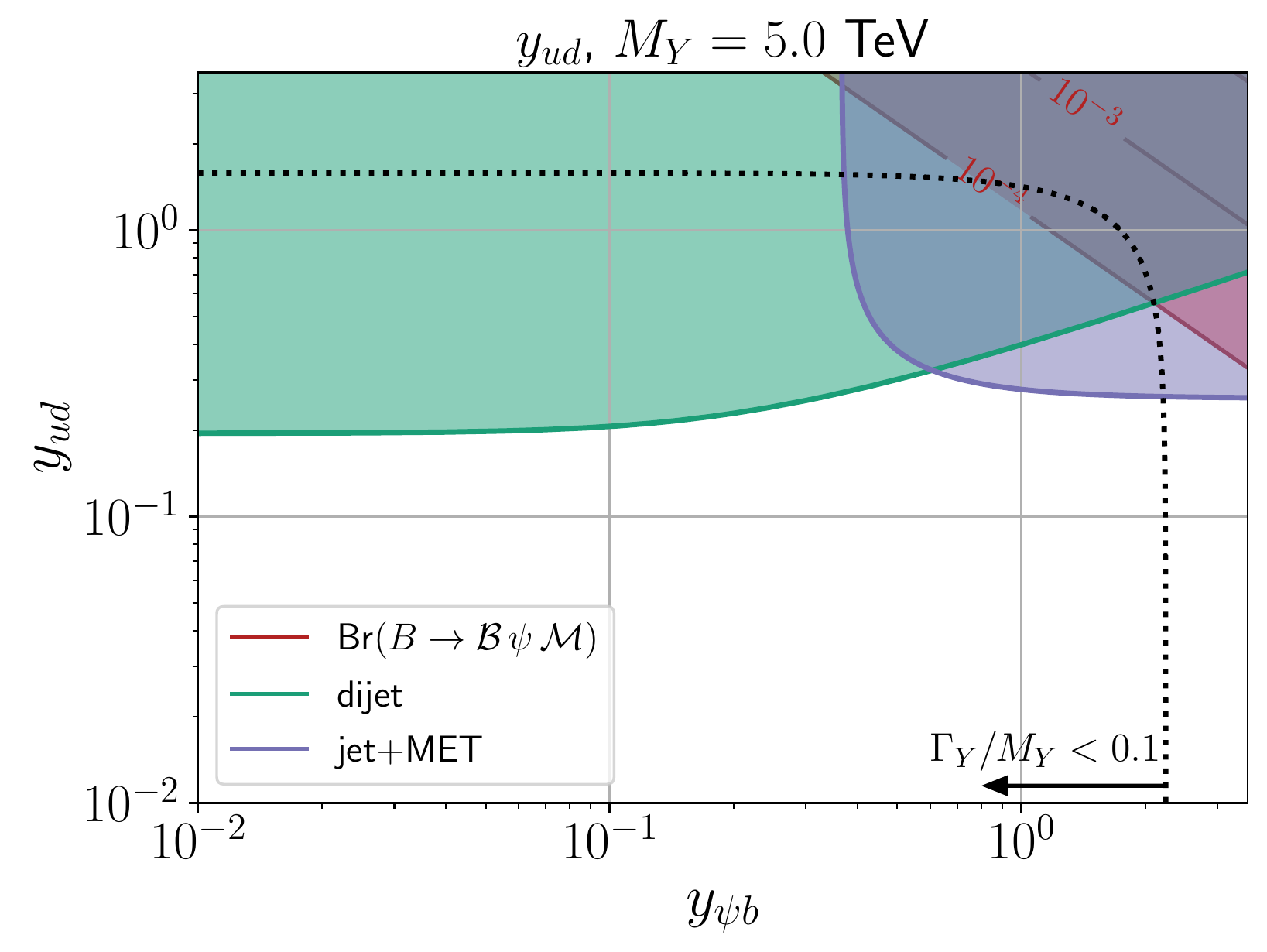}
\\
		\label{fig:yus_1_5TeV}
		\includegraphics[width=0.26\textwidth]{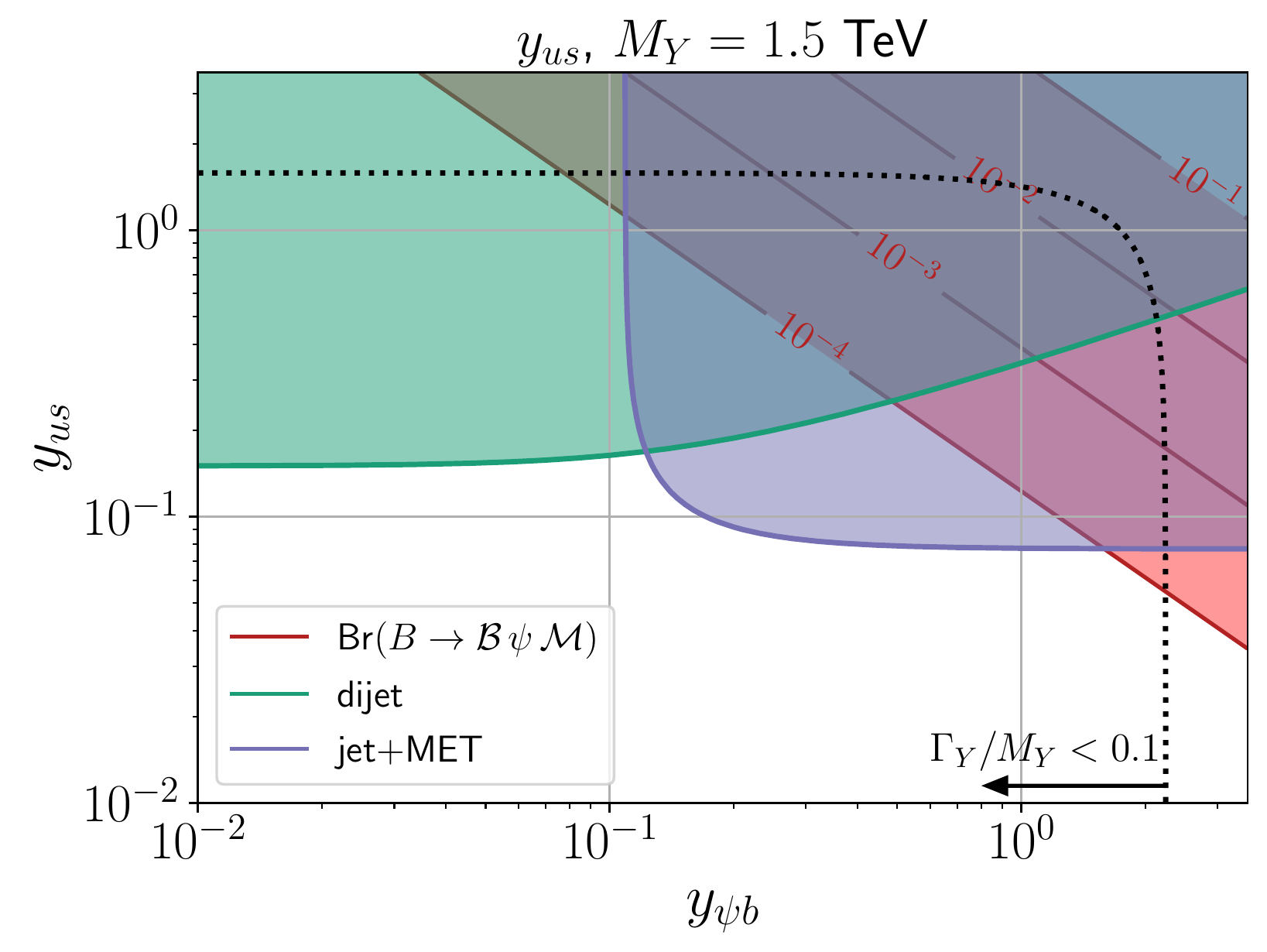}
&
		\label{fig:yus_3TeV}
		\includegraphics[width=0.26\textwidth]{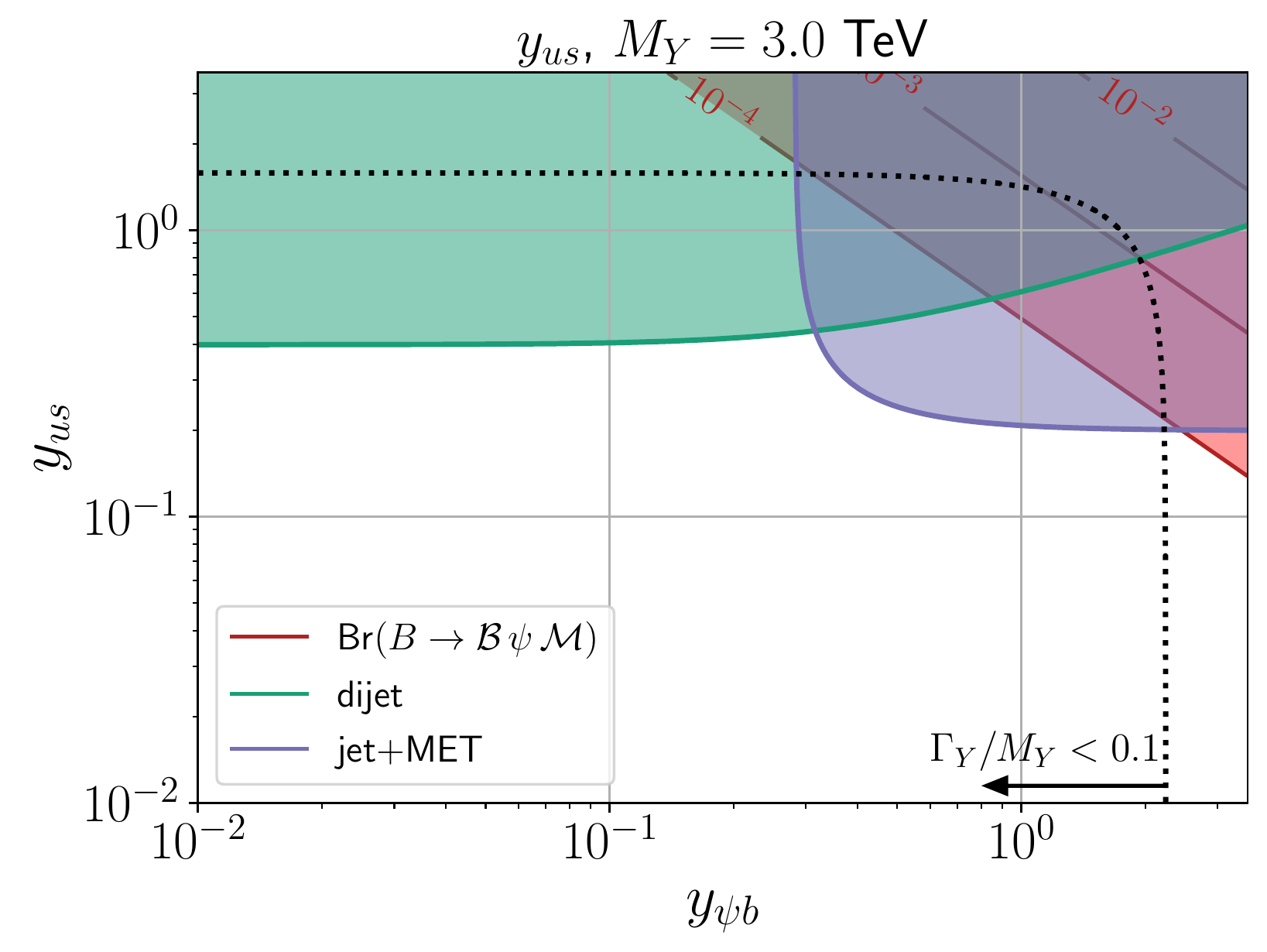}
&
		\label{fig:yus_5TeV}
		\includegraphics[width=0.26\textwidth]{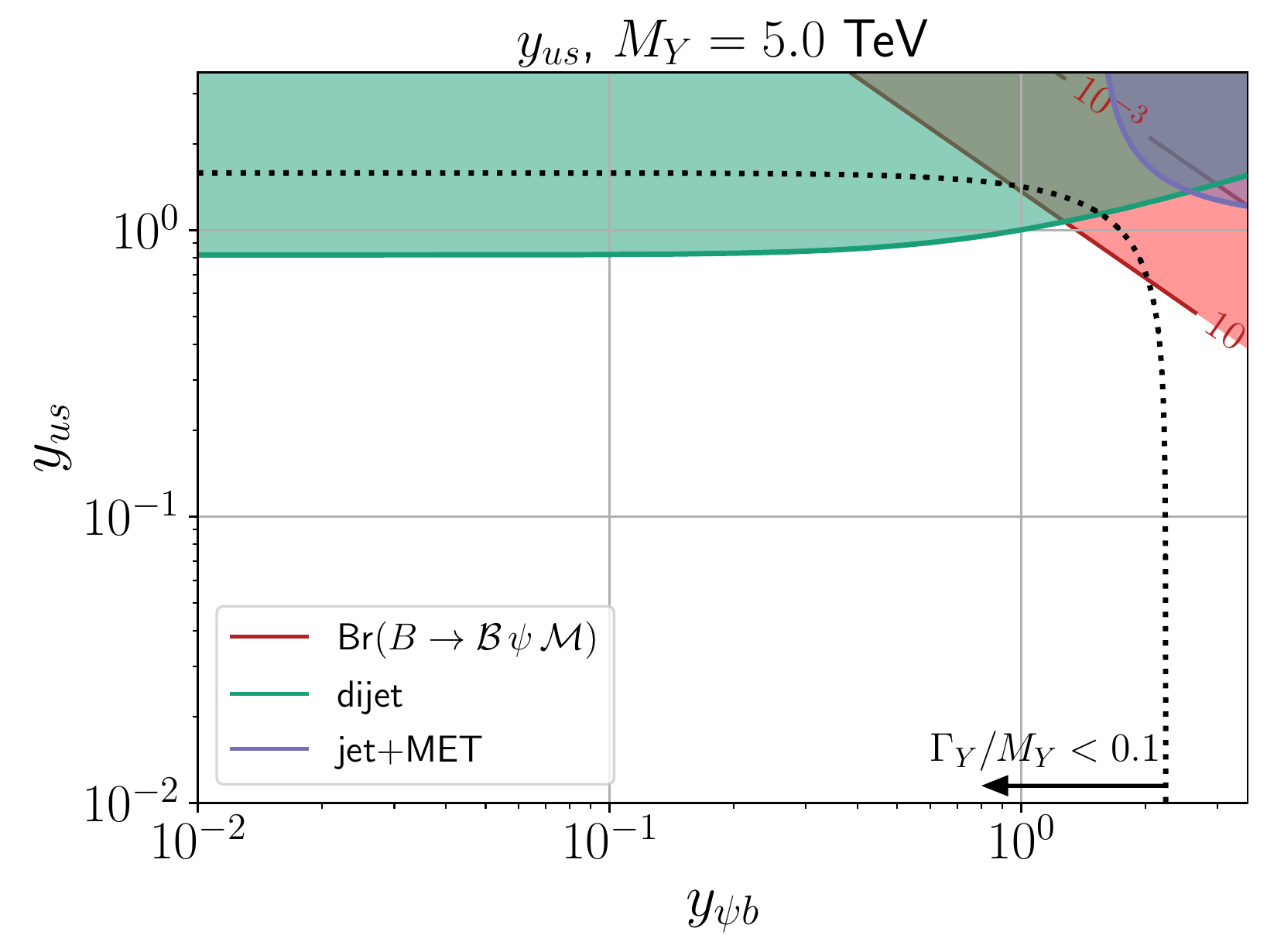}
\\
		\label{fig:yub_1_5TeV}
		\includegraphics[width=0.26\textwidth]{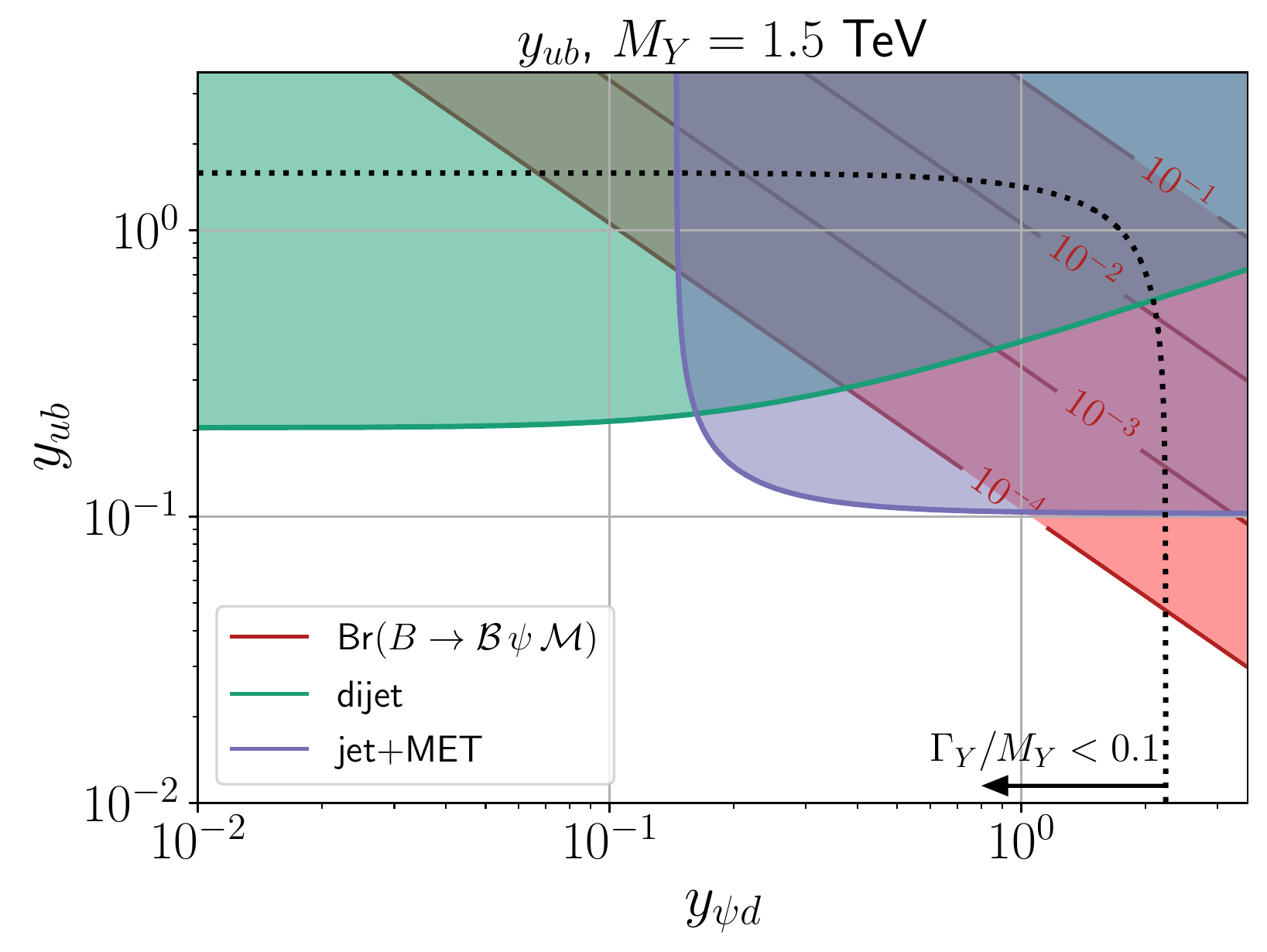}
&
		\label{fig:yub_3TeV}
		\includegraphics[width=0.26\textwidth]{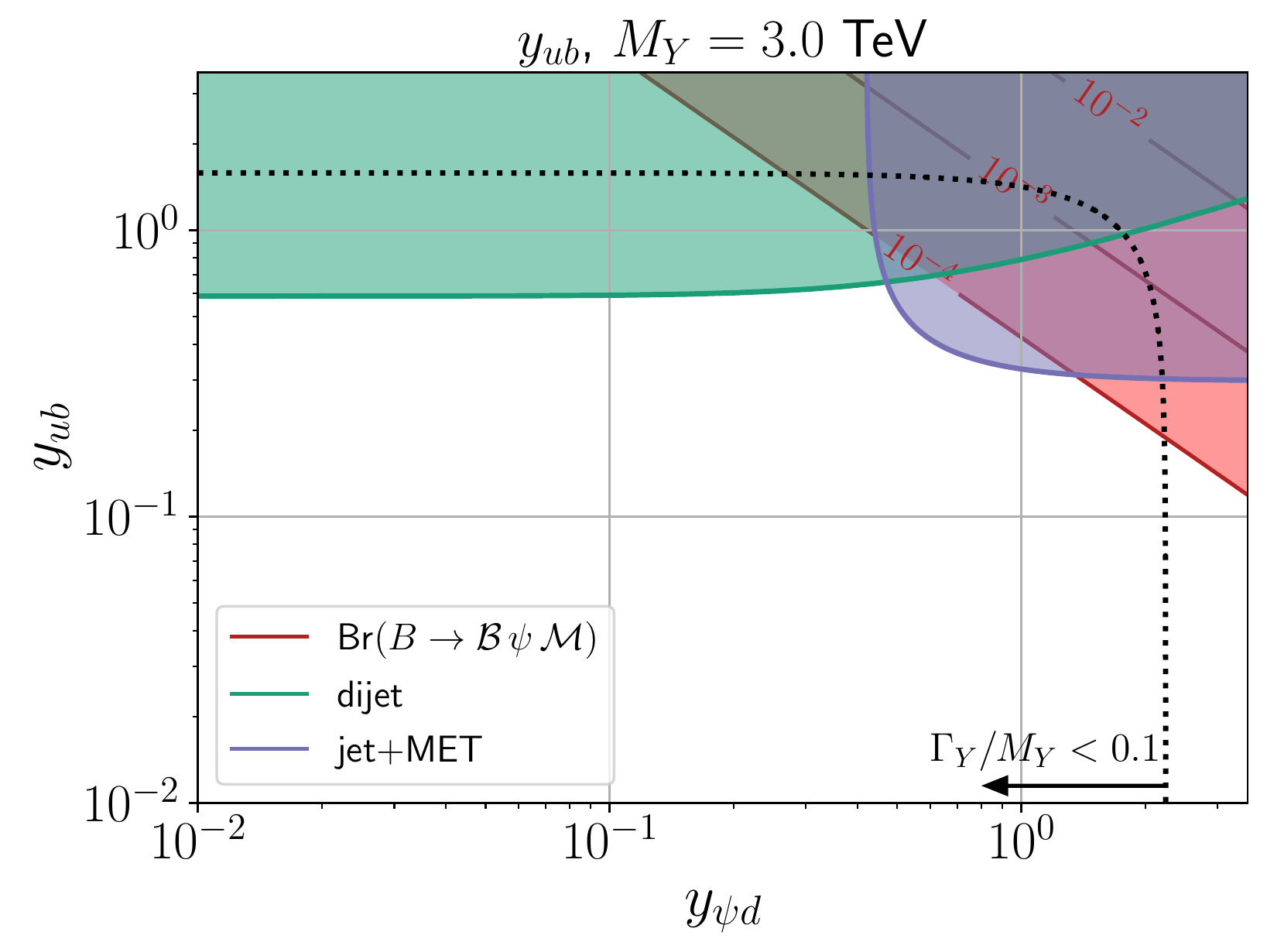}
&
		\label{fig:yub_5TeV}
		\includegraphics[width=0.26\textwidth]{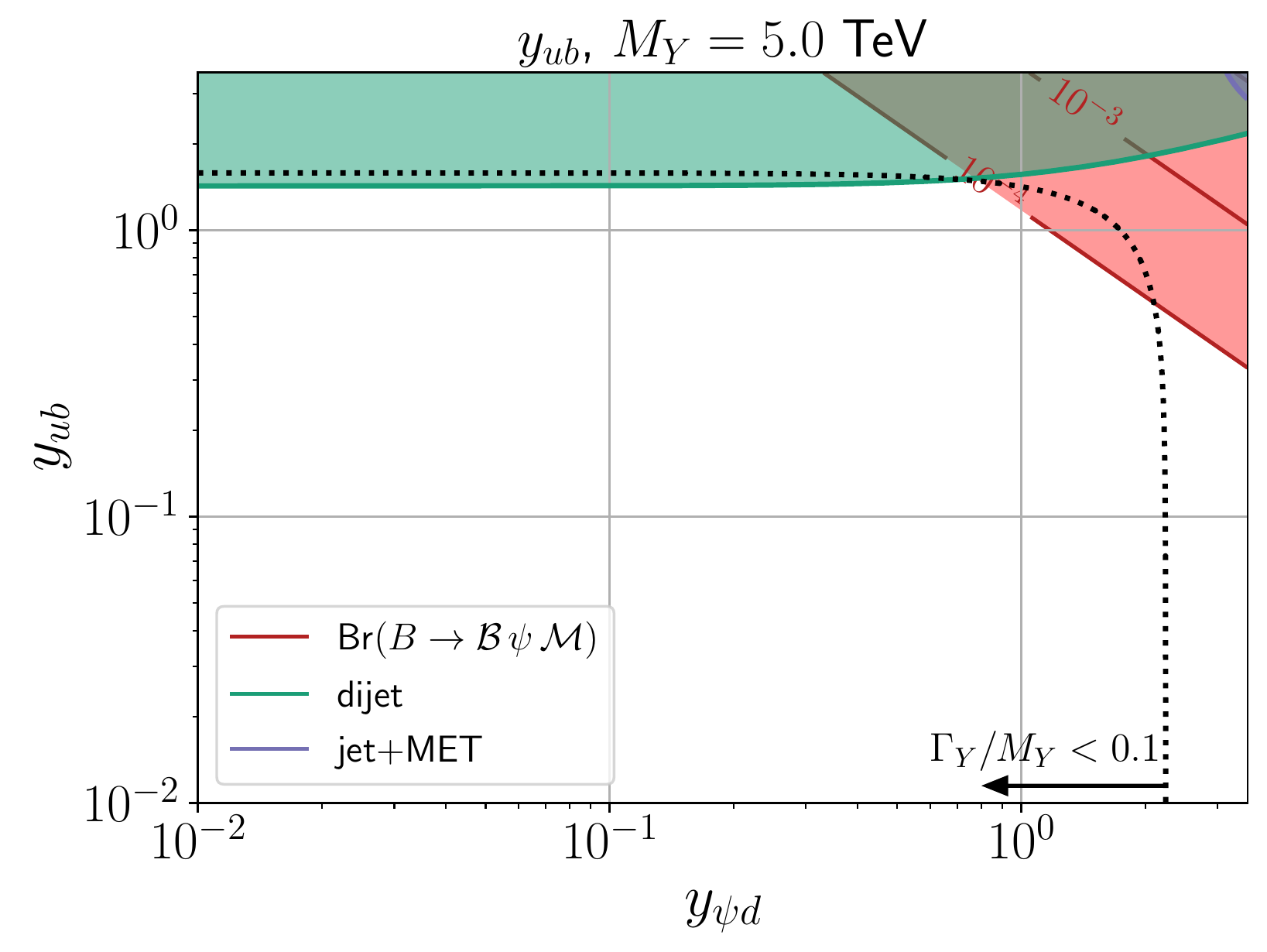}
\\
		\label{fig:ycd_1_5TeV}
		\includegraphics[width=0.26\textwidth]{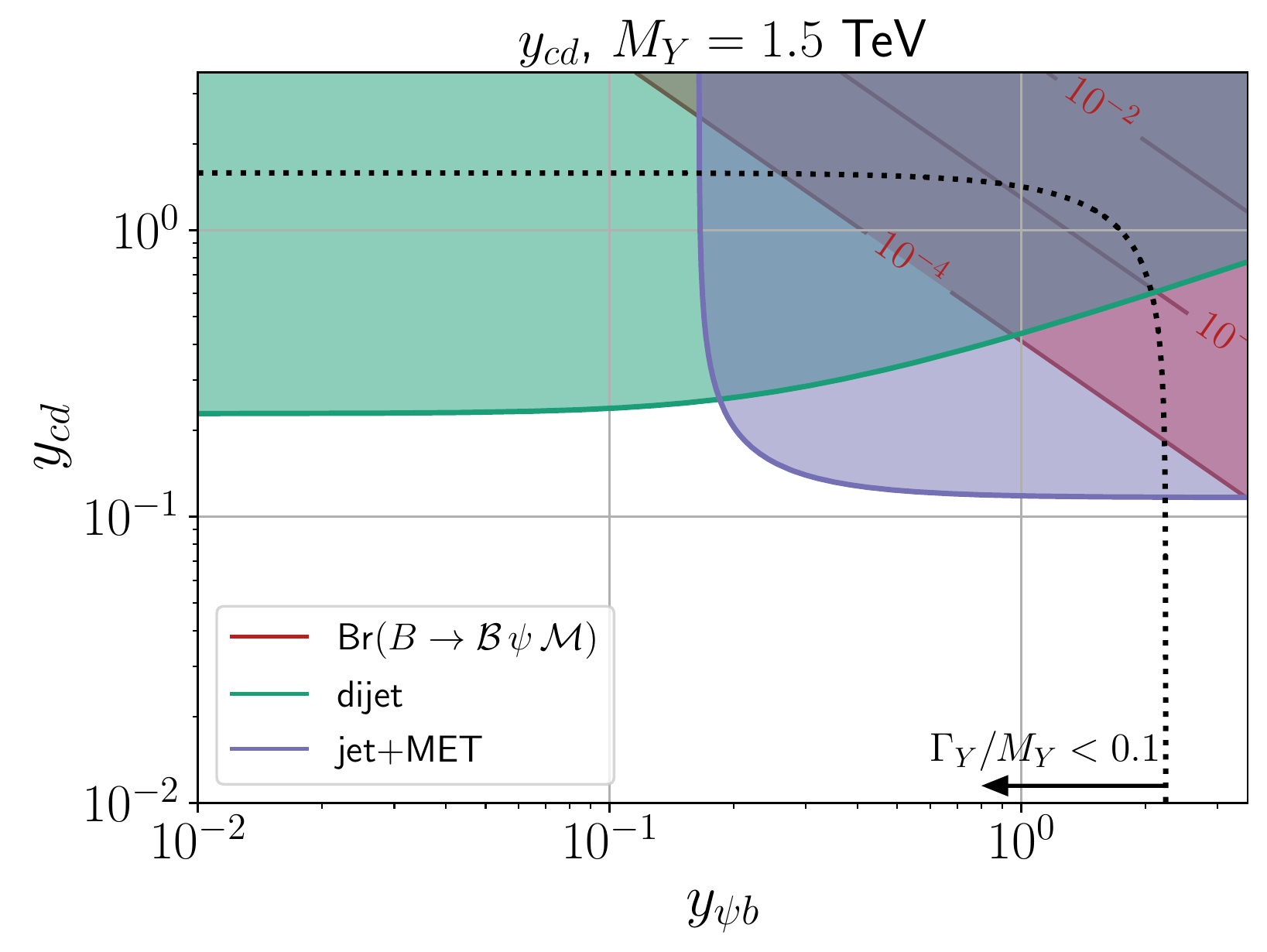}
&
		\label{fig:ycd_3TeV}
		\includegraphics[width=0.26\textwidth]{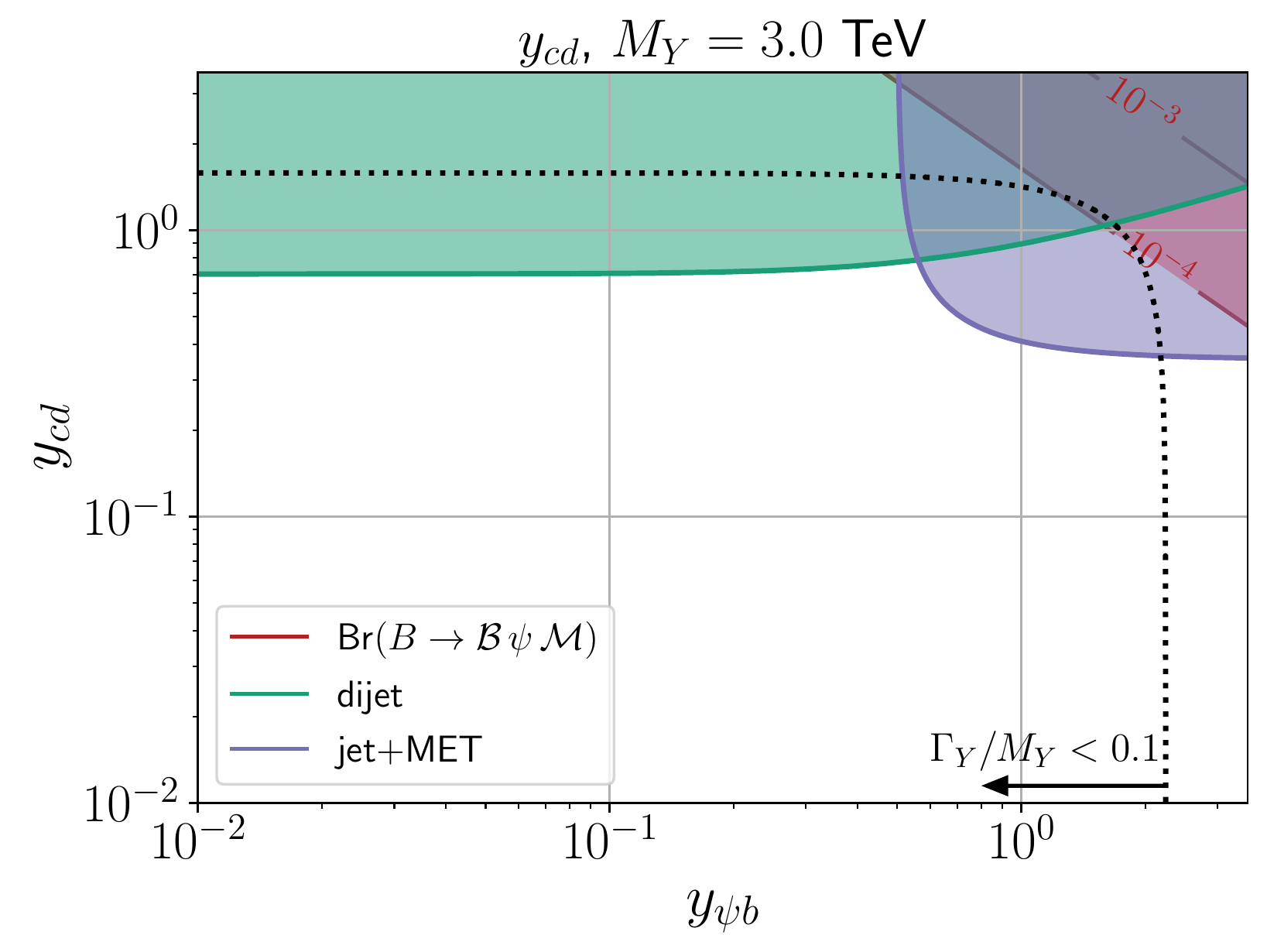}
&
		\label{fig:ycd_5TeV}
		\includegraphics[width=0.26\textwidth]{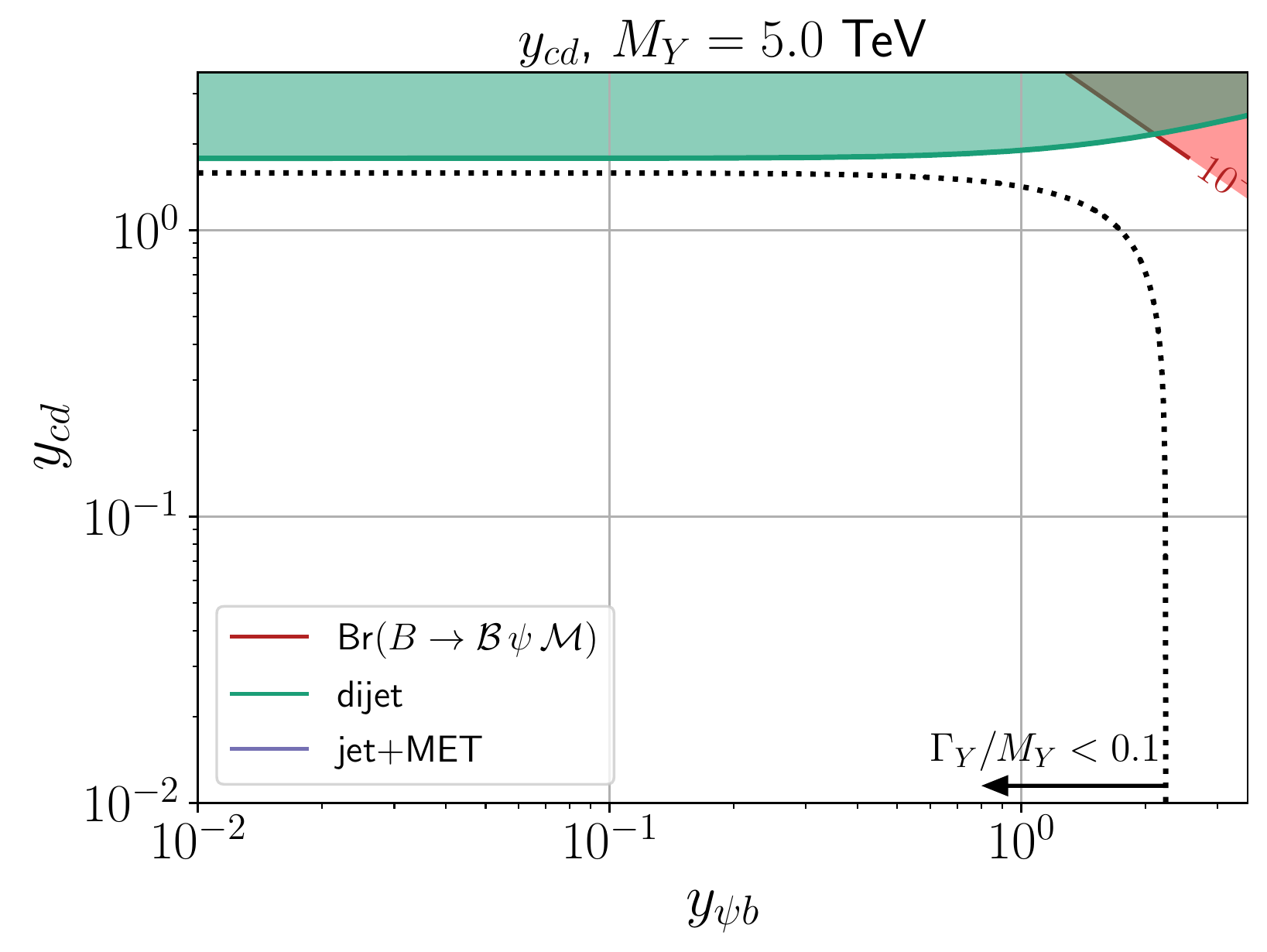}
\\
		\label{fig:ycs_1_5TeV}
		\includegraphics[width=0.26\textwidth]{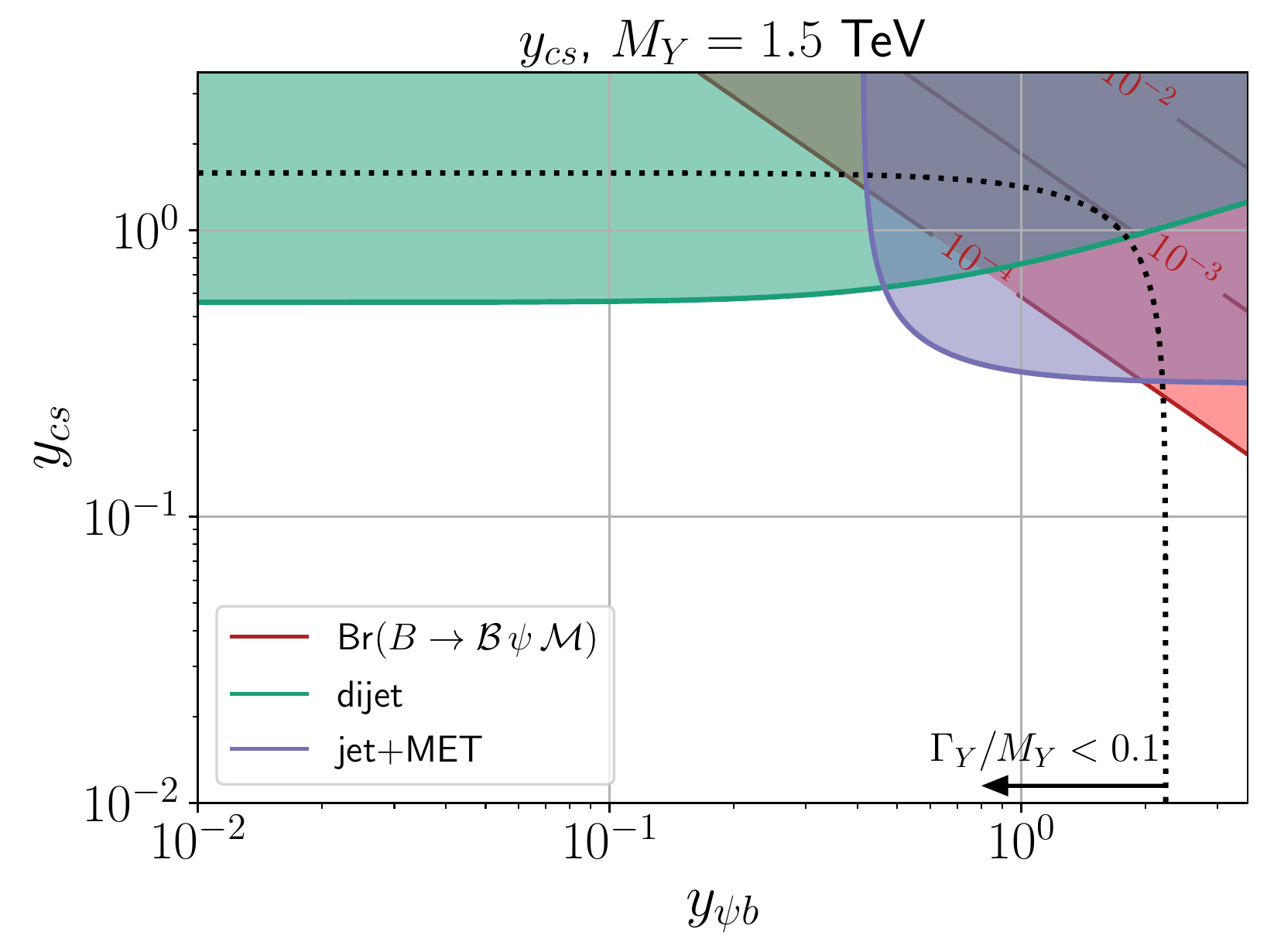}
&
		\label{fig:ycs_3TeV}
		\includegraphics[width=0.26\textwidth]{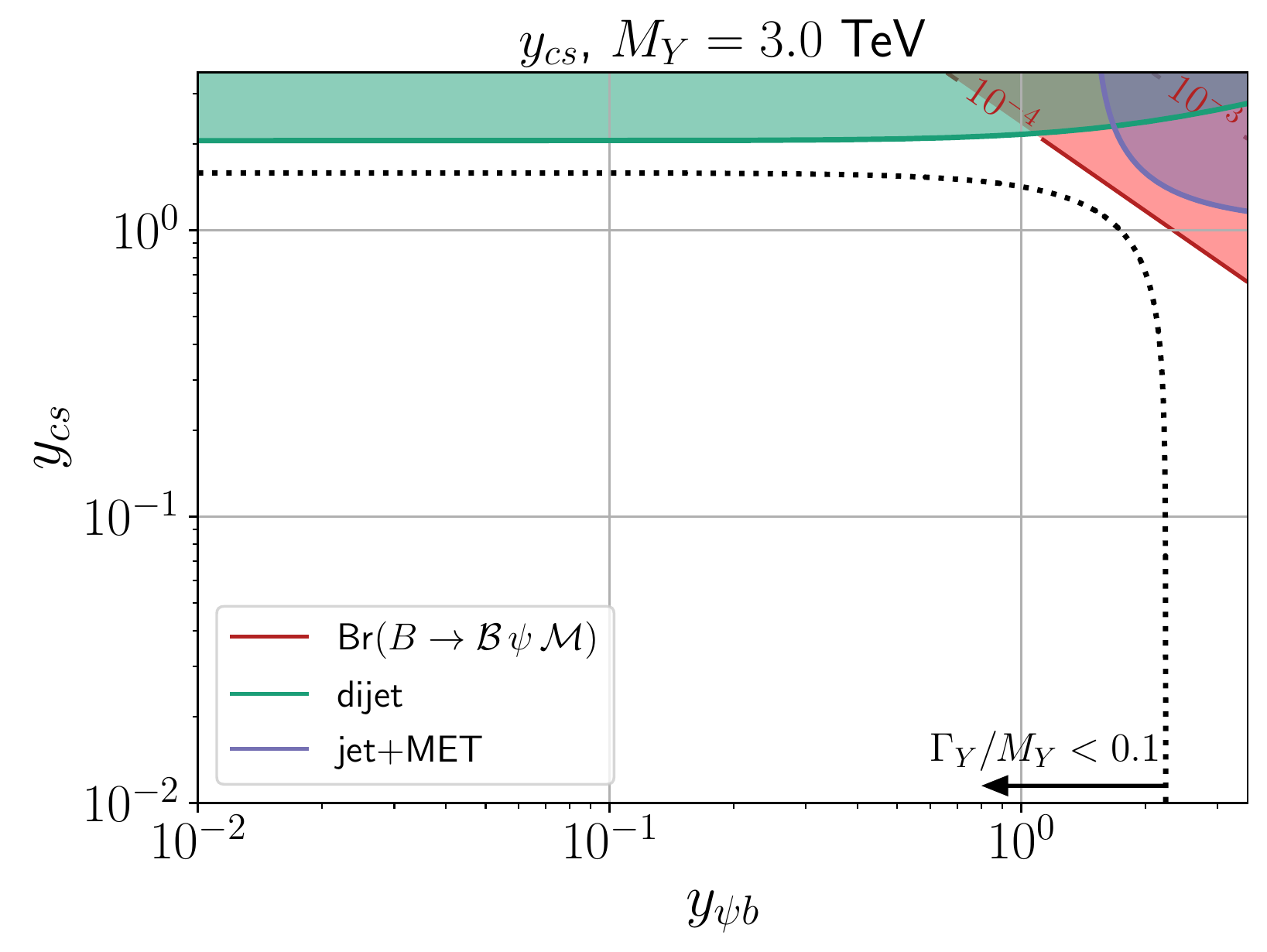}
&
		\label{fig:ycs_5TeV}
		\includegraphics[width=0.26\textwidth]{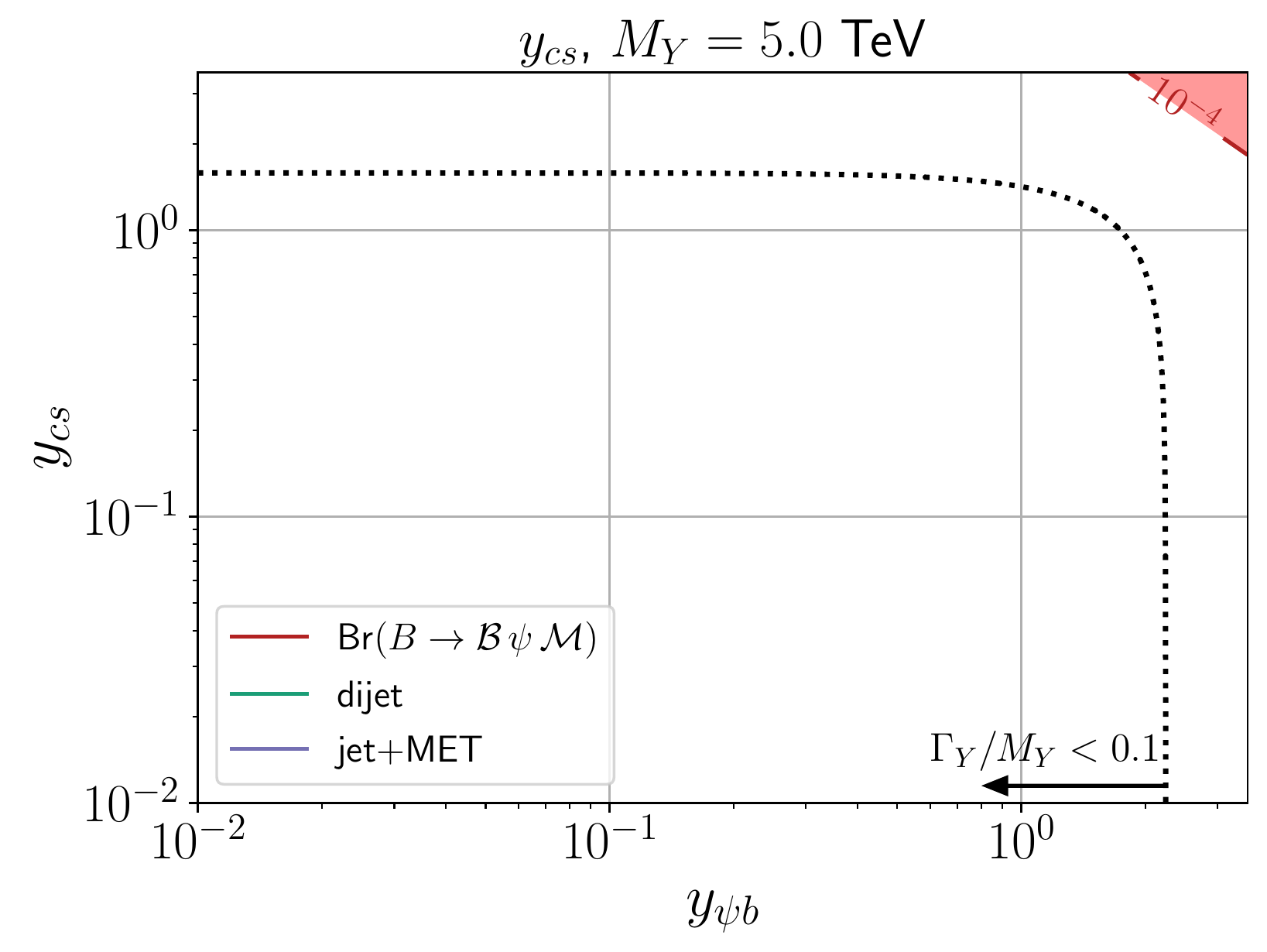}
\\
		\label{fig:ycb_1_5TeV}
		\includegraphics[width=0.26\textwidth]{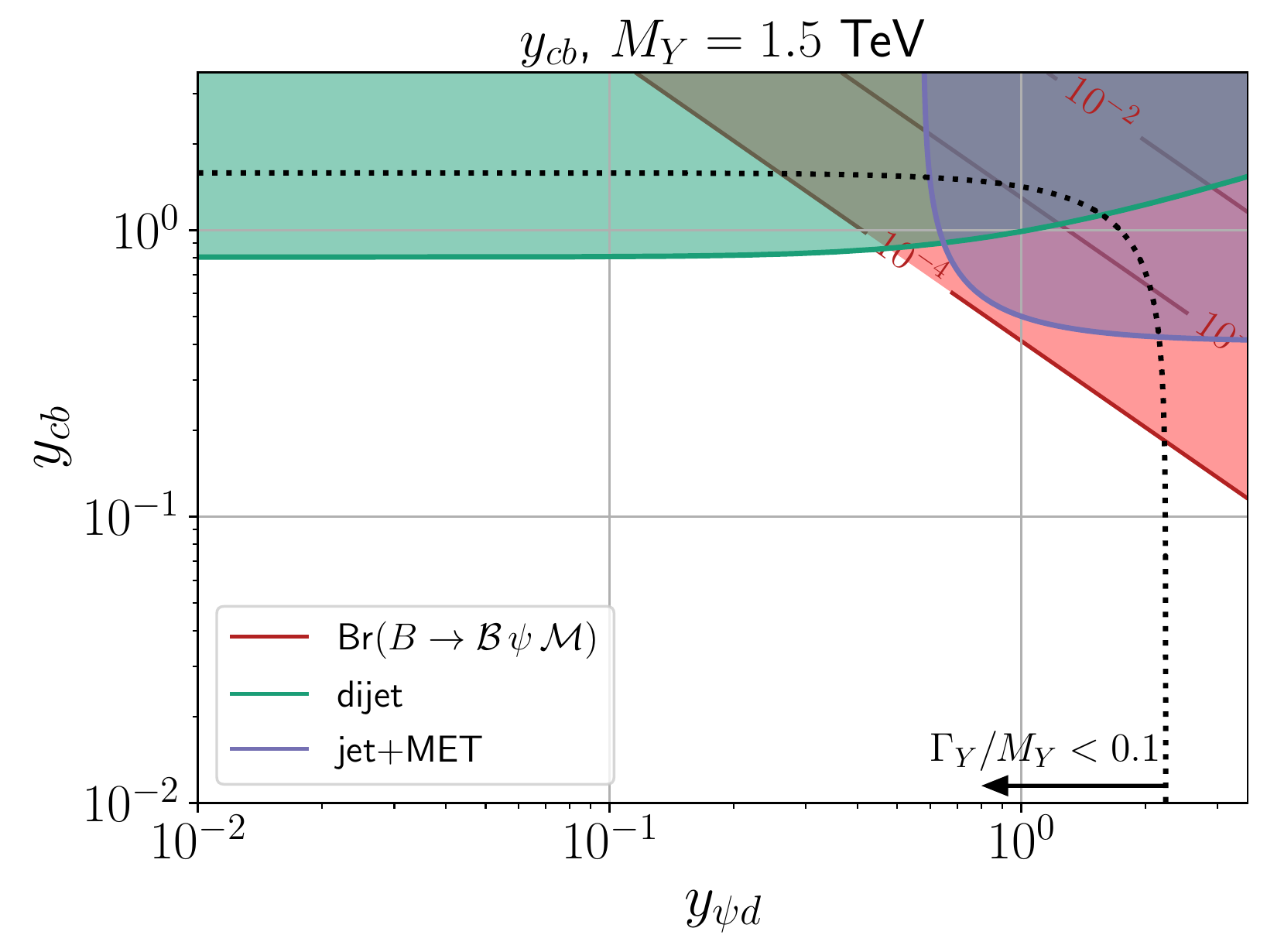}
&
		\label{fig:ycb_3TeV}
		\includegraphics[width=0.26\textwidth]{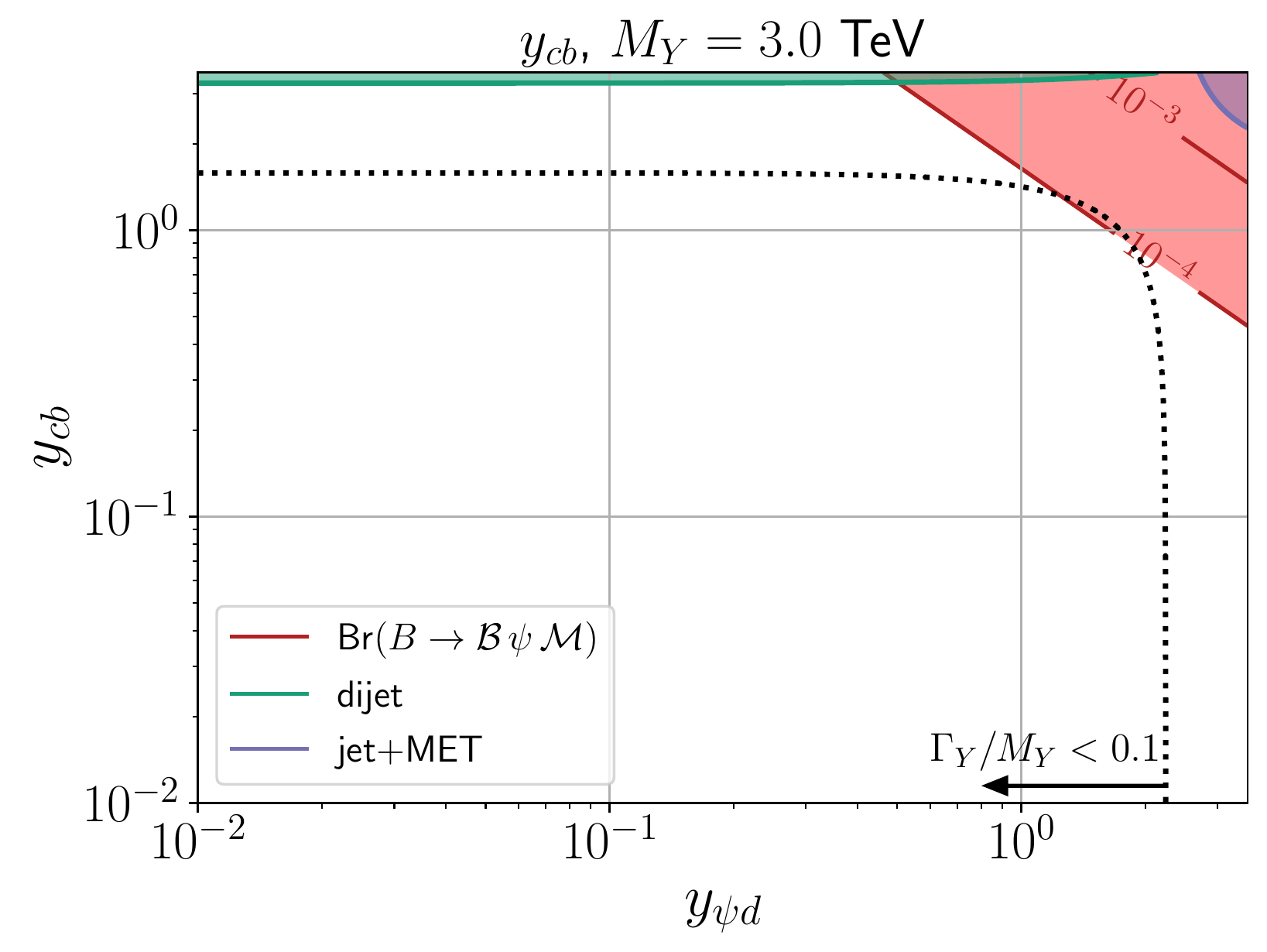}
&
		\label{fig:ycb_5TeV}
		\includegraphics[width=0.26\textwidth]{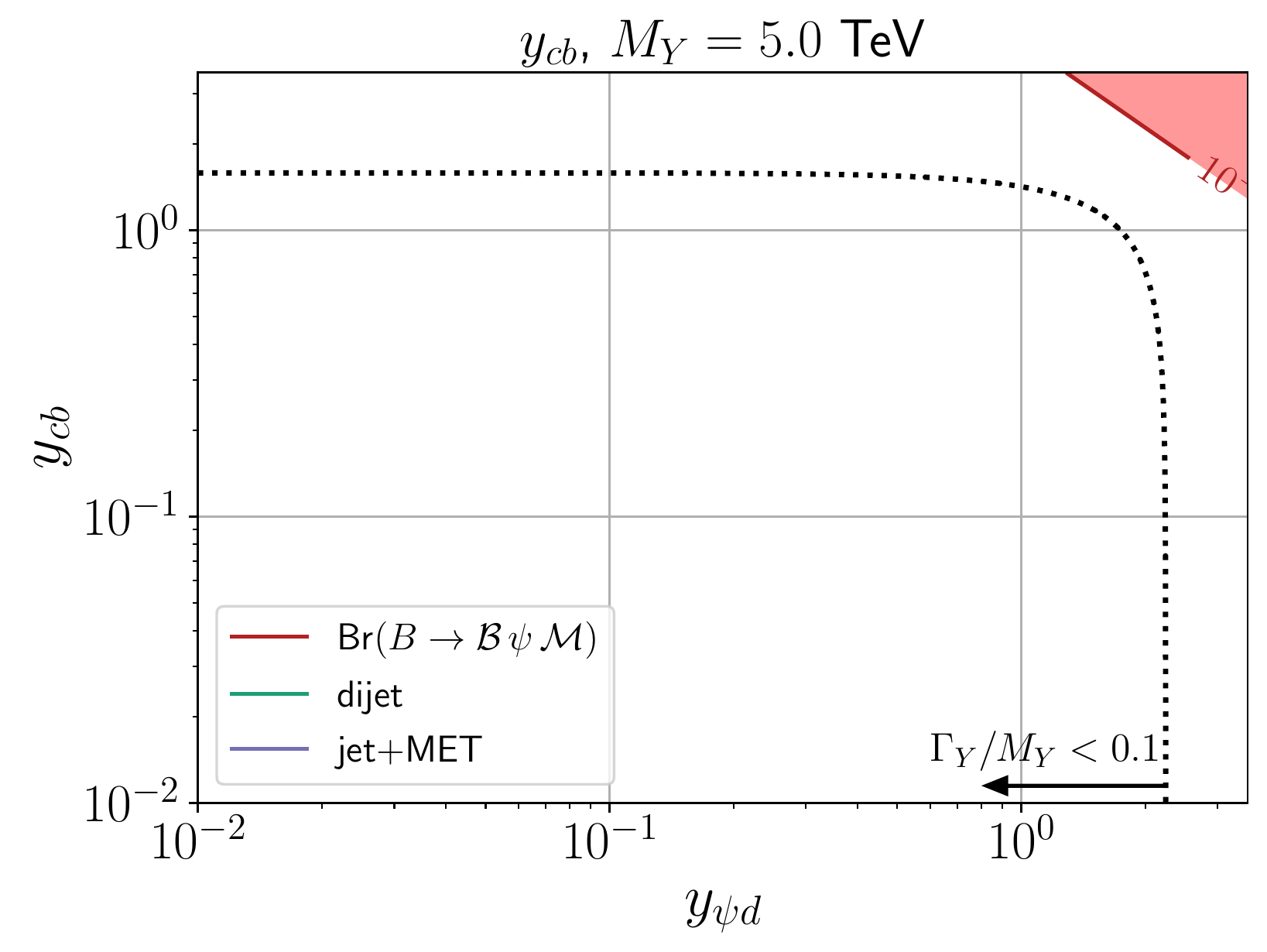}
\end{tabular}
\vspace{-0.2cm}
\caption{LHC limits on the couplings of the color-triplet scalar defined in Eq.~\eqref{eq:LflavorUV}. The green (blue) region is excluded by dijet (jet+MET) searches. The red band band corresponds to successful baryogenesis and encompasses $0.1> \br > 10^{-4}$ for $m_\psi = 1.5 $~GeV. In black dashed we show the region of parameter space in which $\Gamma_Y/M_Y = 0.1$. In blue dashed line we show the reach of the high luminosity LHC. We have restricted the couplings to be $<\sqrt{4\pi}$.
}
\label{fig:LHC_combined_plots}
\end{figure*}

\begin{figure*}[t]
\centering
\setlength{\tabcolsep}{10pt}
\renewcommand{\arraystretch}{1}
\begin{tabular}{ccc}
		\label{fig:ydb_ypu_1_5TeV}
		\includegraphics[width=0.26\textwidth]{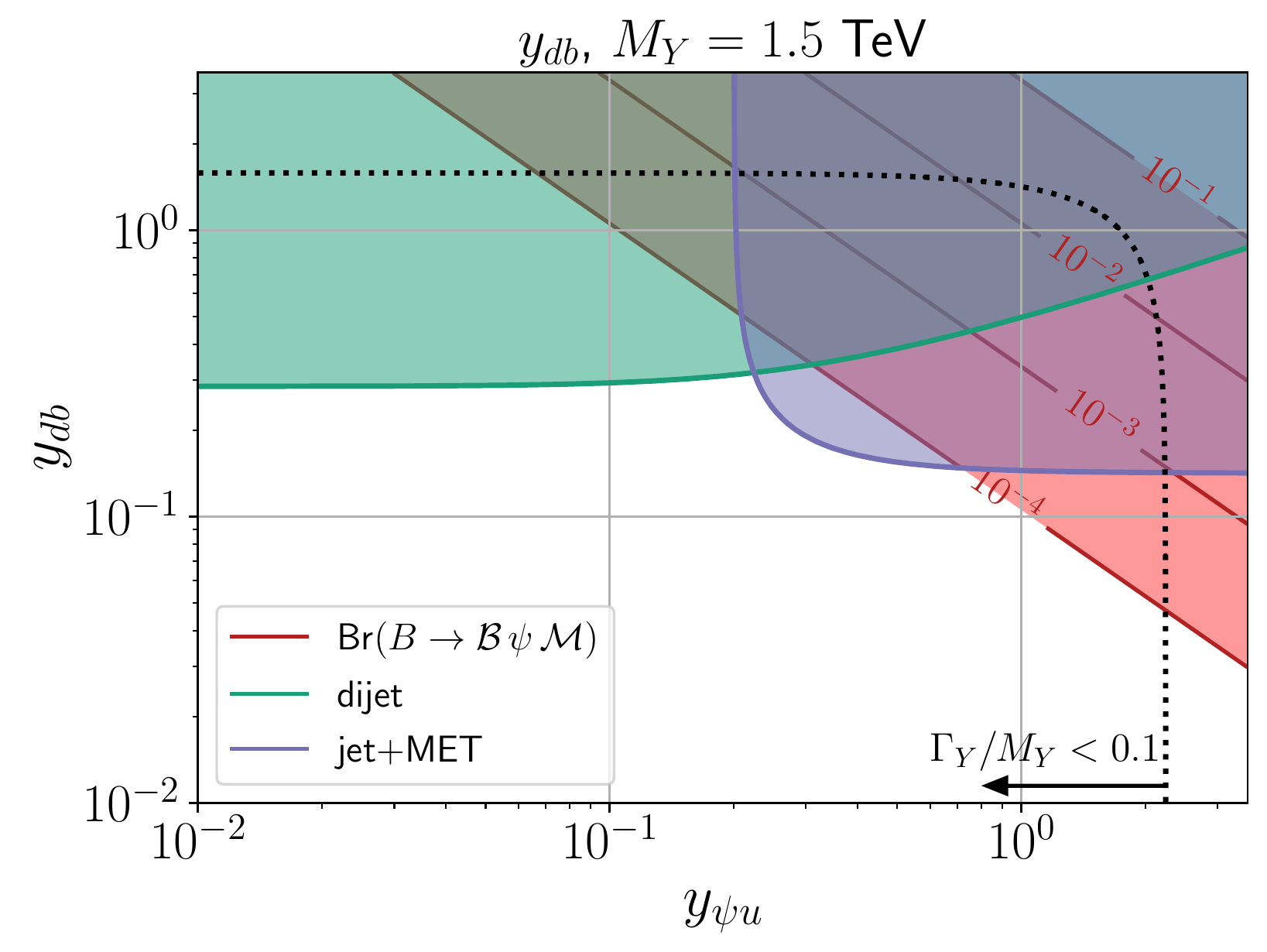}
&
		\label{fig:ydb_ypu_3TeV}
		\includegraphics[width=0.26\textwidth]{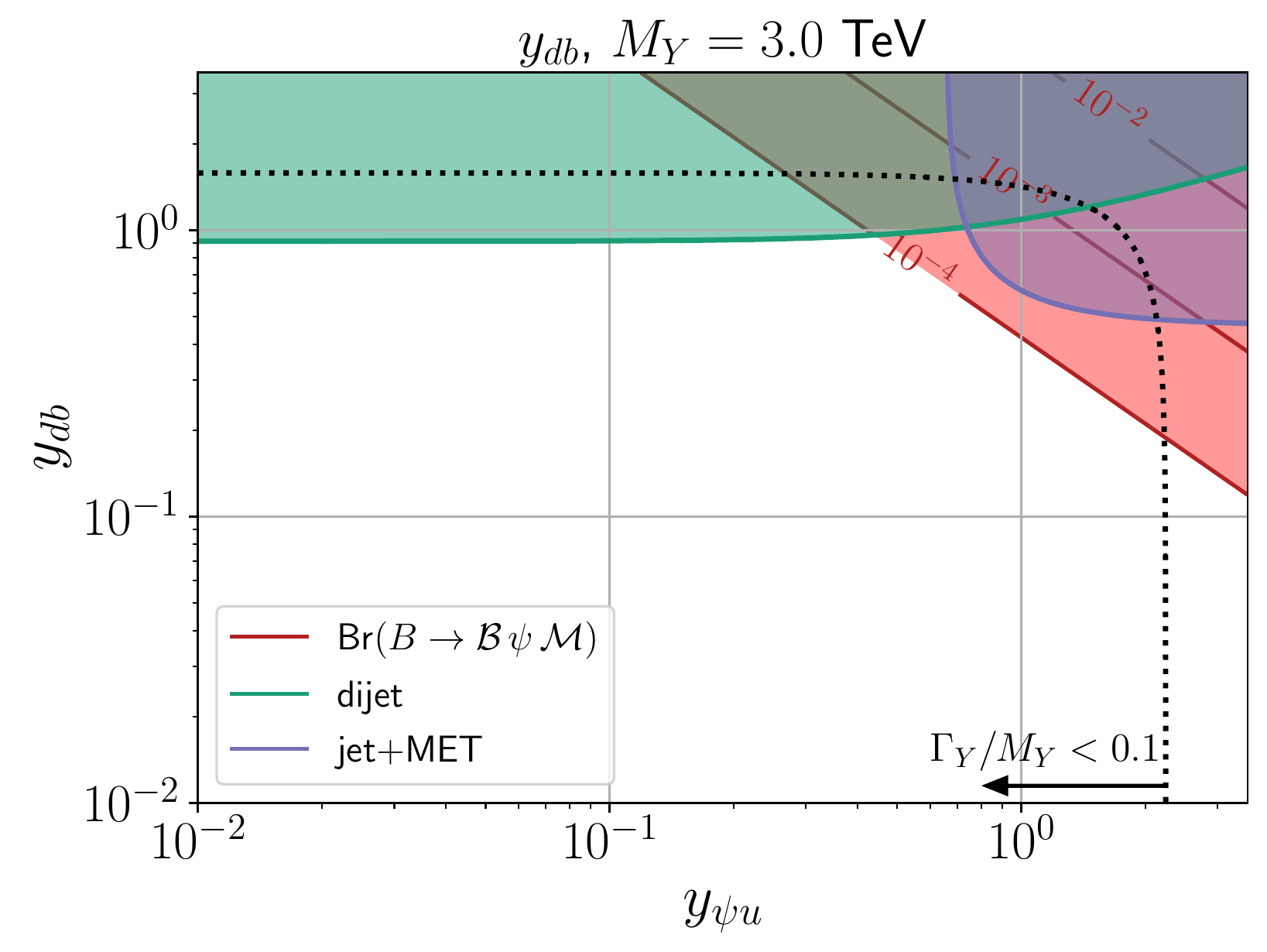}
&
		\label{fig:ydb_ypu_5TeV}
		\includegraphics[width=0.26\textwidth]{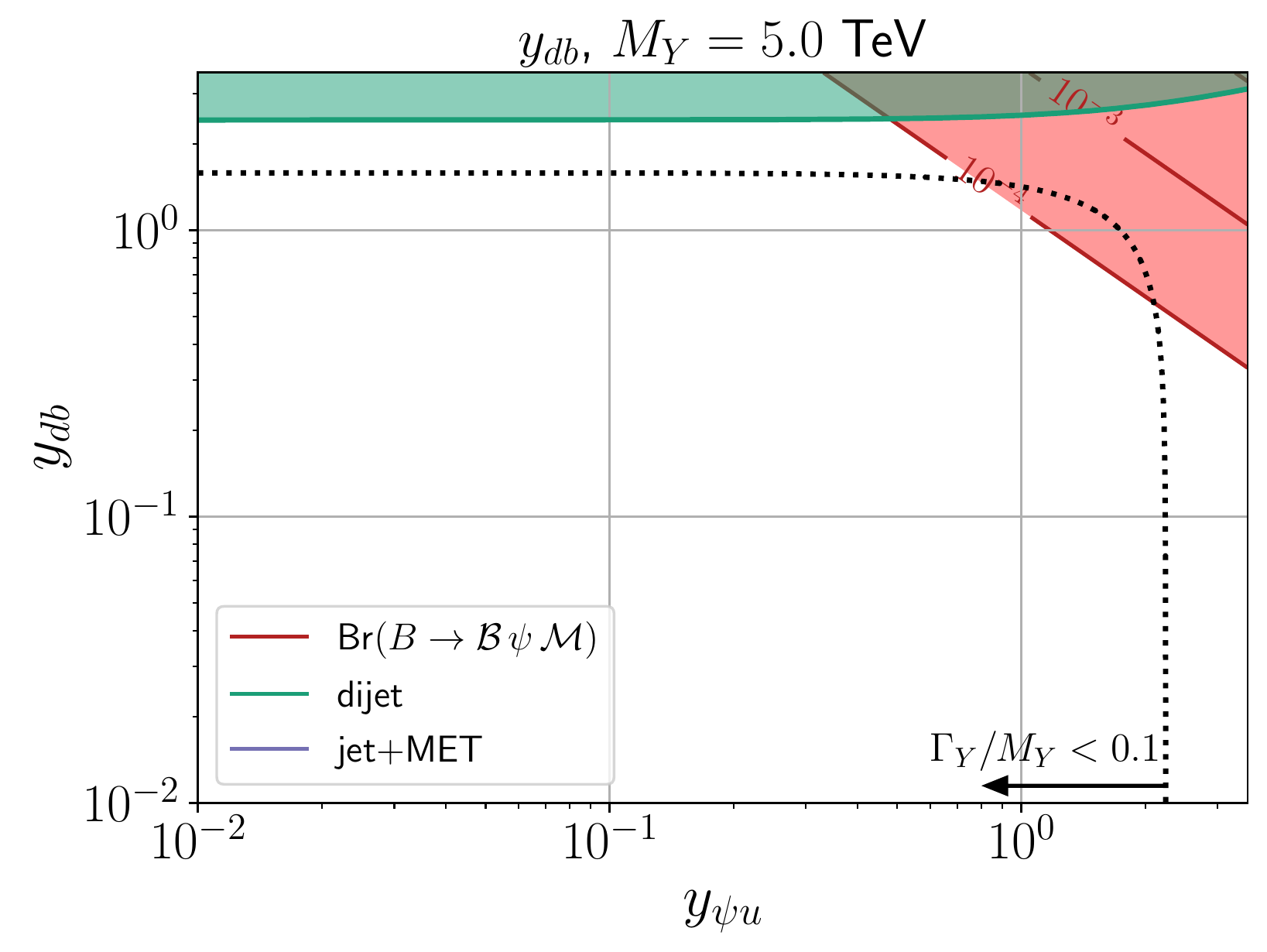}
\\
		\label{fig:ydb_ypc_1_5TeV}
		\includegraphics[width=0.26\textwidth]{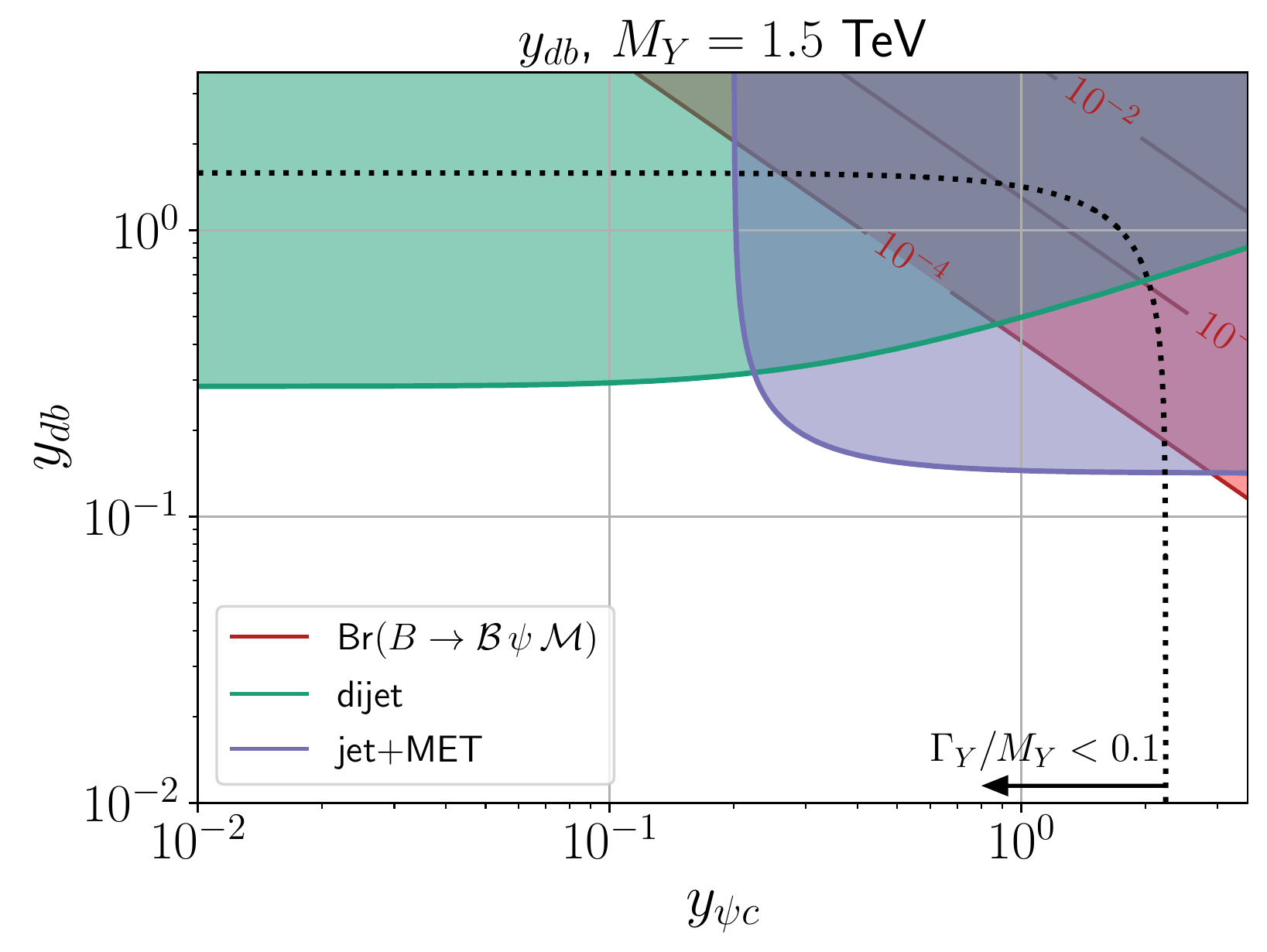}
&
		\label{fig:ydb_ypc_3TeV}
		\includegraphics[width=0.26\textwidth]{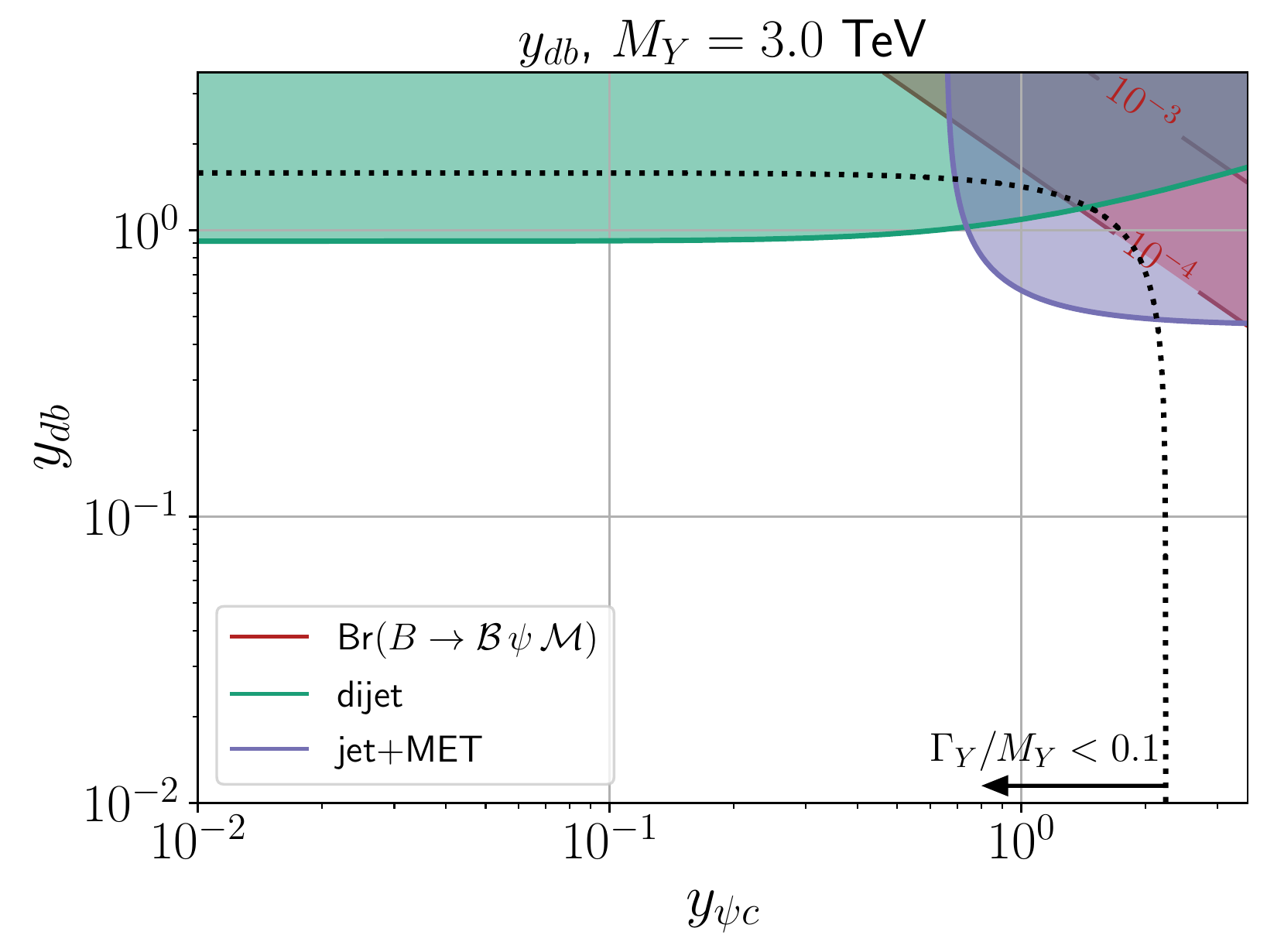}
&
		\label{fig:ydb_ypc_5TeV}
		\includegraphics[width=0.26\textwidth]{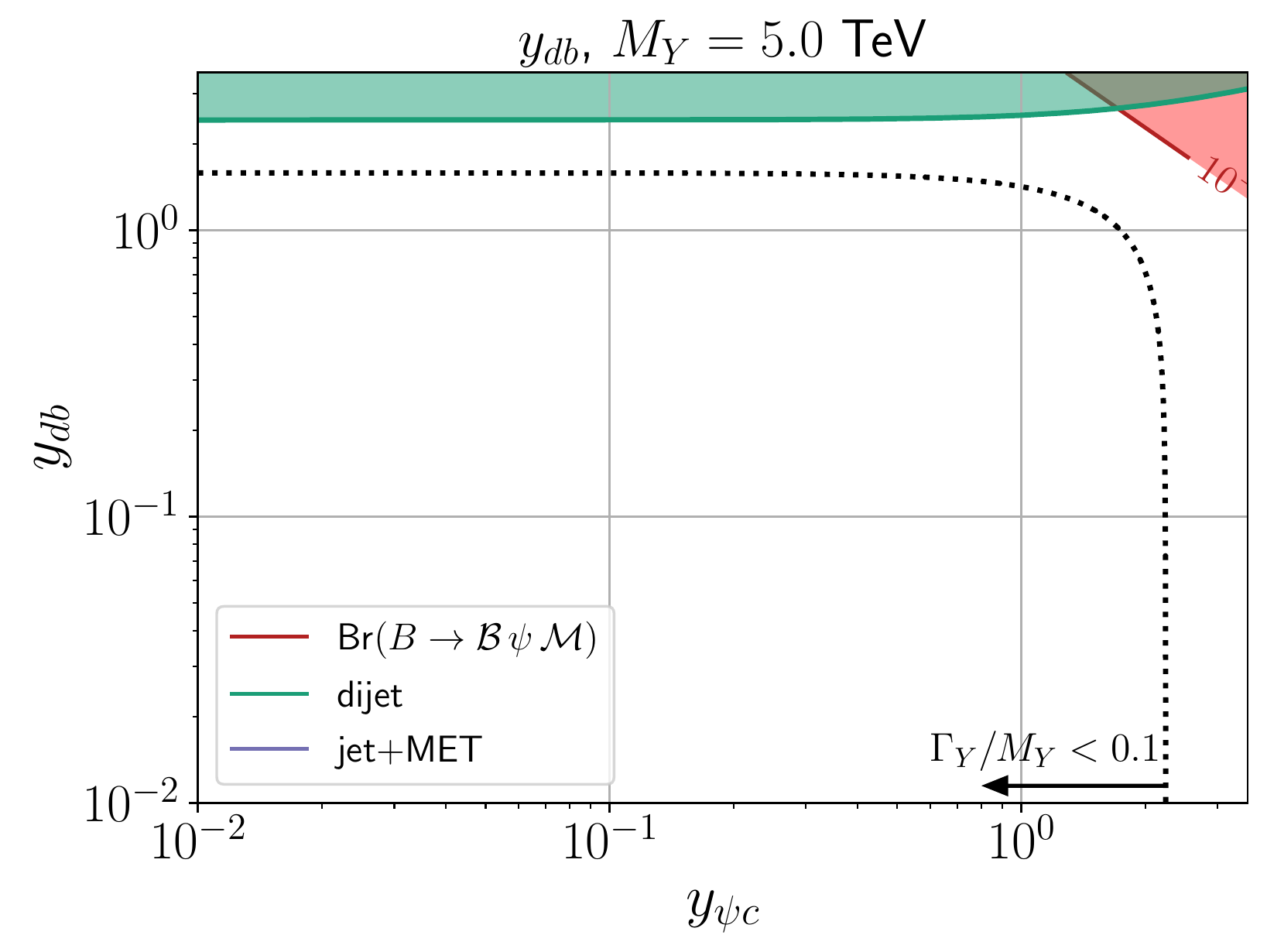}
\\
		\label{fig:ysb_ypu_1_5TeV}
		\includegraphics[width=0.26\textwidth]{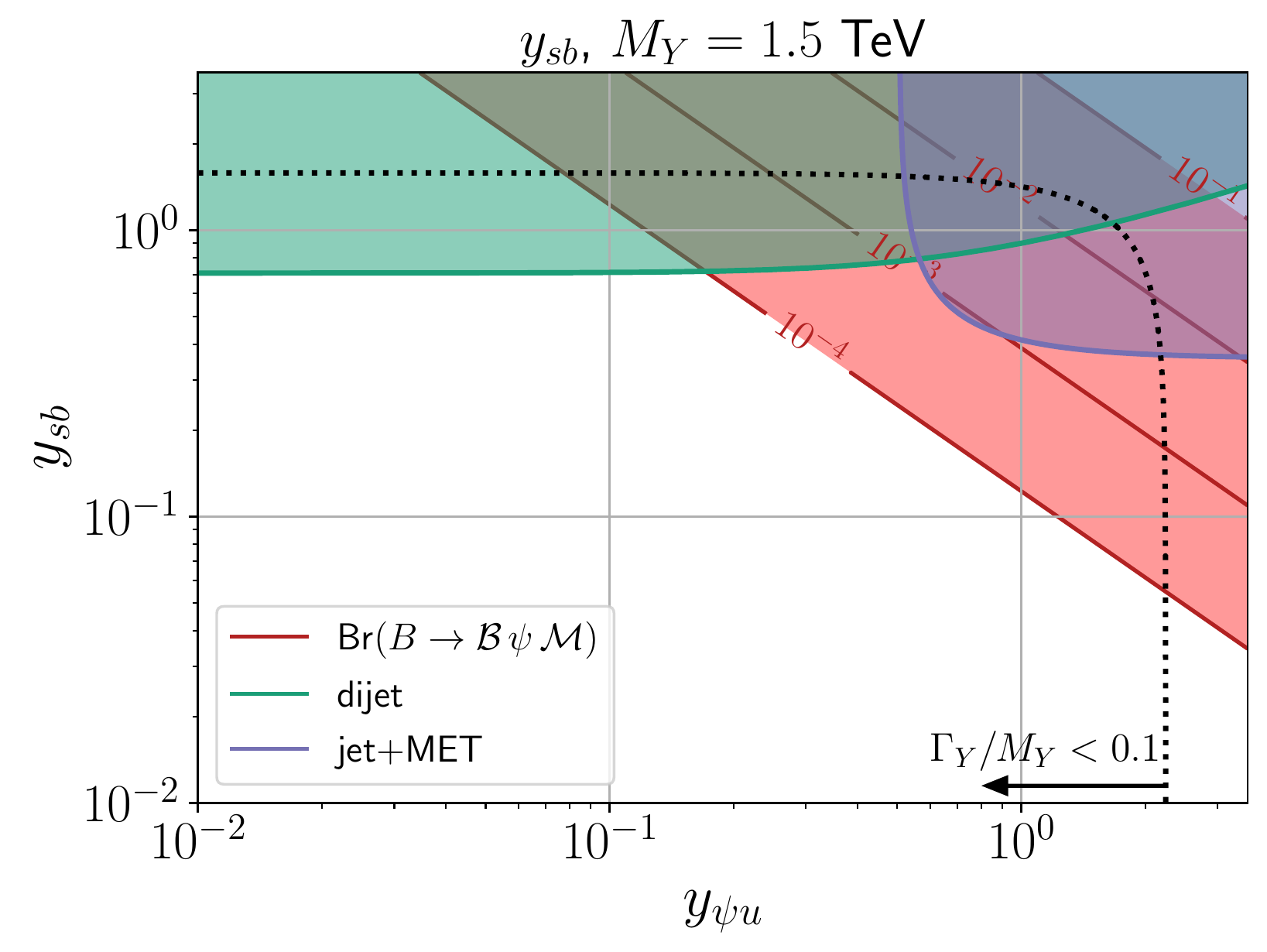}
&
		\label{fig:ysb_ypu_3TeV}
		\includegraphics[width=0.26\textwidth]{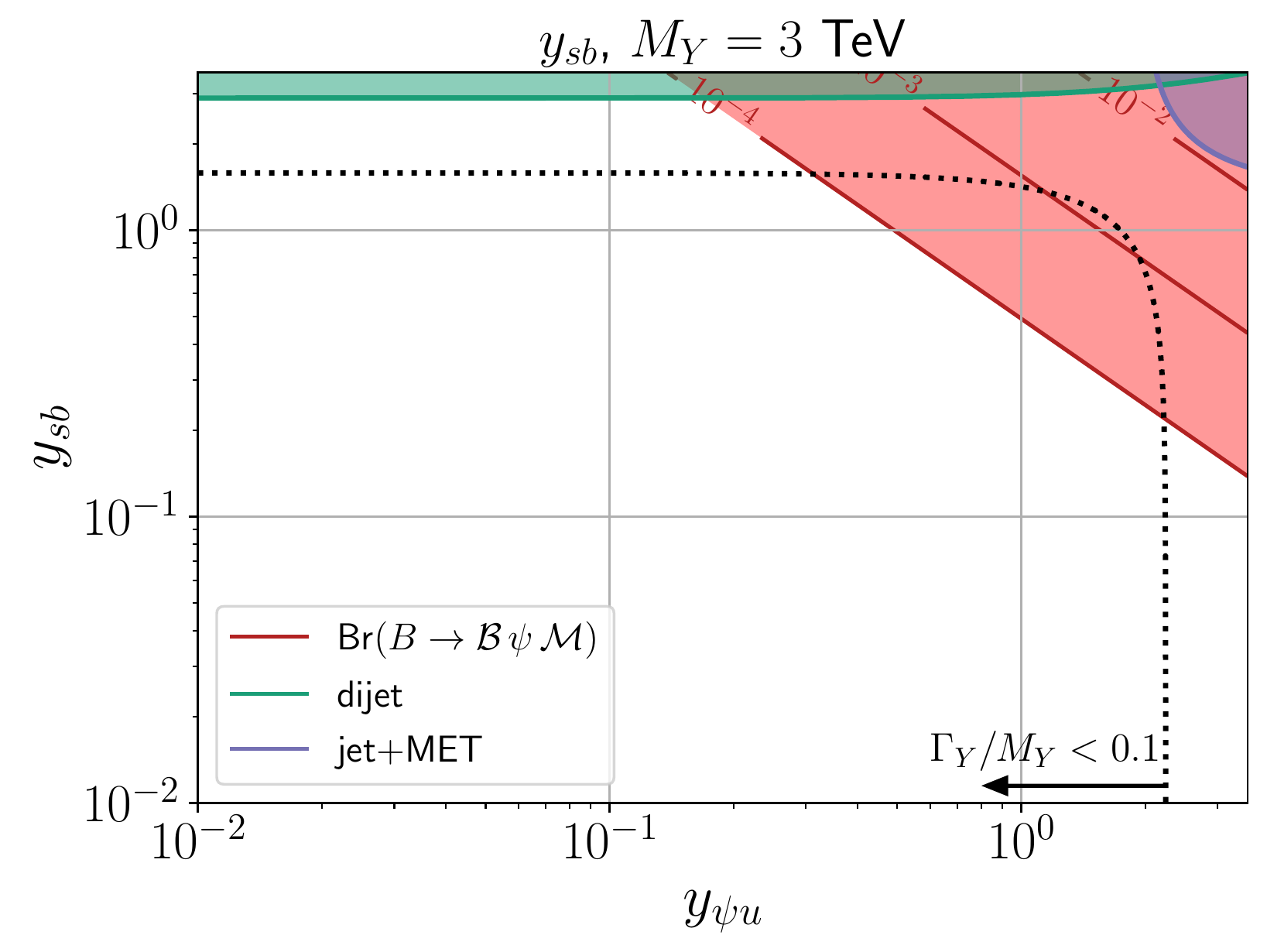}
&
		\label{fig:ysb_ypu_5TeV}
		\includegraphics[width=0.26\textwidth]{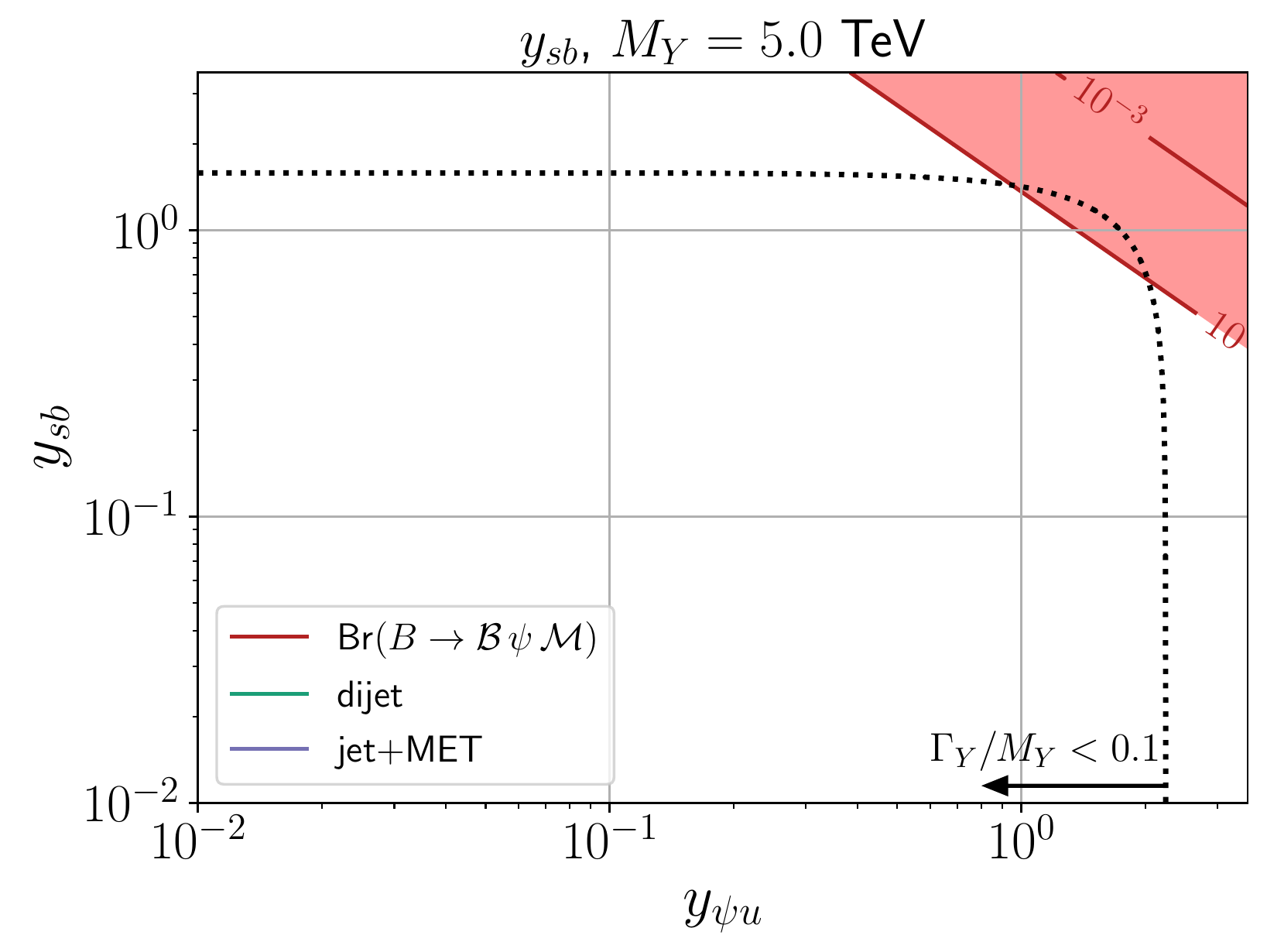}
\\
		\label{fig:ysb_ypc_1_5TeV}
		\includegraphics[width=0.26\textwidth]{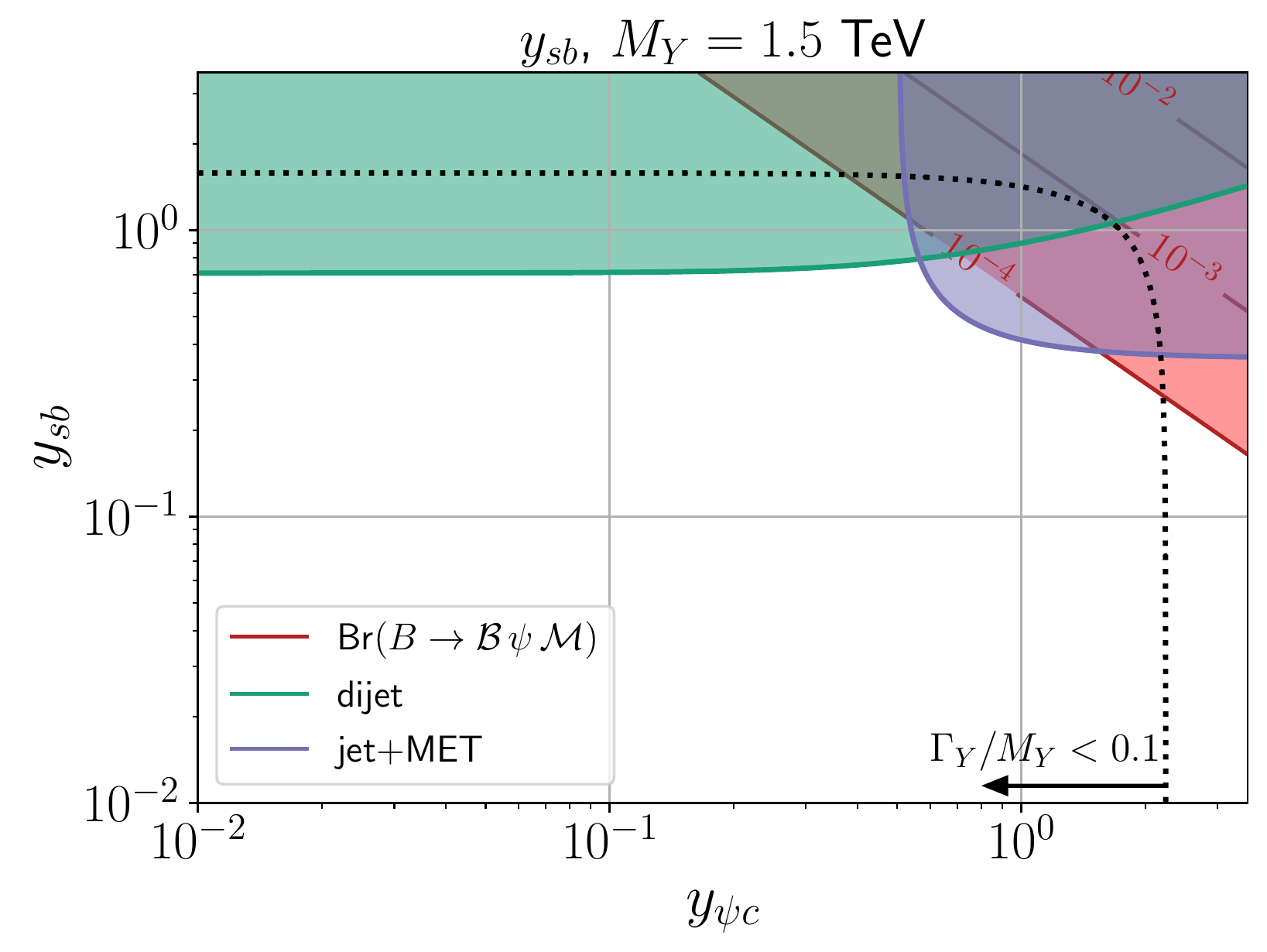}
&
		\label{fig:ysb_ypc_3TeV}
		\includegraphics[width=0.26\textwidth]{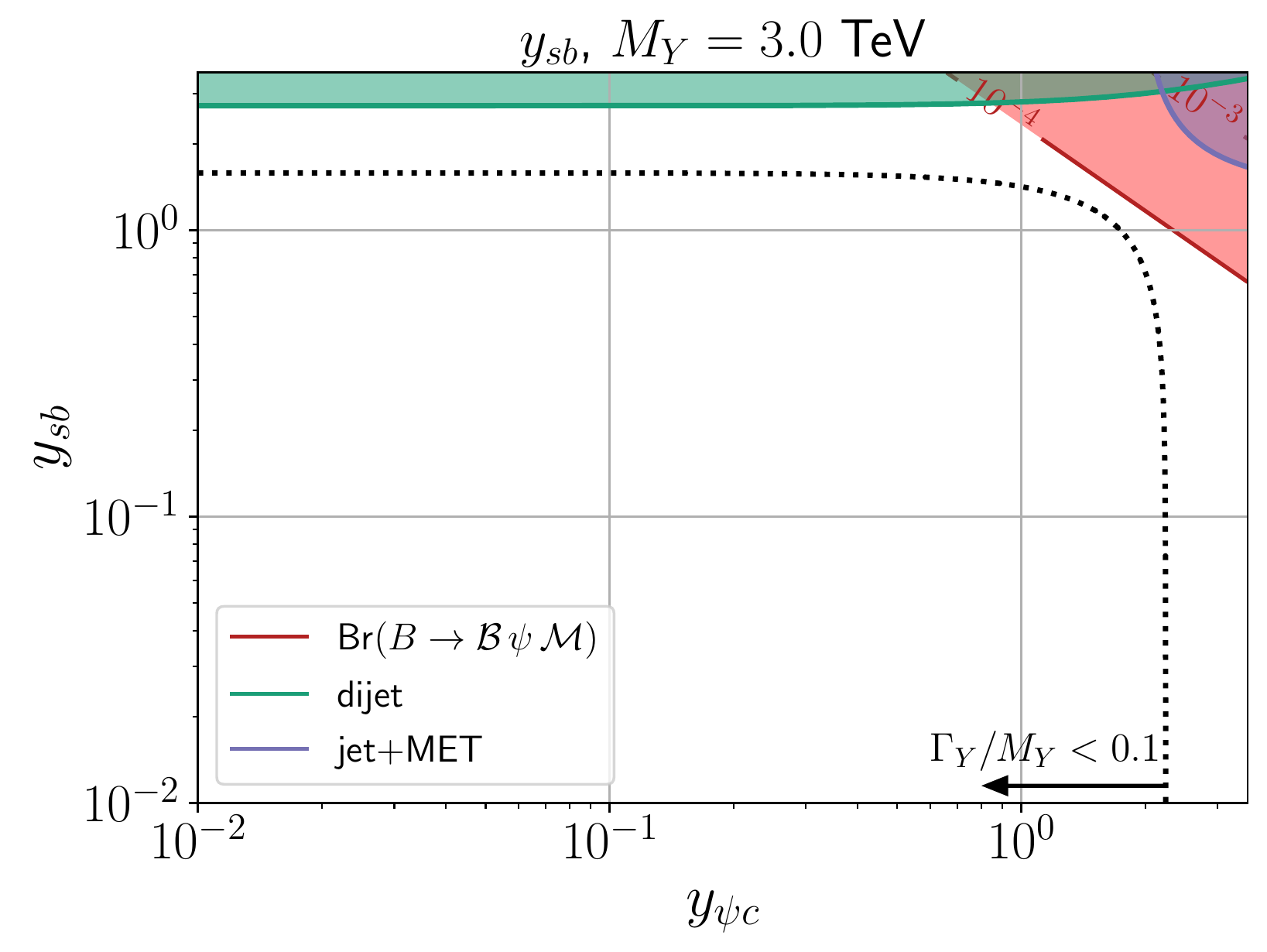}
&
		\label{fig:ysb_ypc_5TeV}
		\includegraphics[width=0.26\textwidth]{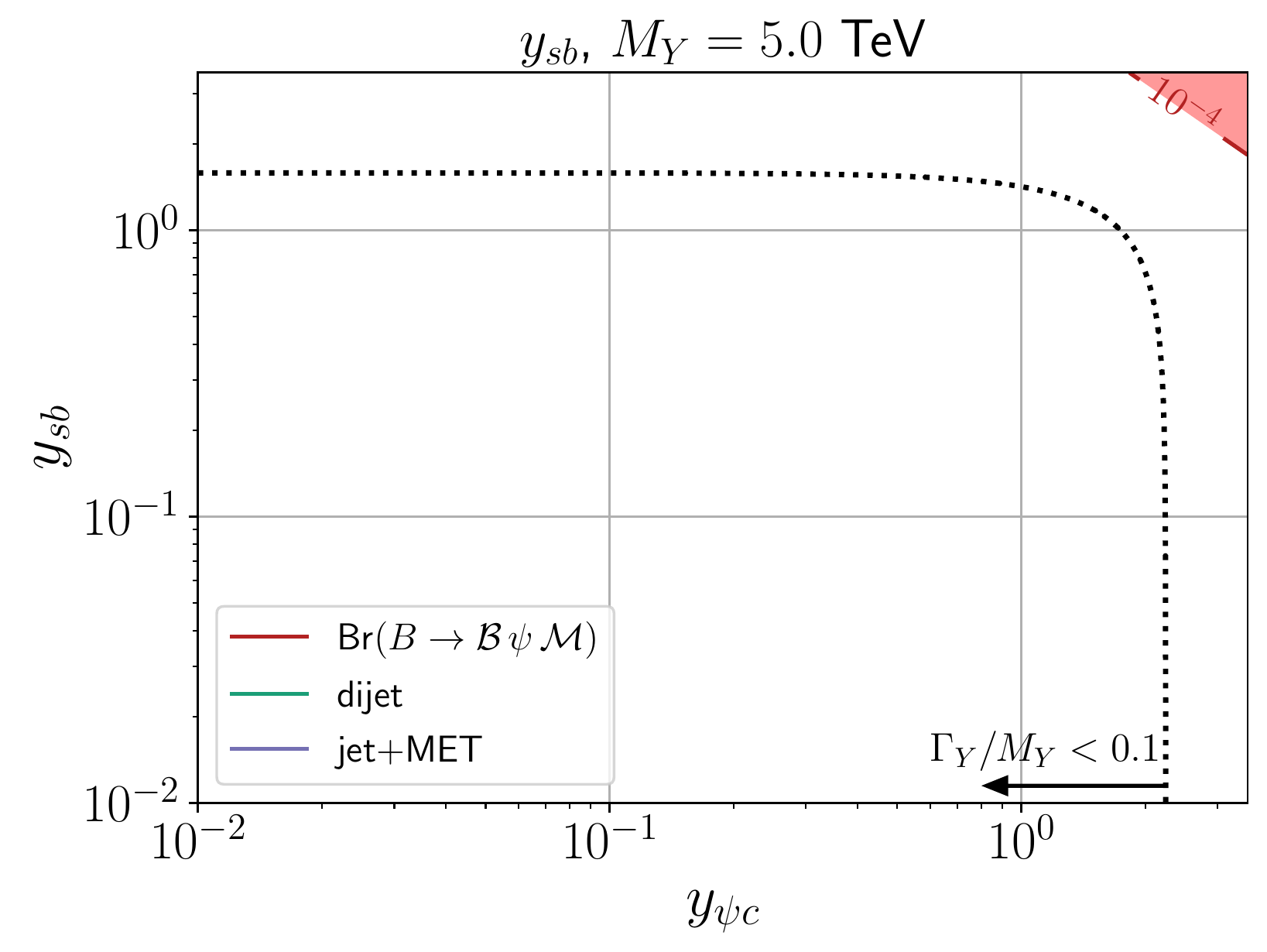}
\\
\end{tabular}
\vspace{-0.2cm}
\caption{Same as Fig.~\ref{fig:LHC_combined_plots} but for a color-triplet scalar with SM gauge quantum numbers $Y \sim (3,1,+2/3)$ described by Lagrangian~\eqref{eq:LflavorUV_23}.}
\label{fig:LHC_combined_plots_23}
\end{figure*}

For the hypercharge $-1/3$ case, the combined limits on the $y_{ud}$ vs. $y_{\psi d}$ parameter space are shown in Fig~\ref{fig:LHC_combined_plots} for three benchmark values of $M_Y=1.5,\,3,\,\mathrm{and}\, 5\,\mathrm{TeV}$ and each of the quark flavor variations in $y_{ud}$ (given that no flavor-tagging techniques are employed in the analyses, the limits are insensitive to the quark flavor of the $y_{\psi d}$ coupling).
Each of the panels in Fig.~\ref{fig:LHC_combined_plots} shows the dijet exclusion in green and the jet+MET limits in blue, and the thin dotted line indicates where the narrow width approximation $\Gamma_Y\ll M_Y$ is expected to break down.
The red band corresponds to the range in which $B$-Mesogenesis is possible.
The position of this band in each panel depends on the quark flavors involved: in the presence of a $c$ quark, the $B$ decays into a charmed baryon and $\Delta m$ in Eq.~\eqref{eq:Baryogenesis_range} becomes smaller.
As a result, the band shifts to larger couplings as compared to the cases involving only $u$, $d$, and $s$ quarks.
This is the reason why we distinguish between $y_{\psi u}$ and $y_{\psi c}$ in the hypercharge $2/3$ case shown in Fig.~\ref{fig:LHC_combined_plots_23}, where we plot the limits in the $y_{dd'}$ vs. $y_{\psi u}$ planes.

\subsection{New physics in neutral meson mixing}\label{sec:MesonMixing}
The mass difference of neutral mesons is modified by the presence of a color-triplet scalar with the couplings given in~\eqref{eq:LflavorUV}.
As a consequence, the flavor structure of the coupling constants $y_{u_id_j}$ / $y_{d_id_j}$ and $y_{\psi d_k}$ / $y_{\psi u_k}$ can be constrained with neutral meson mixing data.
The new contributions to the mass difference can be expressed as
\begin{equation}
\Delta M^{\mathrm{NP}} = 2\left( \sum_{i=1}^5 C_i M_i + \sum_{i=1}^3 \tilde{C}_i \tilde{M}_i \right),
\end{equation}
where we have split the contributions into a partonic amplitude $C_i$ and a hadronic matrix element $M_i$.
Our labelling follows the conventions of~\cite{Bagger:1997gg}.
The $Y$ particle can mediate the two new kinds of box diagrams depicted in Fig.~\ref{fig:box_diagrams}, which yield contributions to $\tilde{C}_1$, $C_4$, and $C_5$.
The corresponding matrix elements can be expressed as

\begin{align}
\tilde{M}^M_1 &= \frac{1}{3}f_M^2m_M B^M_1, \\ \nonumber
M^M_4 &= \left[\frac{1}{24} + \frac{1}{4}\left( \frac{m_M}{m_{q_1}+m_{q_2}} \right)^2 \right] f_M^2 m_M B^M_4 \, , \\ \nonumber
M^M_5 &= \left[\frac{1}{8} + \frac{1}{12}\left( \frac{m_M}{m_{q_1}+m_{q_2}} \right)^2 \right] f_M^2 m_M B^M_5 ,
\end{align}
In terms of the masses of the meson $M$ and its constituent quarks $q_1,\, q_2$, the decay constant $f_M$, and the bag parameters $B^M_i$.
The partonic amplitudes for the diagrams involving internal SM quarks read
\begin{align}
\tilde{C}_1 &= \frac{1}{128\pi^2} \frac{1}{m_Y^2} \sum_{i,j} y_{q_iq_1}^* y_{q_iq_2} y_{q_jq_1}^* y_{q_jq_2} \tilde{S}_1(x_{q_i},x_{q_j}),\\
C_4 &= -C_5,\\
C_5 &= \frac{1}{128\pi^2} \frac{g^2}{m_Y^2} \sum_{i,j} y_{q_iq_2} y_{q_jq_1}^*V_{q_jq_1} V_{q_iq_2}^* \tilde{S}_5(x_{q_i},x_{q_j}),
\end{align}
while the one with the internal dark fermion $\psi$ results in
\begin{align}
\tilde{C}_1 &= \frac{1}{128\pi^2} \frac{1}{m_Y^2} \left( y_{\psi q_1}^* y_{\psi q_2}\right)^2 \tilde{S}_1(x_{\psi},x_{\psi}).
\end{align}
Here, $x_{f} = m^2_{f} / m^2_Y$ and we have defined
\begin{align}
\tilde{S}_1(x_{q_i},x_{q_j}) &= I_4(x_{q_i},x_{q_j},1,1), \\
\tilde{S}_5(x_{q_i},x_{q_j}) &= \sqrt{x_{q_i} x_{q_j}} \Big( I_2(x_{q_i},x_{q_j},x_W,1)  \nonumber \\ & \qquad\, + \frac{1}{4x_W} I_4(x_{q_i},x_{q_j},x_W,1)  \Big),
\end{align}
with the loop integrals
\begin{equation}
I_n (x,y,w,z) = \int_{0}^{\infty} \frac{t^{n/2} \diff t}{(t+x)(t+y)(t+w)(t+z)}.
\end{equation}

\bibliography{biblio}

\end{document}